\documentclass[twocolumn, amssymb,nobibnotes,aps,prb,nopacs]{revtex4}

\usepackage{amsmath}
\usepackage{graphicx}
\usepackage{subfig}
\newcommand \dd[1]  { \,\textrm d{#1} }   
\usepackage[justification=RaggedRight,font=footnotesize]{caption}
\usepackage{multirow}

\begin{document}
\title{Fully kinetic simulation study of ion-acoustic solitons \\ in the presence of trapped electrons}
\author{S. M. Hosseini Jenab\footnote{Email: Mehdi.Jenab@nwu.ac.za}}
\author{F. Spanier \footnote{Email: Felix@fspanier.de}}
\affiliation{Centre for Space Research, North-West University, Potchefstroom Campus, Private Bag X6001, Potchefstroom 2520, South Africa}
\date{\today}

\begin{abstract}
The nonlinear fluid theory developed by Schamel
suggests a modified KdV equation to describe the temporal evolution of ion acoustic (IA) solitons
in the presence of trapped electrons.
The validity of this theory is studied here
by verifying solitons' main characteristic,
i.e., stability against successive mutual collisions.
We have employed a kinetic model as a more comprehensive theory than the fluid one, 
and utilized a fully kinetic simulation approach (both ions and electrons are treated based on the Vlasov equation).
In the simulation approach, these solitons are excited self-consistently 
by employing the nonlinear process of IA solitons formation from an initial density perturbation (IDP).
The effect of the size of IDPs on the chain formation is proved by the simulation code as a benchmark test.
It is shown that the IA solitons,
in presence of trapped electrons,
can retain their features (both in spatial and velocity direction) after successive mutual collisions.
The collisions here include encounters of IA solitons with the same trapping parameter, while differing in size.
Kinetic simulation results reveal a complicated behavior 
during a collision between IA solitons in contrast to the fluid theory predictions and simulations.
In the range of parameters considered here
two oppositely propagating solitons rotate around their collective center in the phase space during a collision,
independent of their trapping parameters. 
Furthermore, they exchange some portions of their trapped populations.
\end{abstract}
\maketitle

\section{Introduction} \label{Sec_Introduction}
The ion-acoustic (IA) solitons were first discovered in the context of nonlinear fluid theory by Washimi and Taniuti \cite{Washimi1966996}. 
These nonlinear modes, localized structures, possess two characteristics:
\begin{itemize}
 \item propagation without change in their features such as velocity, shape and size (e.g. width and height)
 \item stability against (theoretically infinite number of) mutual collisions.
\end{itemize}
The IA solitons have been observed both in laboratory experiments \cite{ikezi1970formation} 
and in a wide range of space plasma observations
\cite{pickett2003solitary,hobara2008cluster,pickett2004solitary}. 
Fluid theory predicts the existence condition for solitons called nonlinear dispersion relation (NDR) which is,
in fact, 
a sensitive relationship among different features of solitons, e.g. velocity and size. 
Any localized structure with values other than what is dictated by NDR,
is expected to break into N-solitons in a long-term evolution.
The existence of N-soliton solution for the KdV equation has been proven with different mathematical approaches 
starting by the seminal work of Hirota \cite{Hirota1971,Hirota1972}.
In this paper, by harnessing this phenomenon, i.e. chain formation, 
self-consistent IA solitons are produced from an initial density perturbation (IDP).

Schamel\cite{schamel_3,schamel_4,schamel_5} has developed a modified KdV (mKdV) equation, 
by extending the work of Washimi and Taniuti \cite{Washimi1966996} to include the trapping effect of electrons.
Based on $\beta$ (\textit{trapping parameter}), the distribution function of trapped electrons can take three different types of shapes, 
namely \textit{hollow} ($\beta<0$), \textit{plateau} ($\beta = 0$) and \textit{hump} ($\beta>0$).
Schamel has identified three regimes considering the trapping effect \cite{schamel_3}. 
For $\beta = 1$, the KdV regime recovers from mKdV solutions. 
For $\beta_c<\beta < 1 $ and $\beta<\beta_c$, two modified KdV regimes are proposed with their own distinctive IA solitons, namely Schamel-KdV and Schamel respectively.
$\beta_c$ depends on the amplitude of the IA soliton and stays below $\beta_c<1.0$.

This study is an attempt to verify Schamel's theory and its prediction about IA solitons in presence of trapped electrons, 
based on a fully kinetic simulation approach for the first time
(both electrons and ions dynamics are treated by the Vlasov equation).
The second characteristic of (self-consistently excited) IA solitons
is addressed for different values of $\beta$.
The trapping parameter ($\beta$) ranges from negative to positive values,
covering all the three possible shapes of the distribution function of trapped electrons. 
All the three regimes suggested by Schamel,
are also examined in the chosen range of the trapping parameter. 
Collisions of the IA solitons are analyzed in both spatial and velocity directions
focusing on number density profiles and distribution functions respectively.
However, the study is limited to collisions of IA solitons with the same trapping parameter.

There have been a few simulation studies considering IA solitons. 
Most of these simulations have utilized fluid-based simulations (either KdV or fully fluid)\cite{Kakad2013,Sharma2015},
which can't include the trapping effect accurately. 
In case of hybrid-PIC simulations,
the trapping effect of electrons are mostly ignored by assuming electrons as a Boltzmann's fluid\cite{Qi20153815}. 
Kakad \textit{et al.}\cite{Kakad20145589} have considered the trapping effect in their PIC simulations.
However, they have not studied the trapping effect systematically in the range as wide as the one reported here.
Furthermore, 
the inherit noise in PIC smooths out the details of the distribution function of trapped electrons, 
destroying trapping effect.

The self-consistent approach to create IA solitons,
i.e. chain formation,
is tested for small and large amplitude IDPs in Sec.\ref{size_effect}.
Predictions and simulation results reported by fluid or PIC method are verified\cite{Kakad2013,Kakad20145589,Sharma2015,Qi20153815} as a benchmark test. 
Stability of the IA solitons 
in presence of trapped electrons,
against collisions is addressed in Sec.\ref{section_stability}. 
The kinetic details of the collisions for three different shapes of the trapped electrons distribution functions
are presented in Sec.\ref{section_process}.

\section{Basic Equations and Numerical Scheme} \label{B_equations}
All variables and quantities used in the rest of the text, 
are normalized to dimensionless forms to simplify the equations.
Space and time are normalized by $\lambda_{Di}$ and $\omega^{-1}_{pi}$ respectively,
where $\omega_{pi}  = \sqrt{n_{i0} e^2 /(m_i \epsilon_0) }$ denotes the ion plasma frequency
and $\lambda_{Di} = \sqrt{\epsilon_0 K_B T_i /(n_{i0} e^2) }$ is the characteristic
ion Debye length.
The velocity variable $v$ has been scaled by
the ion thermal speed $v_{th_i} = \sqrt{K_B T_i/m_i}$,
while the electric
field and the electric potential have been reduced by $K_B T_i/(e \lambda_{Di})$
and $K_B T_i/e$ respectively (here, $K_B$ is Boltzmann's constant).
The densities of the two species are normalized by $n_{i0}$, while energy is scaled by $K_B T_i$.
In order to introduce an IDP into the simulation domain, 
the so-called Schamel distribution function\cite{schamel_1,schamel_2} is used as follow:
\begingroup\makeatletter\def\f@size{8.3}\check@mathfonts
\def\maketag@@@#1{\hbox{\m@th\large\normalfont#1}}%
\begin{equation*}
f_{s}(v) =  
  \left\{\begin{array}{lr}
     A \ exp \Big[- \big(\sqrt{\frac{\xi_s}{2}} v_0 + \sqrt{E(v)} \big)^2 \Big]   &\textrm{if}
      \left\{\begin{array}{lr}
      v<v_0 - \sqrt{\frac{2E_{\phi}}{m_s}}\\
      v>v_0+\sqrt{\frac{2E_{\phi}}{m_s}} 
      \end{array}\right. \\
     A \ exp \Big[- \big(\frac{\xi_s}{2} v_0^2 + \beta_s E(v) \big) \Big] &\textrm{if}  
     \left\{\begin{array}{lr}
      v>v_0-\sqrt{\frac{2E_{\phi}}{m_s}} \\
      v<v_0 + \sqrt{\frac{2E_{\phi}}{m_s}} 
      \end{array}\right.
\end{array}\right.
\label{Schamel_Dif}
\end{equation*}\endgroup
in which $A = \sqrt{ \frac{\xi_s}{2 \pi}} n_{0s}$,
and $\xi_s = \frac{m_s}{T_s}$ are amplitude and the normalization factor respectively.
$E(v) = \frac{\xi_s}{2}(v-v_0)^2 + \phi\frac{1}{T_s q_s}$ represents the (normalized) energy of particles.
$v_0$ stands for the velocity of the IA soliton.
In the set of the simulations presented here,
this distribution function has been used to introduce a stationary IDP ($v_0 =0$) at $x_0$:
\begin{equation}
 \phi = \psi \ exp (\frac{x-x_0}{\Delta})^2.
\end{equation}
$\psi$ and  $\Delta$ are the amplitude and width of the stationary IDP respectively.
It is proven that this distribution function 
satisfies the continuity and positiveness conditions
while producing a trapped population in its phase space \cite{schamel_1,schamel_2}.

The simulation method,
employed here,
has been developed by the authors based on the method called \textit{Vlasov-Hybrid Simulation}(VHS),
which was initially proposed by Nunn \cite{nunn1993novel} 
(for details see \cite{jenab2014vlasov,jenab2014multicomponent,jenab2011preventing}).
It follows the trajectories of the 
so called \textit{phase points} \cite{kazeminezhad2003vlasov} in the phase space,
depending on Liouville's theorem as the theoretical framework.
It meets the condition of positiveness of the distribution function during temporal evolution perfectly.
Preserving entropy $(\int f  \ln f  \dd v \dd x)$
and energy stands as one of the major advantage of the method.
In simulations presented in this paper,
each plasma species (i.e. electrons and ions)
is described by the (scaled) Vlasov equation:
\begin{multline}
\frac{\partial f_s(x,v,t)}{\partial t} 
+ v \frac{\partial f_s(x,v,t)}{\partial x} 
\\ +  \frac{q_s}{m_s} E(x,t) \frac{\partial f_s(x,v,t)}{\partial v} 
= 0, \ \ \  s = i,e
\label{Vlasov}
\end{multline}
where $s = i,e$ represents the corresponding species.
The variable $v$ denotes velocity in phase space.
$q_s$ and $m_s$ are normalized by $e$ and $m_i$ respectively.
Densities of the plasma components are calculated through integration as:
\begin{equation}
n_s(x,t) = n_{0s}\int f_s(x,v,t) dv
\label{density}
\end{equation}
which are coupled with Poisson's equation:
\begin{equation}
\frac{\partial^2 \phi(x,t)}{\partial x^2}  = n_e(x,t) - n_i(x,t)
\label{Poisson}
\end{equation}
The equilibrium values $n_{s0}$ are assumed to satisfy 
the quasi-neutrality condition ($n_{e0} = n_{i0} $) at the initial step.

The constant parameters which remain fixed through all
of our simulations are: 
$\frac{m_i}{m_e} = 100$,
time step $d\tau = 0.01$,
$\theta = \frac{T_e}{T_i} = 64$
and $L = 4096$, 
where L is the length of the simulation box.
Perturbation feature are either $\psi = 0.05$ and  $\Delta = 10$ (small IDP) or $\psi = 0.2$ and  $\Delta = 500$ (large IDP).
The values of $\beta$ were modified between
successive simulations in range of $-1.0\leq \beta \leq 1.0$.
We have considered a two-dimensional phase space with one spatial and one velocity axis.
The phase space grid $(N_x, N_v)$ size is $(4096, 4000)$.
The periodic boundary condition is employed on x-direction 
in order to create successive collisions between IA solitons.

\section{Results and Discussion} \label{Results}
Before presenting the simulation results,
a general overview of their temporal evolution is reported.
An initial density perturbation (IDP),
which is selected to be around $x/\lambda_{Di} = 512$, 
is produced in the simulation domain using the Schamel distribution function.
Firstly, this IDP breaks into two oppositely drifting density perturbations (DDPs)
due to the symmetry in the velocity direction. 
As the temporal progression continues,
each of the DDPs emit
their own Langmuir wavepackets ahead of itself. 
These wavepackets are much faster than the ionic structures such as DDPs. 
Therefore, they quickly get separated from the IDPs. 
Furthermore, 
the DDPs forms one or more IA solitons with different velocities and sizes, but with the same $\beta$ as the stationary IDP\cite{jenab2016trapping}.
The number of IA solitons depends on the value of the trapping parameter as well as the amplitude of IDPs. 
IA solitons with higher amplitude possess higher velocities,
hence the IA solitons are aligned in spatial direction based on their height.
Behind the DDPs, ion-acoustic wavepackets are created which are slightly slower than the DDPs. 
The IA wavepackets are produced independent from the DDPs. 

The solitons and DDPs created from the chain formation are not stationary BGK states.
However, they are produced self-consistently which makes them free from any approximation used in the 
Schamel's theory derivation of mKDV equations such as small amplitude.
Furthermore, this process resembles 
the experimental approach used in Double Plasma (DP) device to excite solitons \cite{ikezi1970formation}.

Since the periodic boundary condition is employed on the spatial direction,
the wavepackets never leave the simulation domain.
In long-term simulations,
they resonate with each other and cause numerical instabilities
so that
particles are pushed out of the simulation domain.
Hence the conservation laws are violated.
Simulation results presented here are long before the time of the resonance 
and the conservation laws are checked for deviations to stay below one percent constantly.
This effect restricts the trapping parameter ($\beta$) range,
reported in this study. 
The further the trapping parameter deviates from zero, the stronger wavepackets would appear. 
On the other hand, since the focus is on the collision of the IA solitons,
the periodic boundary condition can't be removed. 
Therefore here the study is limited to $-1.0<\beta<1.0$.

\begin{figure}
  \subfloat{\includegraphics[width=0.5\textwidth]{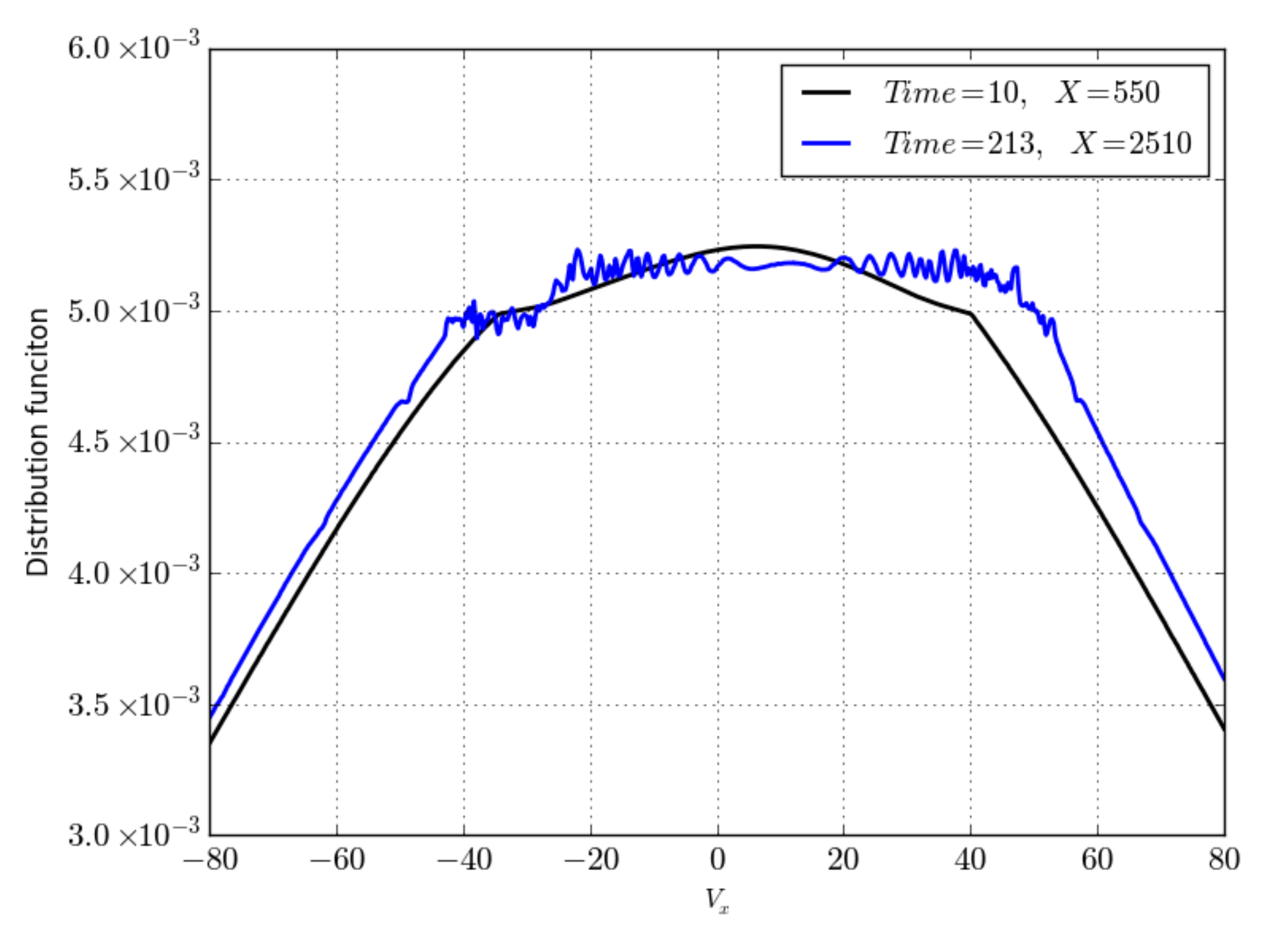} } \\
  \subfloat{\includegraphics[width=0.5\textwidth]{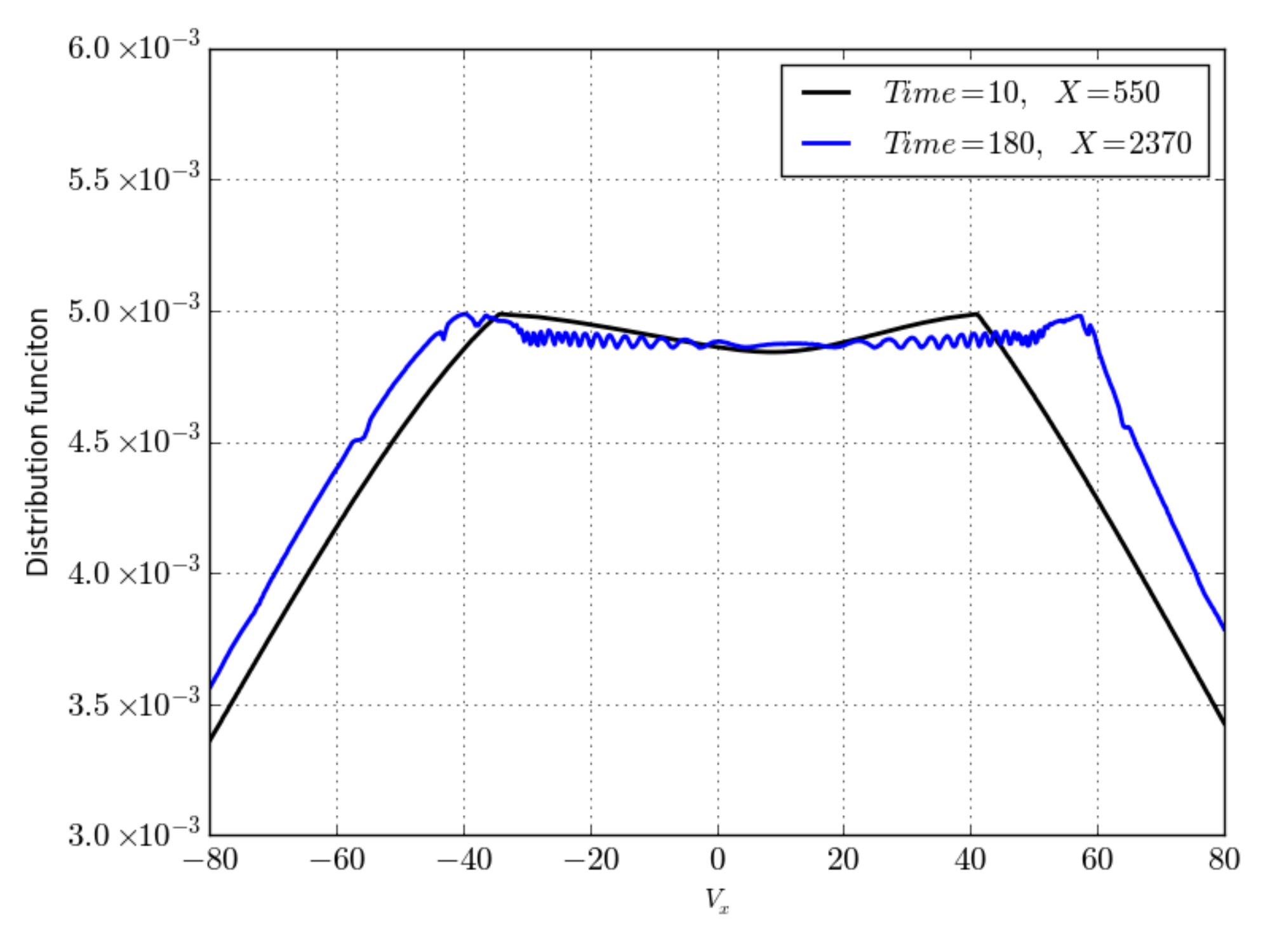} }
  \caption{The evolution of the distribution function versus velocity is shown for $\beta=0.2$ (top figure)
  and $\beta=-0.1$ (bottom figure). 
  At $\tau = 10$ the distribution function is presented at $x = 550$,
  middle of the DDP,
  which shows hump (hollow) for $\beta=0.2$ ($\beta=-0.1$) by a black line, which is smooth and soft.
  Blue line, noisy and filamented, presents the distribution function at the middle of the first soliton just before its first collision.
  Filamented structures can be witness to grow inside the distribution function of trapped population.}
 \label{cross_DF}
\end{figure}

In the early stage of the evolution, before any collisions, the internal structure of solitons in the phase space 
shows filamentation structures which become finer during temporal progression (Fig.\ref{cross_DF}). 

  \subsection{The effect of IDP size} \label{size_effect}
\begin{figure}
 \subfloat{\includegraphics[width=0.5\textwidth]{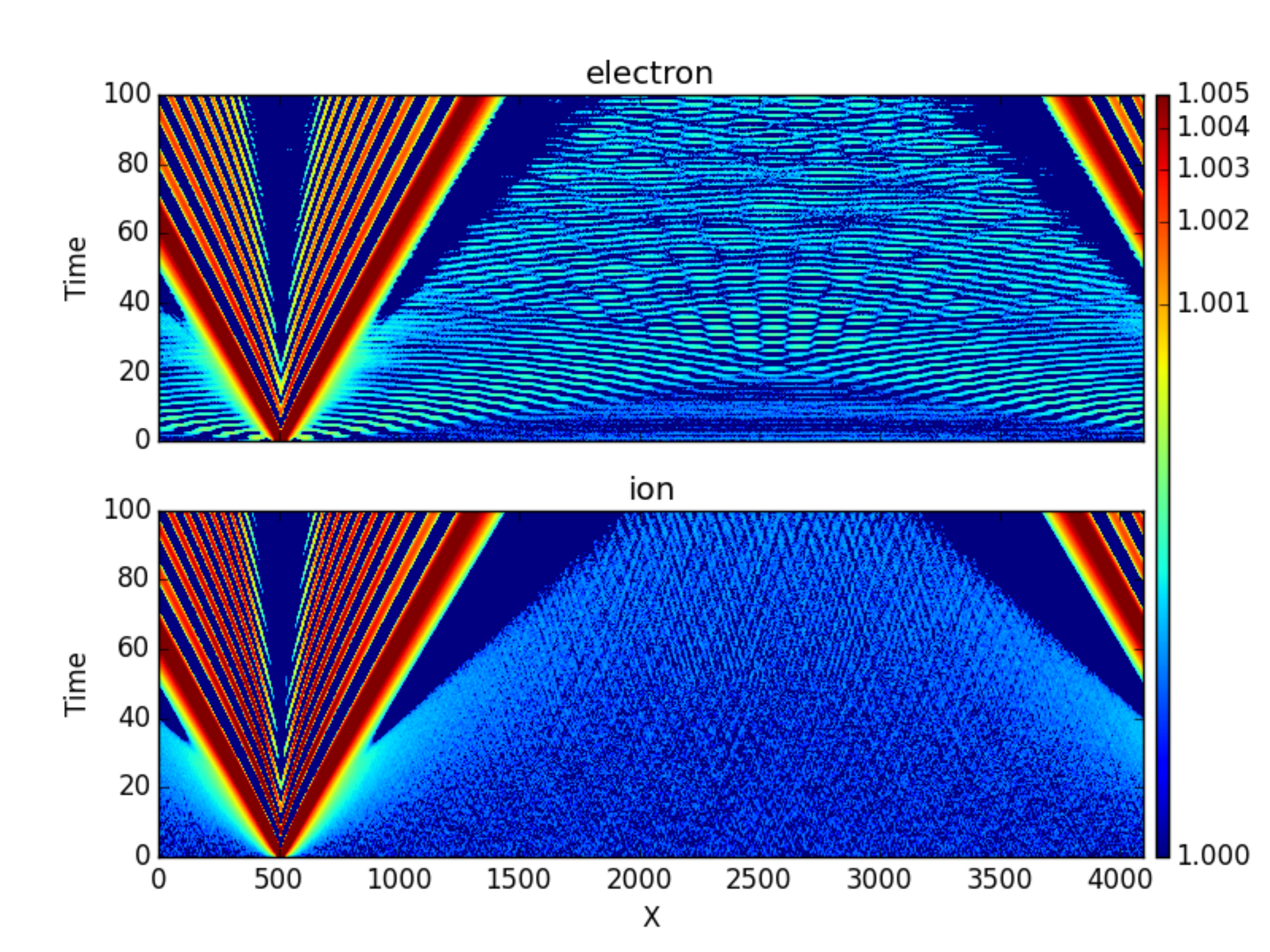}\label{small_B0}} 
 \caption{The temporal evolution of electrons/ions number density,
 for the case of a small IDP ($\psi = 0.05$, $\Delta=10$) with $\beta=0$,
 is shown in top/bottom figures.
 The propagation of both Langmuir and IA wavepackets can be witnessed.
 The number density are shown with color covering values from $1.0$ to $1.005$.
 Note that the colors are arranged based on a power law distribution so the small amplitude Langmuir wavepacket in the electron number density 
 can be recognizable. }
 \label{small}
\end{figure}

When the amplitude of the DDPs are small enough,
they don't disintegrate into a number of IA solitons.
Each of the DDPs forms just one IA soliton. 
Fig.\ref{small} shows the simulation results for the case of small IDP ($\psi = 0.05$ and $\Delta = 10$) with $\beta = 0$.
The existence of two wavepackets, namely Langmuir and ion-acoustic, can be observed. 
The Langmuir wavepacket is recognizable in the electron number density,
as waves propagating faster than the DDPs. 
For the ion number density in the same area, a small-amplitude noise can be recognized. 
The noise is coming from the effect of Langmuir waves on ions.
Since ions don't resonate/participate in propagation of Langmuir waves,
there is no wave pattern in the noise. 
On the other hand, the IA wavepacket appears behind the DDPs which is also independent from them.
The IA wavepacket starts from the remanence of the initial perturbation when the DDPs have already left it.

\begin{figure}
 \subfloat{\includegraphics[width=0.5\textwidth]{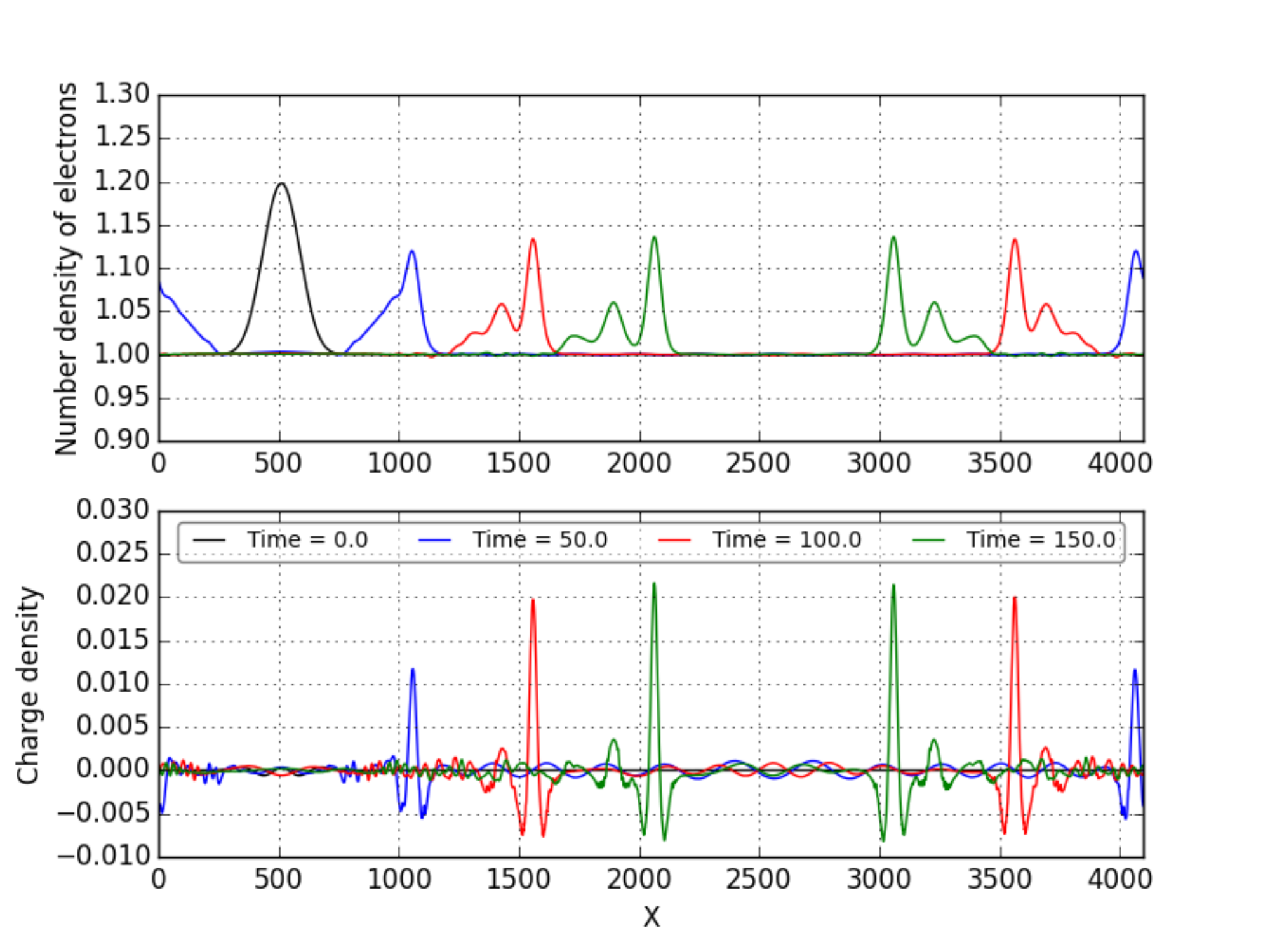}}
 \caption{Profiles of electrons number density and charge density are shown in top and bottom figures respectively, for a large IDP ($\psi=0.2$, $\Delta=500$) with $\beta = -0.1$.
 Four different times during disintegration process, 
 namely $\tau = 0, 50, 100, 150$,
 are presented with black (around $x=500$), blue (around $x=1000$), red (around $x=1500$) and green (around $x=2000$) respectively.
 The disintegration process of each of DDPs into three IA solitons can be seen in details.} 
 \label{large}
\end{figure}

Fig.\ref{large} presents the results for a large IDP ($\psi = 0.2$ and $\Delta = 500$) with $\beta = -0.1$. 
The DDPs disintegrate into three solitons.
Firstly the initial stationary IDP breaks into two oppositely DDPs.
Then,
each of the DDPs steepens on their propagation side due to nonlinearity. 
Furthermore, 
three IA solitons start surfacing.
The earlier they appear (since they are faster), the more dominant/taller they are.
As Fig.\ref{large} shows, the breaking between the first and the associated DDP happens around $\tau=100$, 
and the breaking between the second and the third IA solitons takes place later around $\tau=150$.
Note the difference between Fig.\ref{small} and Fig.\ref{large}, 
in which the ion-acoustic wavepackets are created independent from the DDPs,
but the IA solitons are created from the bulk of the DDPs.
These results, i.e.  
 a) the existence of wavepackets, 
 b) the breaking of an stationary IDP into two oppositely DDPs, 
 c) disintegration of a DDP into one or more IA solitons,
have been reported in both fluid and PIC simulations as well \cite{Kakad2013,Kakad20145589,Sharma2015,Qi20153815}.

  \subsection{Stability against mutual collisions}\label{section_stability}
\begin{figure*}
 \subfloat{\includegraphics[width=0.7\textwidth]{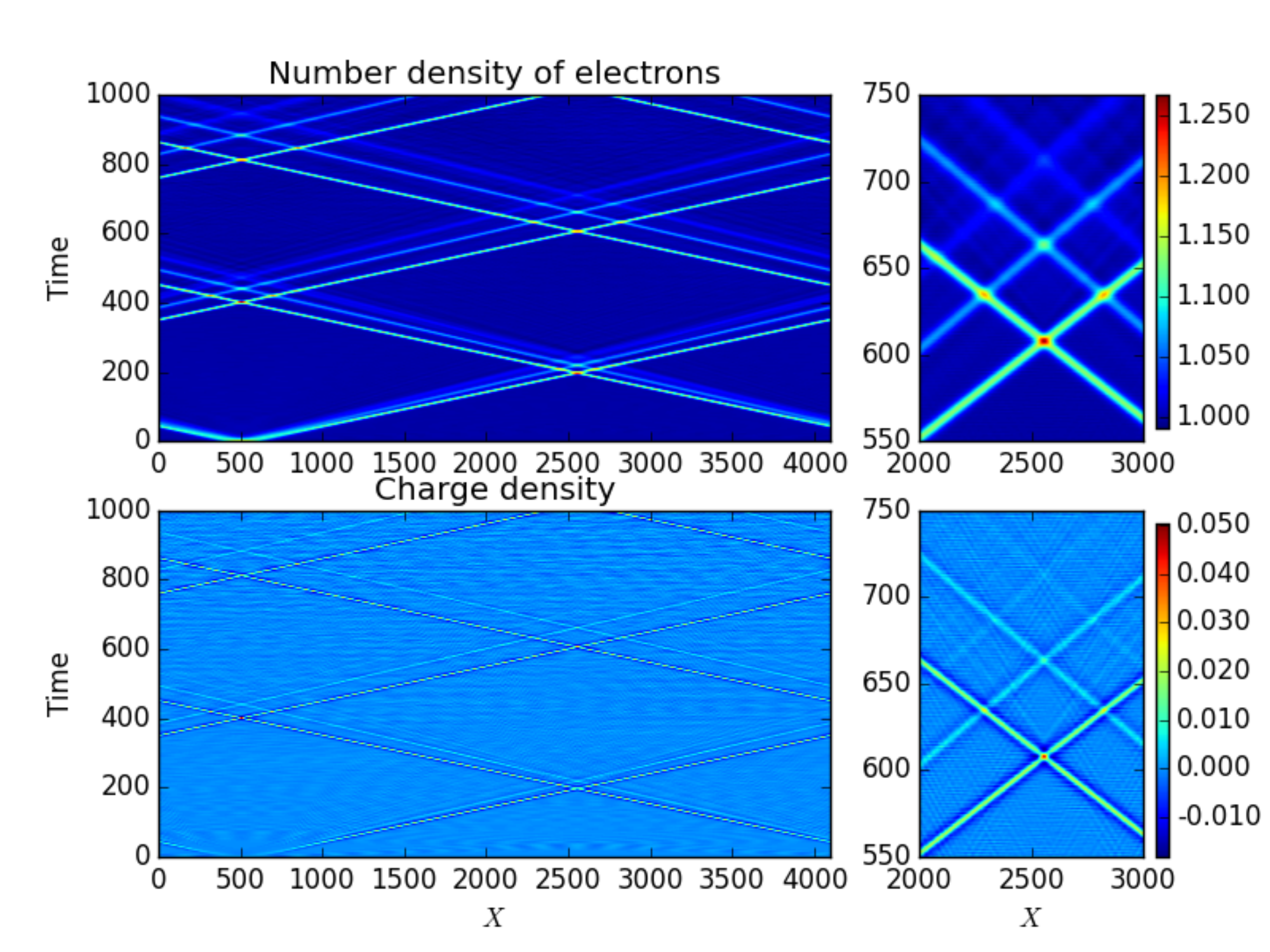}\label{B-01}} \\
 \subfloat{\includegraphics[width=0.7\textwidth]{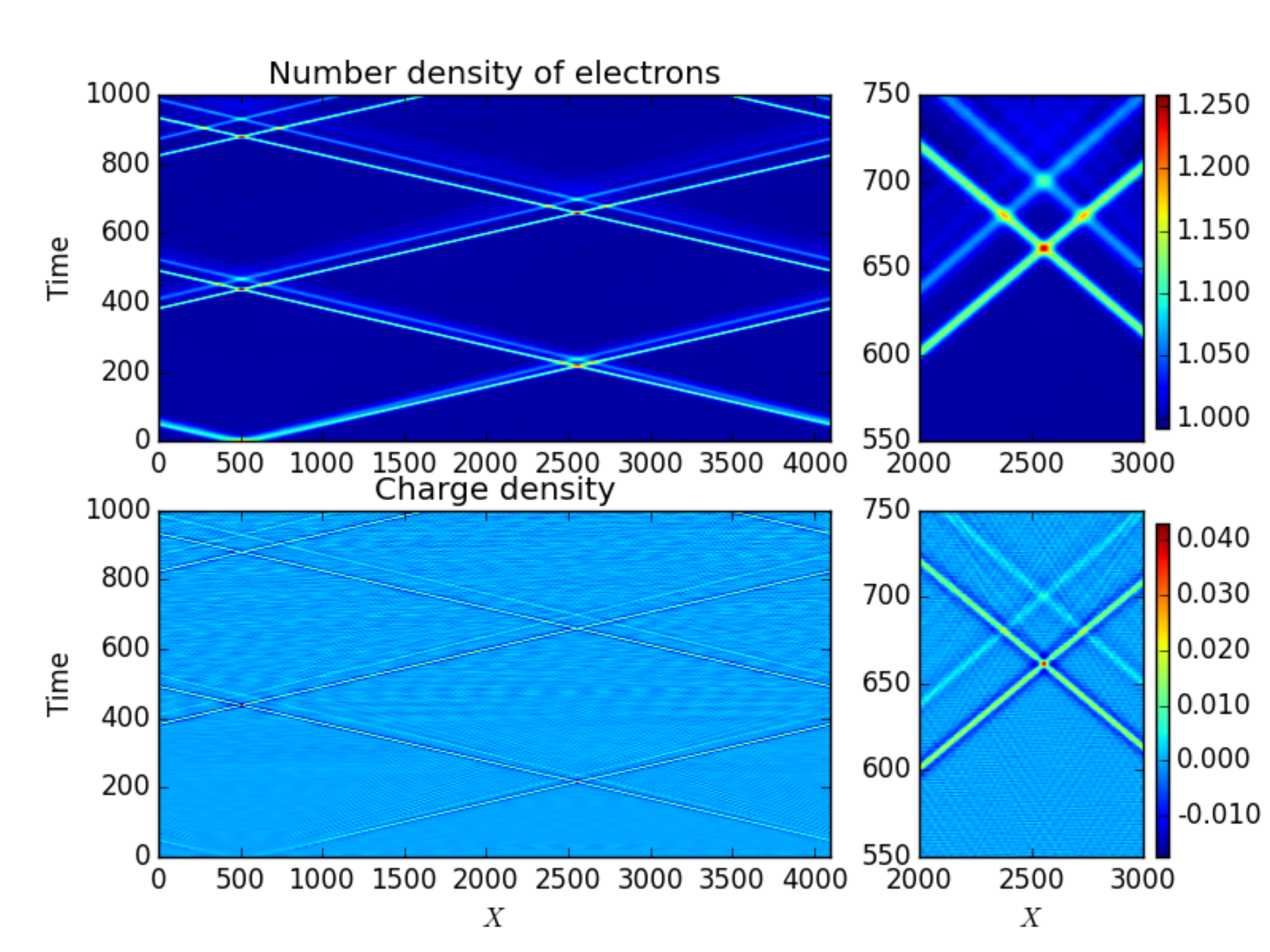}\label{B021}} 
 \caption{The temporal evolution of electrons number density and charge density are shown for a large IDP ($\psi=0.2$,$\Delta=500$) and two values of $\beta$.
 Twelve (eight) collisions are observed for each of six (four) IA solitons in case of $\beta = -0.1$ ($\beta =0.2$) 
 in two top (bottom) figures.
 Zoomed-in figures on the right side of each figures display the details of three successive collisions around the time $550<\tau<750$.
 Since the trajectories of solitons don't change by this collisions, hence their velocities are conserved.}
 \label{collision}
\end{figure*}

Fig.\ref{collision} demonstrates successive mutual collisions between IA solitons for two cases $\beta = -0.1$ and $\beta = 0.0$ for large IDPs ($\psi = 0.2,\Delta=500$). 
In case of $\beta = -0.1$, each of the DDPs break down into three IA solitons. 
Each of the these IA solitons has gone through 12 collisions up to $\tau = 1000$.
These collisions take place between IA solitons of different sizes with same trapping parameter $\beta$.
The 12 collisions happen in 4 sets of triple collisions during $180<\tau<250$, $380<\tau<450$, $580<\tau<650$ and $780<\tau<900$. 
In case of $\beta = 0.2$, two IA solitons emerge from each of the DDPs, and there are 8 collisions.

In order to study the stability of these IA solitons during mutual collisions,
different features of them have been considered.
Two categories of features have been studied here, i.e.  
(a) spatial features such as amplitude, width and shape in the number density profiles and the velocity of propagation, 
and (b) velocity-direction features like width and shape in phase space.

\begin{figure}
  \subfloat{\includegraphics[width=0.5\textwidth]{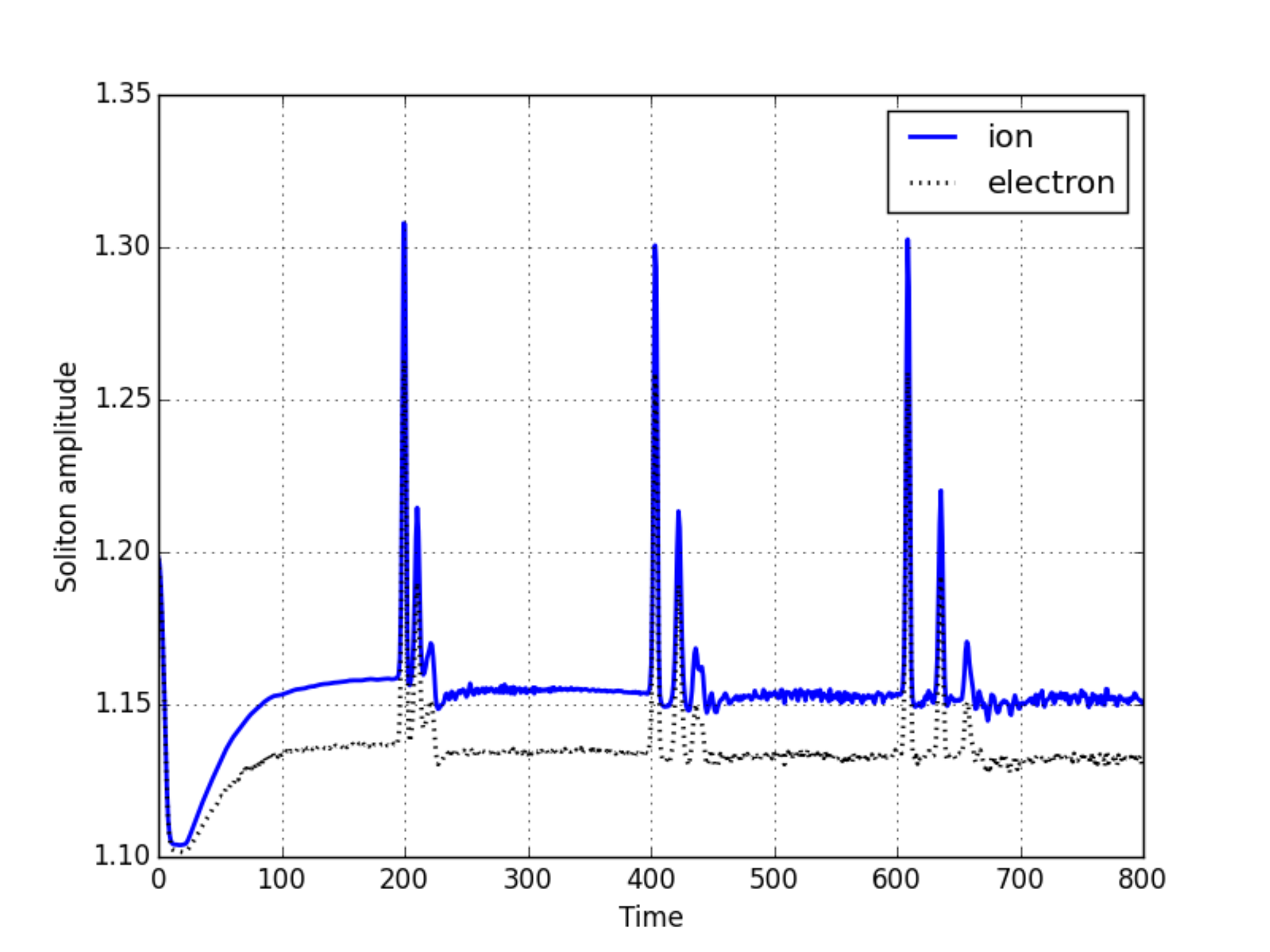} } \\
  \subfloat{\includegraphics[width=0.5\textwidth]{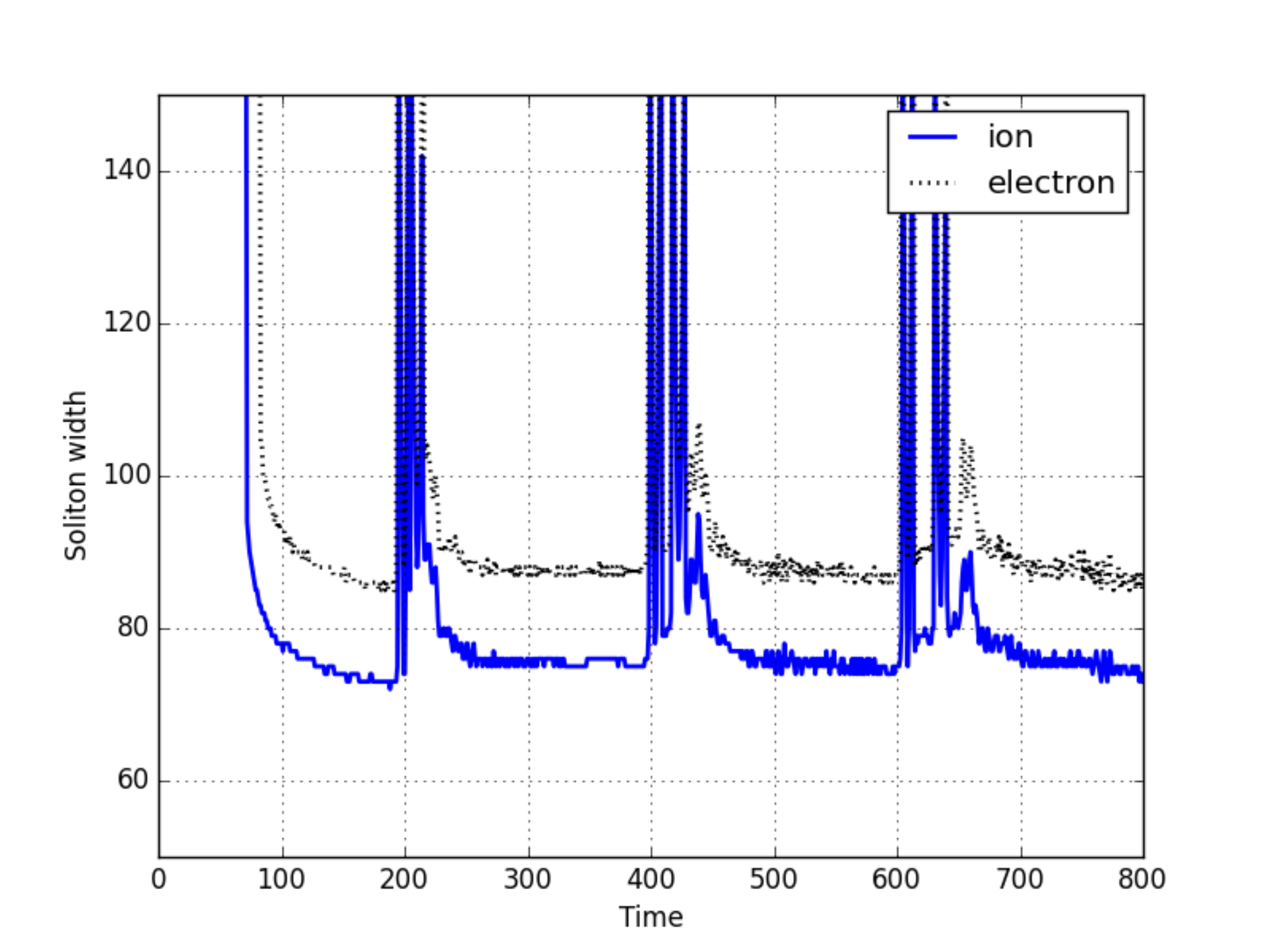} } \\
  \subfloat{\includegraphics[width=0.5\textwidth]{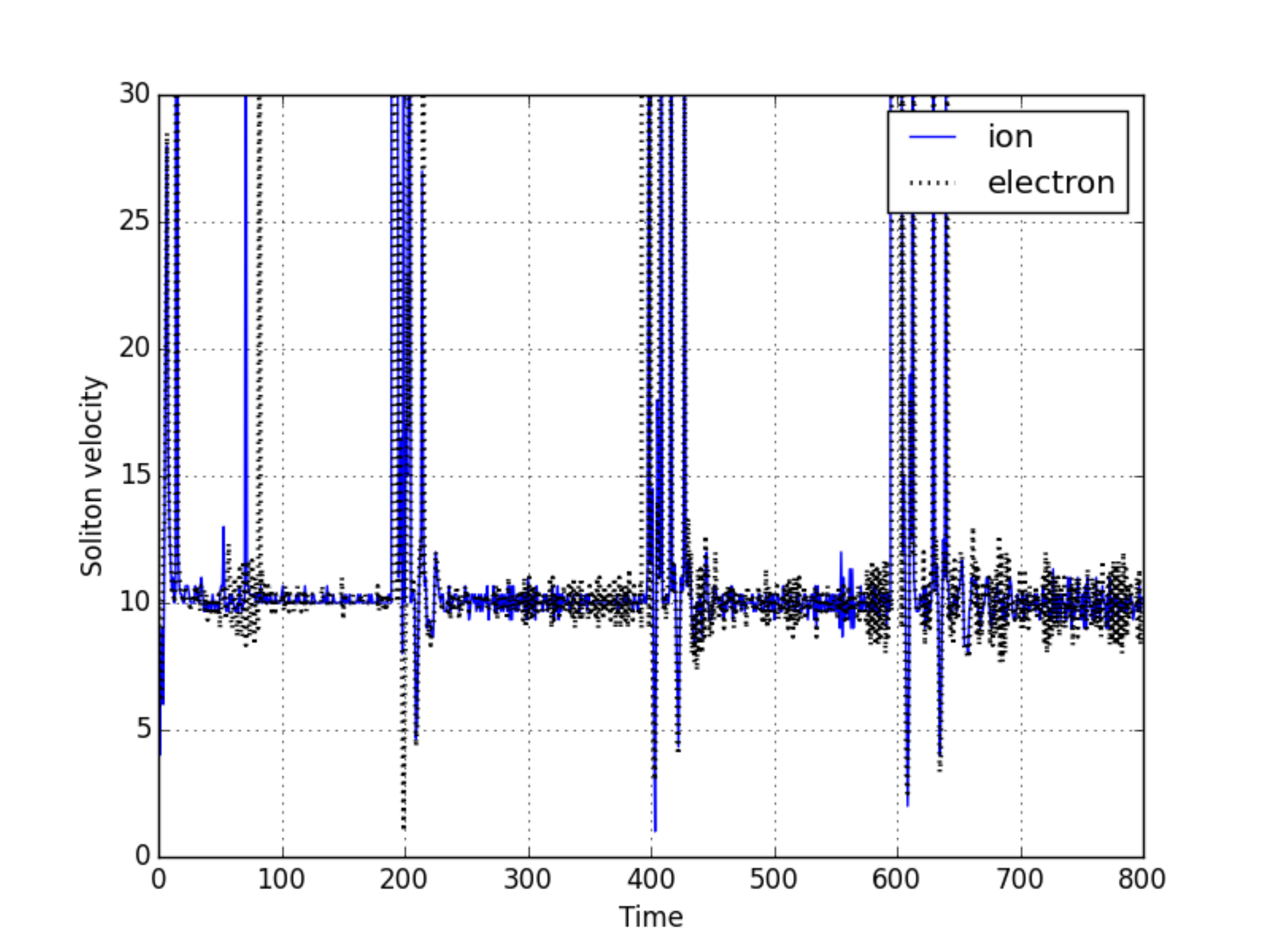} }
  \caption{ The temporal evolution of features such as amplitude (top),  width (middle) and velocity (left) 
  of ion (blue solid line) and electron (black dotted line) number density
  are shown for the right-propagating first/dominant soliton in case of $\beta = -0.1, \psi=0.2$ and $\Delta=500$.
  Anomalies take place in the measurement during collision times,
  i.e. $180<\tau<250$, $380<\tau<450$, $580<\tau<650$.
  The fluctuation of the values around the average for propagation times (excluding the collision intervals)
  are less than $1\%$, $5\%$ and $8\%$ for amplitude ($1.12<a_e<1.14, 1.14<a_i<1.16$), 
  width ($85<w_e<95$, $70<w_i<80$) and velocity ($11.0<v_i=v_e<9.0$) respectively.}
 \label{Soliton_features}
\end{figure}

\begin{figure}
 \subfloat{\includegraphics[width=0.5\textwidth]{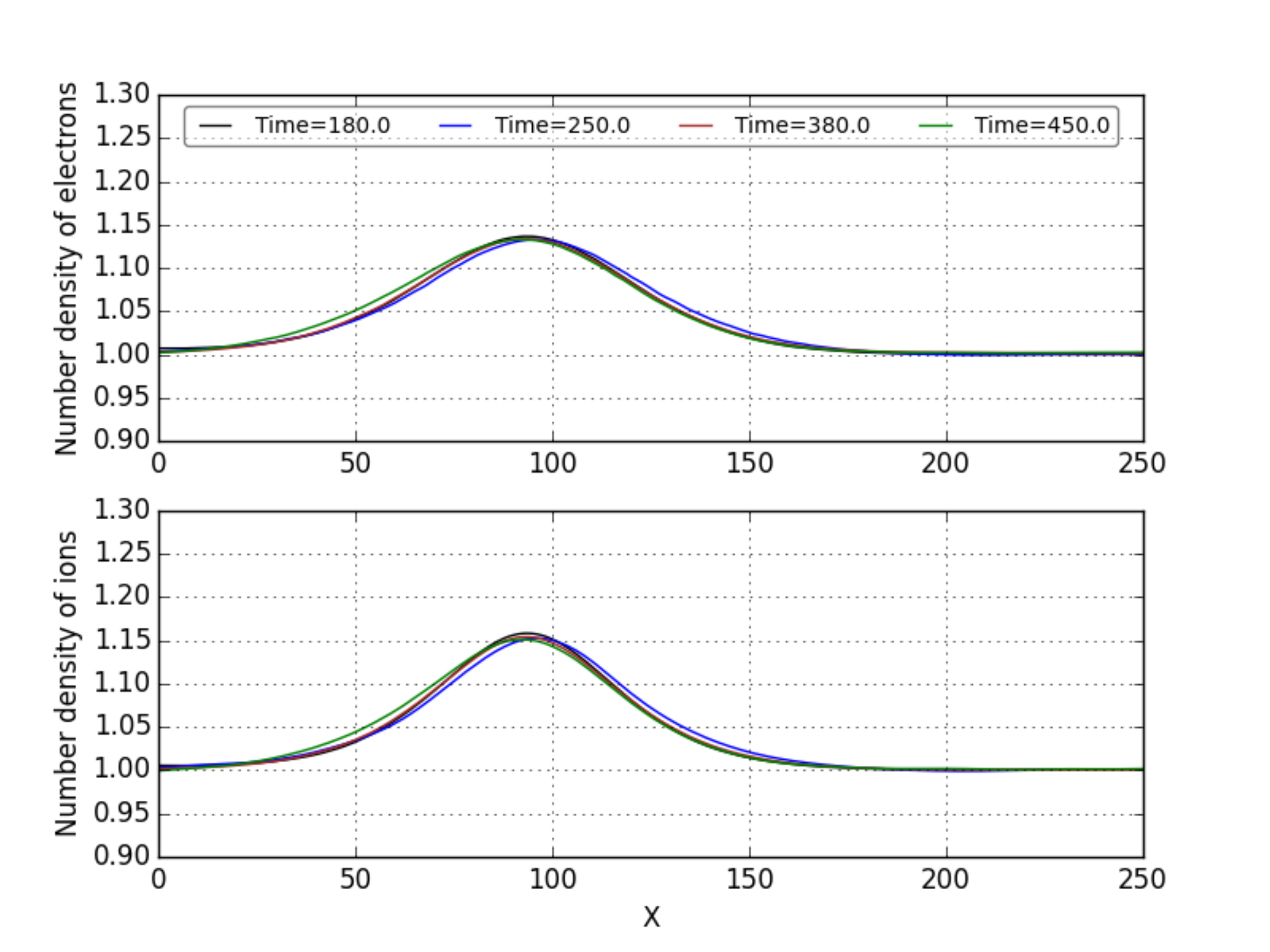}}  \\
 \subfloat{\includegraphics[width=0.5\textwidth]{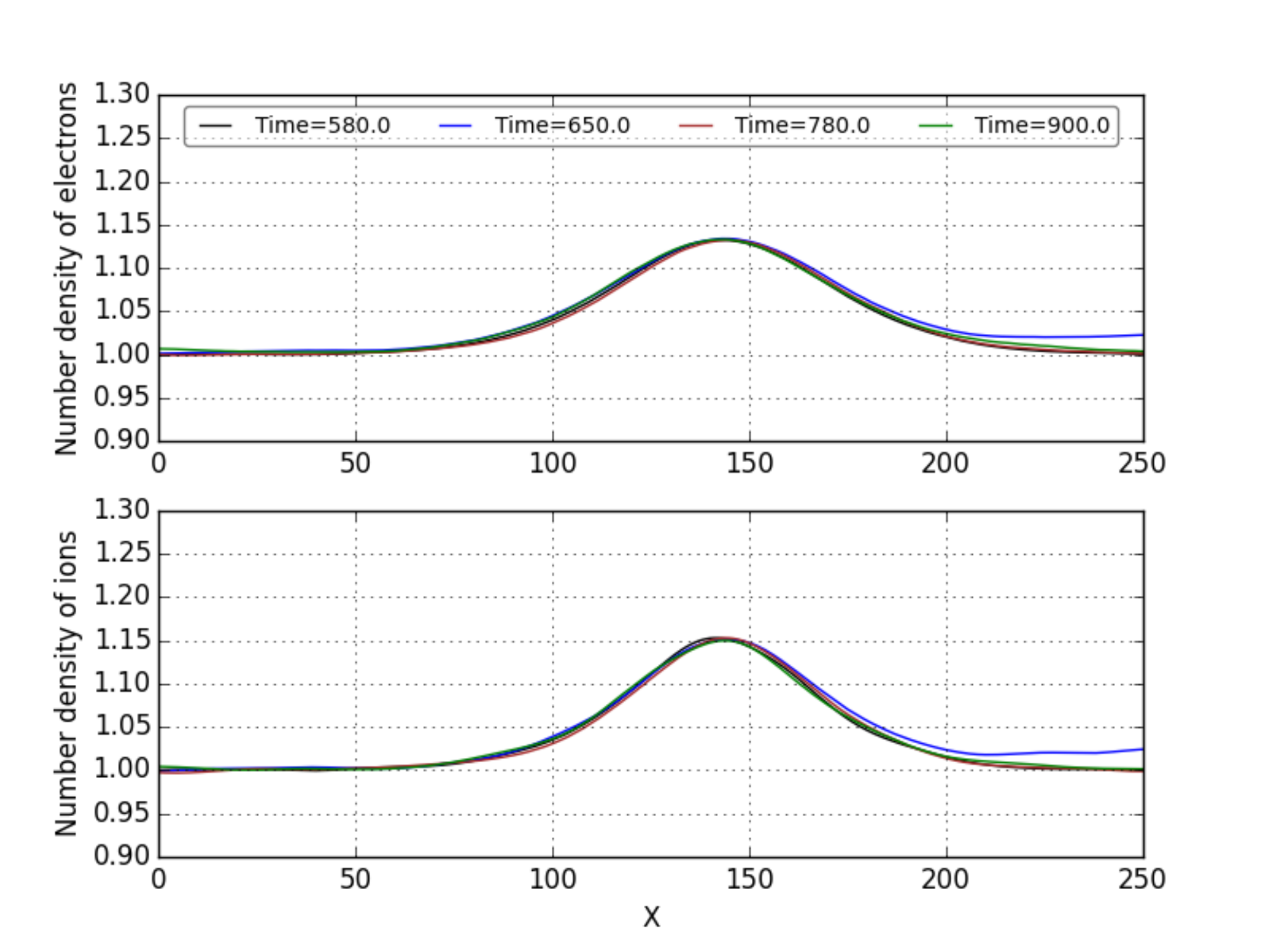}}  
 \caption{For a large IDP ($\psi=0.2$, $\Delta=500$) with $\beta = -0.1$, 
 the number density profiles of the first/dominant IA soliton are presented for different times. 
 In top figure the number density profiles before and after 
 the first ($\tau = 180$, $\tau = 250$) and 
 the second ($\tau = 380$, $\tau = 450$) triple collision are shown.
 In bottom figure the same is presented for the third ($\tau = 580$, $\tau = 650$) and 
 the fourth ($\tau = 780$, $\tau = 900$) triple collisions. }
 \label{Fig_amplitude}
\end{figure}

Fig.\ref{Soliton_features} focuses on three of these spatial features,
i.e. amplitude, width and velocity for the case of trapping parameters with negative values.
It shows that they stay the same after the mutual collisions within acceptable margin of error.
The collision intervals can be easily recognized within all the three figures.
What's more, one can observe the initial break-up of IDP ($\tau<25$) and then steeping of the DDPs before breaking into 
number of solitons ($25<\tau<70$) in the figure reporting temporal evolution of amplitude.
Fig.\ref{Fig_amplitude} presents the number density profiles of 
the first IA soliton for the case $\beta = -0.1$,
hence focusing on shape of solitons and its stability.
The electron/ion number density profile of the first right-propagating IA soliton is shown for eight different times before and after each of the triple collisions. 
The stability of the IA soliton can be observed clearly as its size (including height and width) and shape in the spatial direction don't change. 
These two figures confirm the stability of spatial features against mutual collisions.

The number densities of plasmas species, 
as fluid approximation of their distribution functions
$N~=~\int~f~\dd~v$, 
serve as a starting point for the fluid theory.
Therefore, the stability of their features against mutual collisions prove their fluid-level stability.
In other words, kinetic effects such as electron trapping,
do not alter the stability and the propagation features of IA solitons.

\begin{figure}
\begin{tabular}{c}
\subfloat{\includegraphics[width=0.5\textwidth]{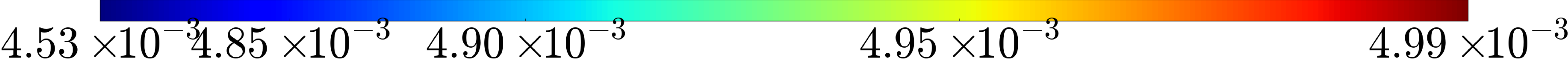} } \\
 \subfloat{\includegraphics[width=0.24\textwidth]{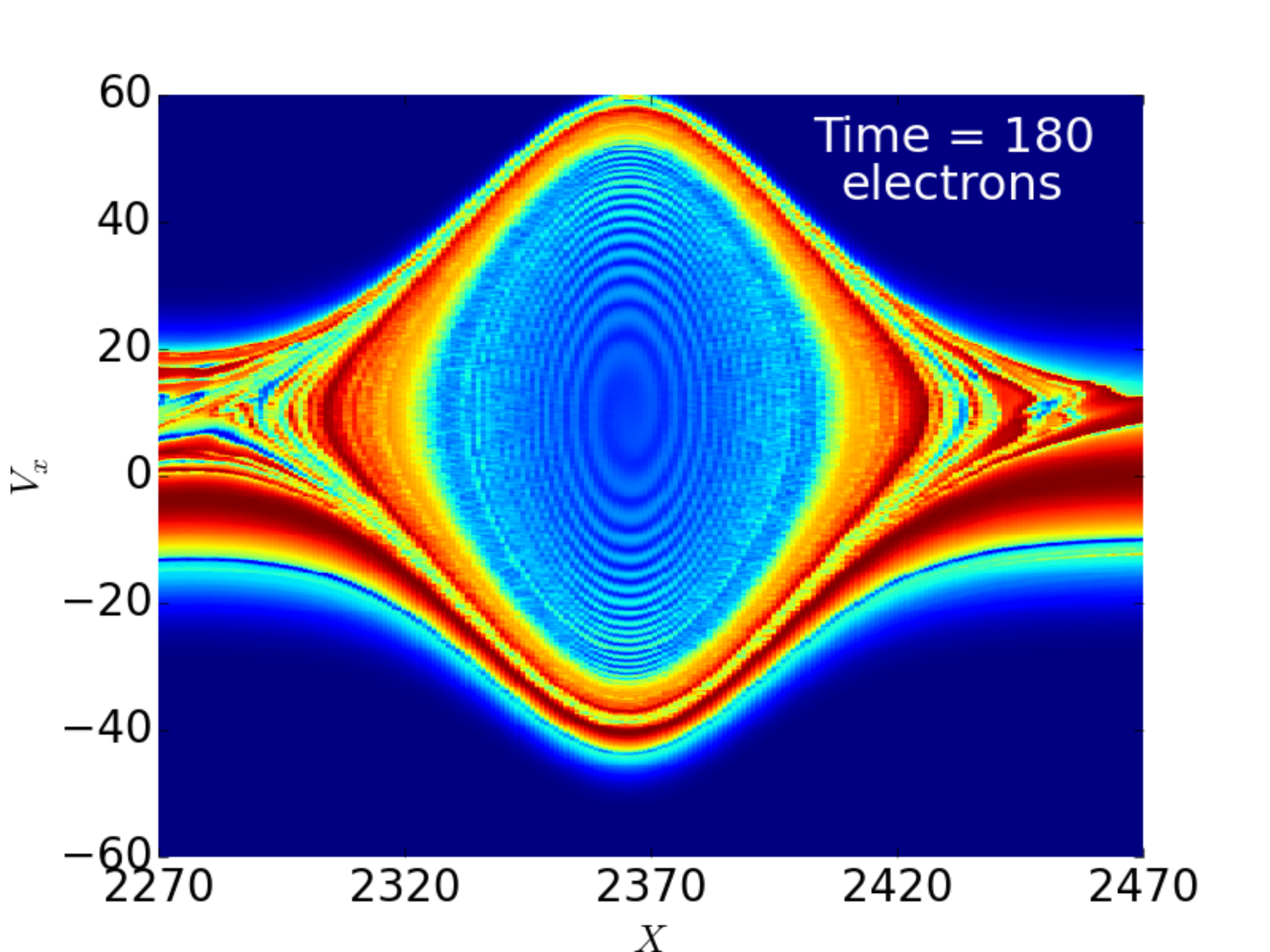} }
 \subfloat{\includegraphics[width=0.24\textwidth]{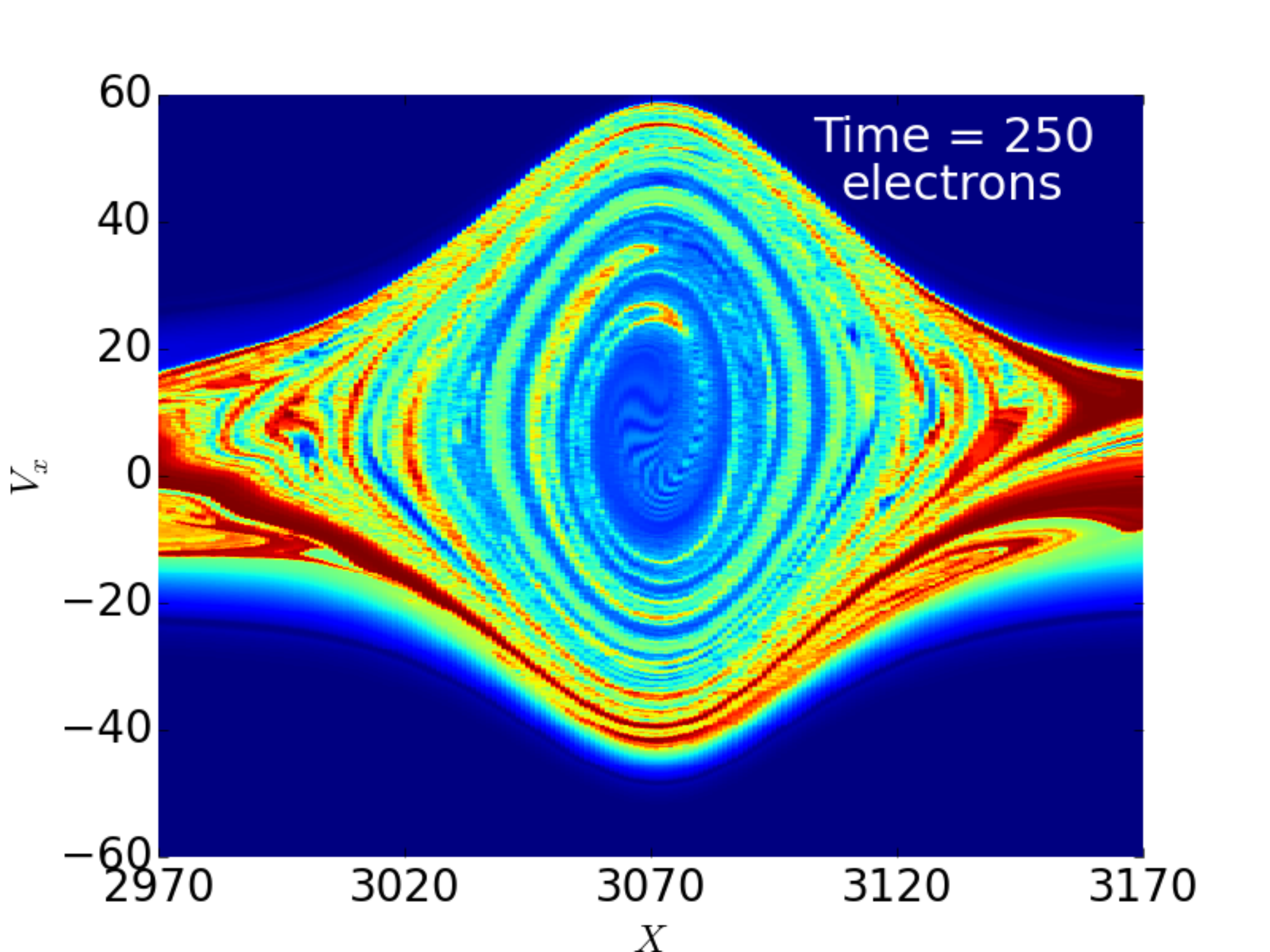} }\\
 \subfloat{\includegraphics[width=0.24\textwidth]{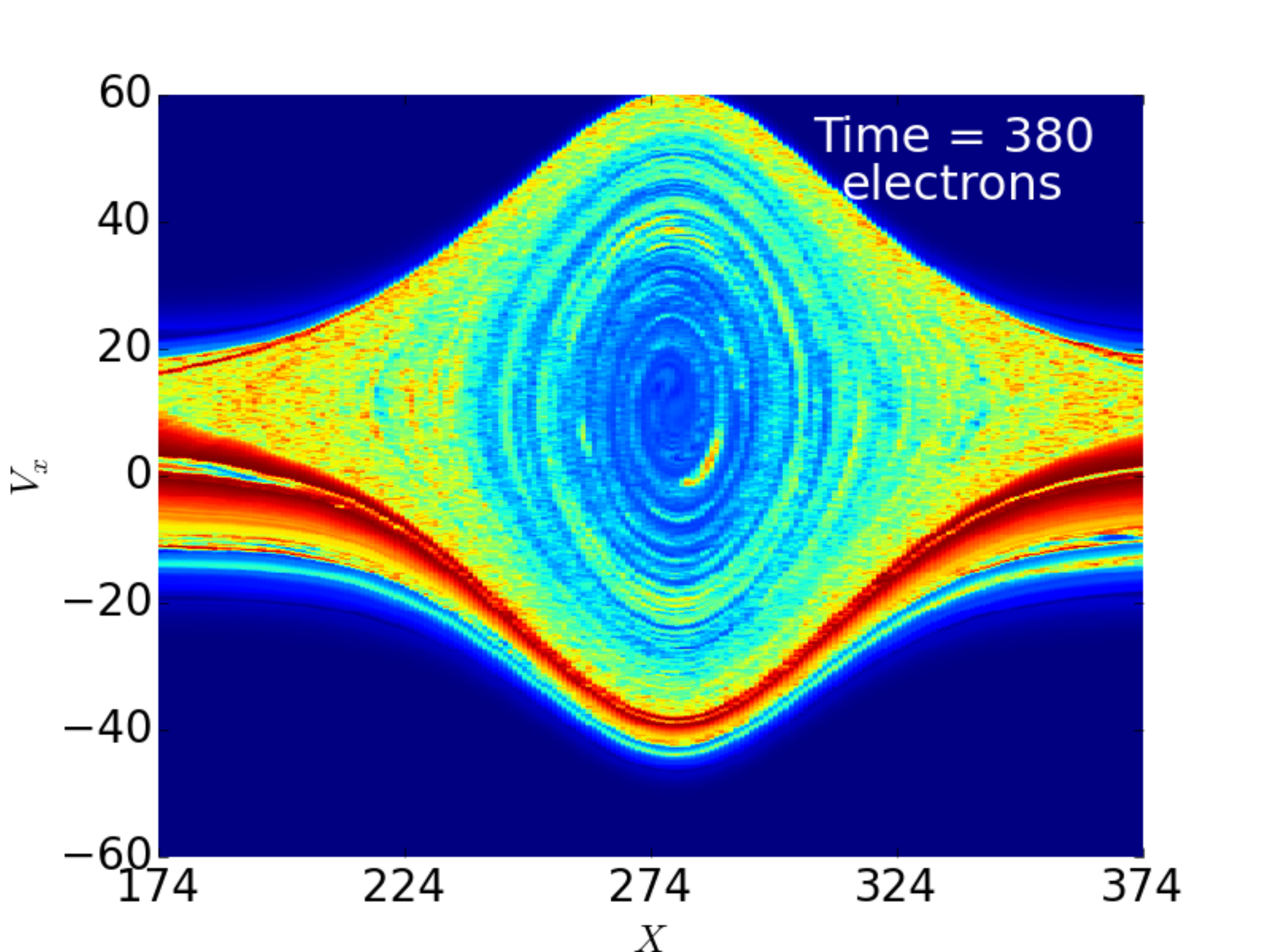} } 
 \subfloat{\includegraphics[width=0.24\textwidth]{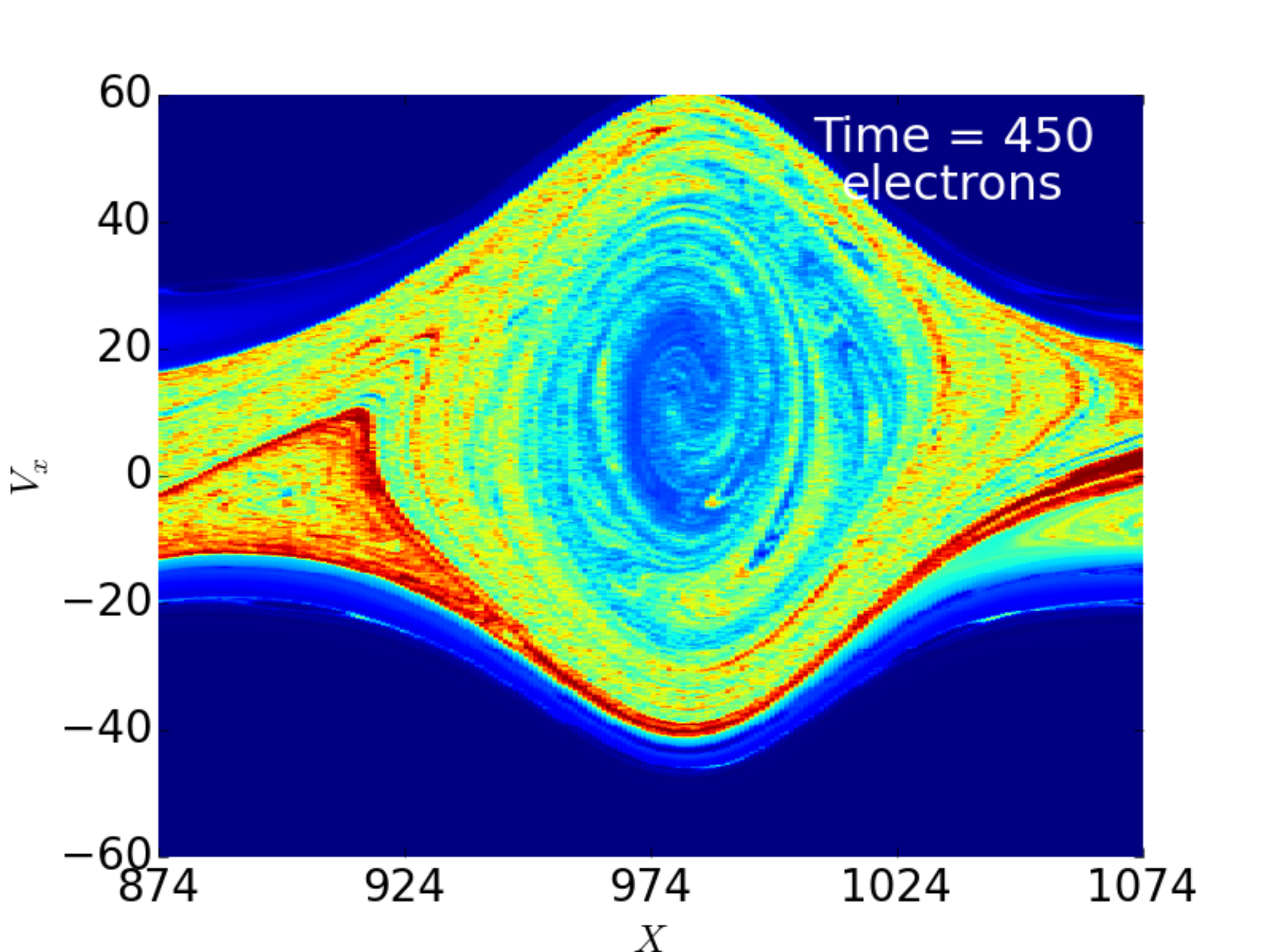} } \\
 \subfloat{\includegraphics[width=0.24\textwidth]{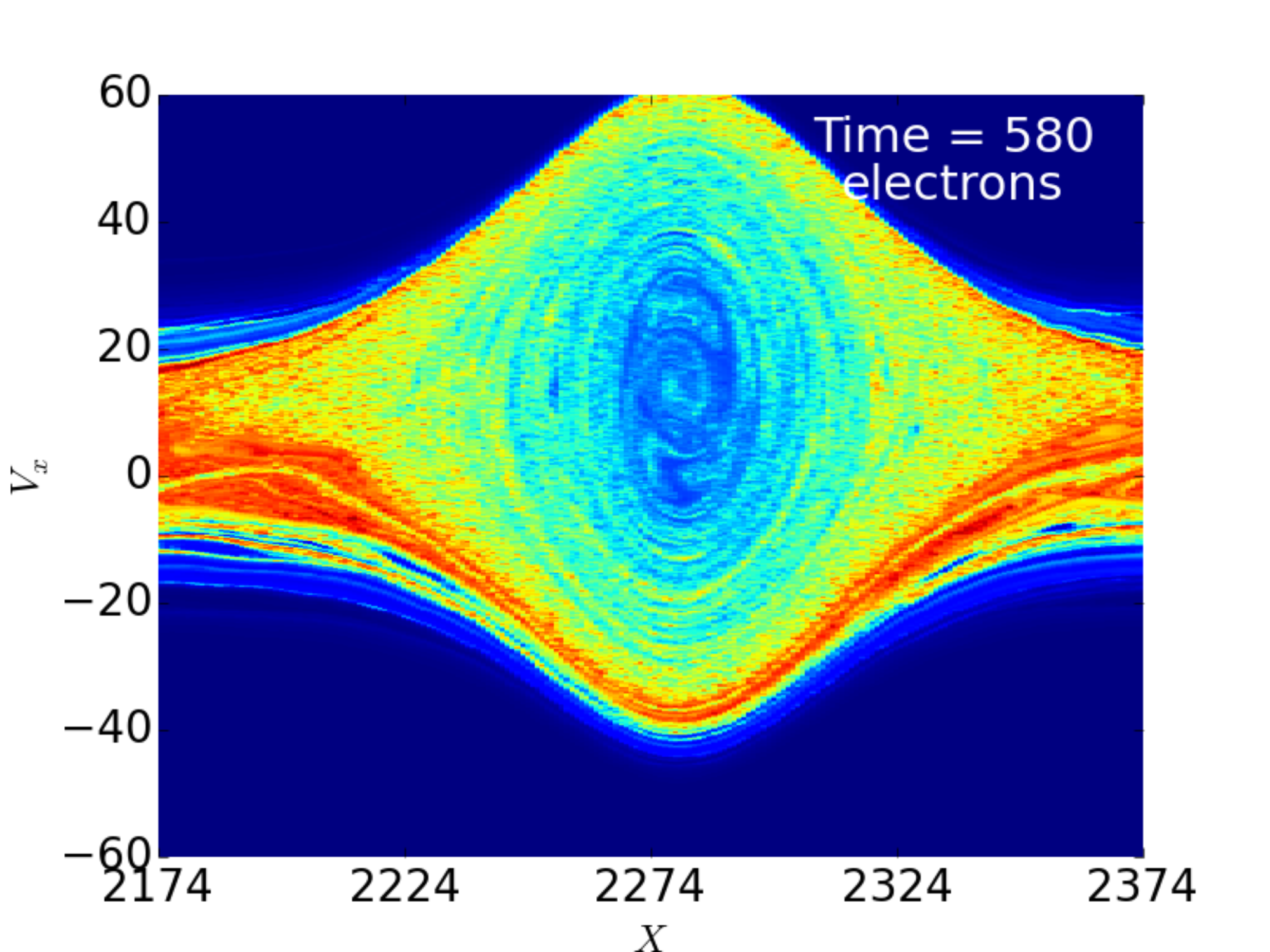} } 
 \subfloat{\includegraphics[width=0.24\textwidth]{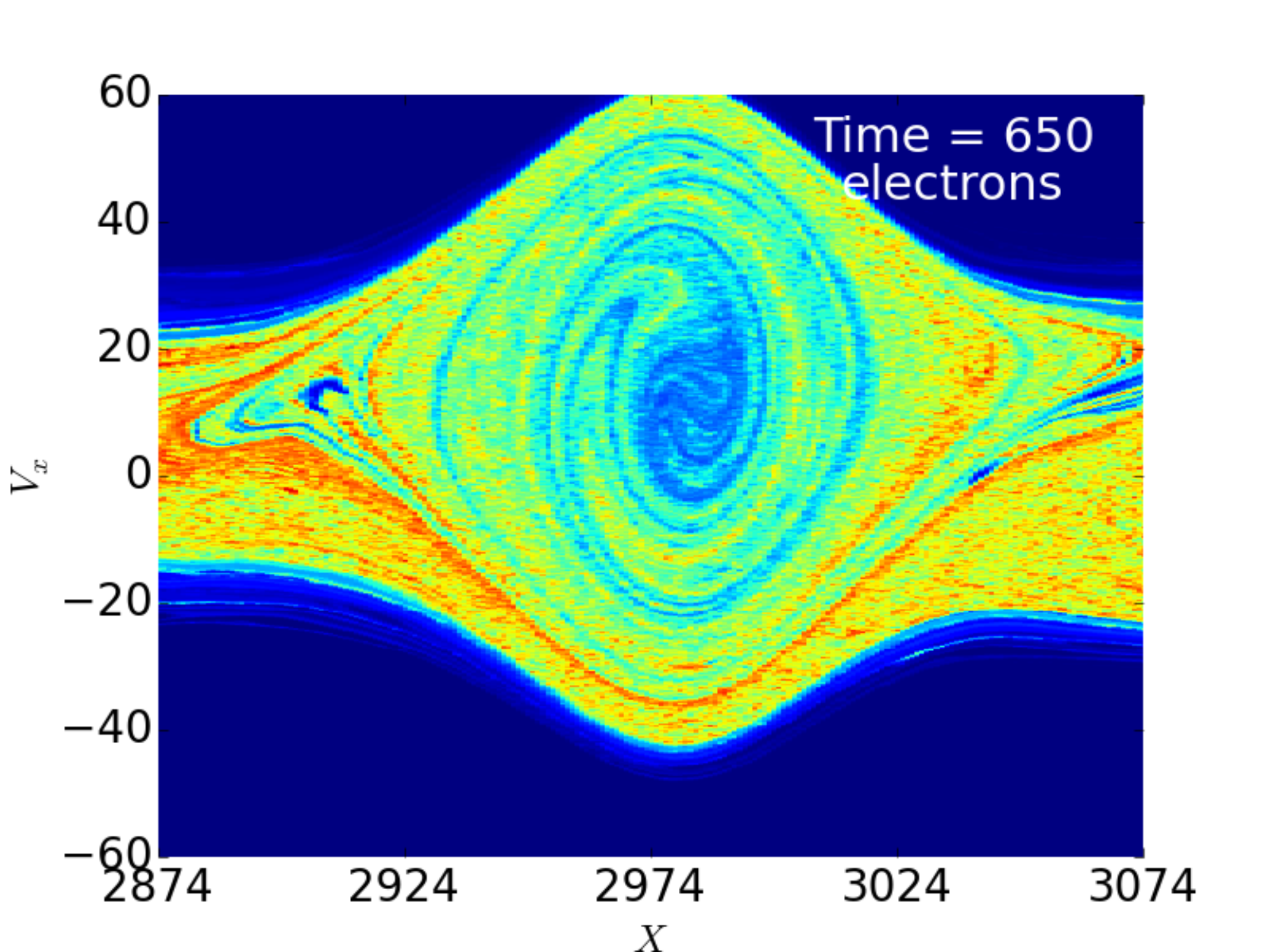} } \\
 \subfloat{\includegraphics[width=0.24\textwidth]{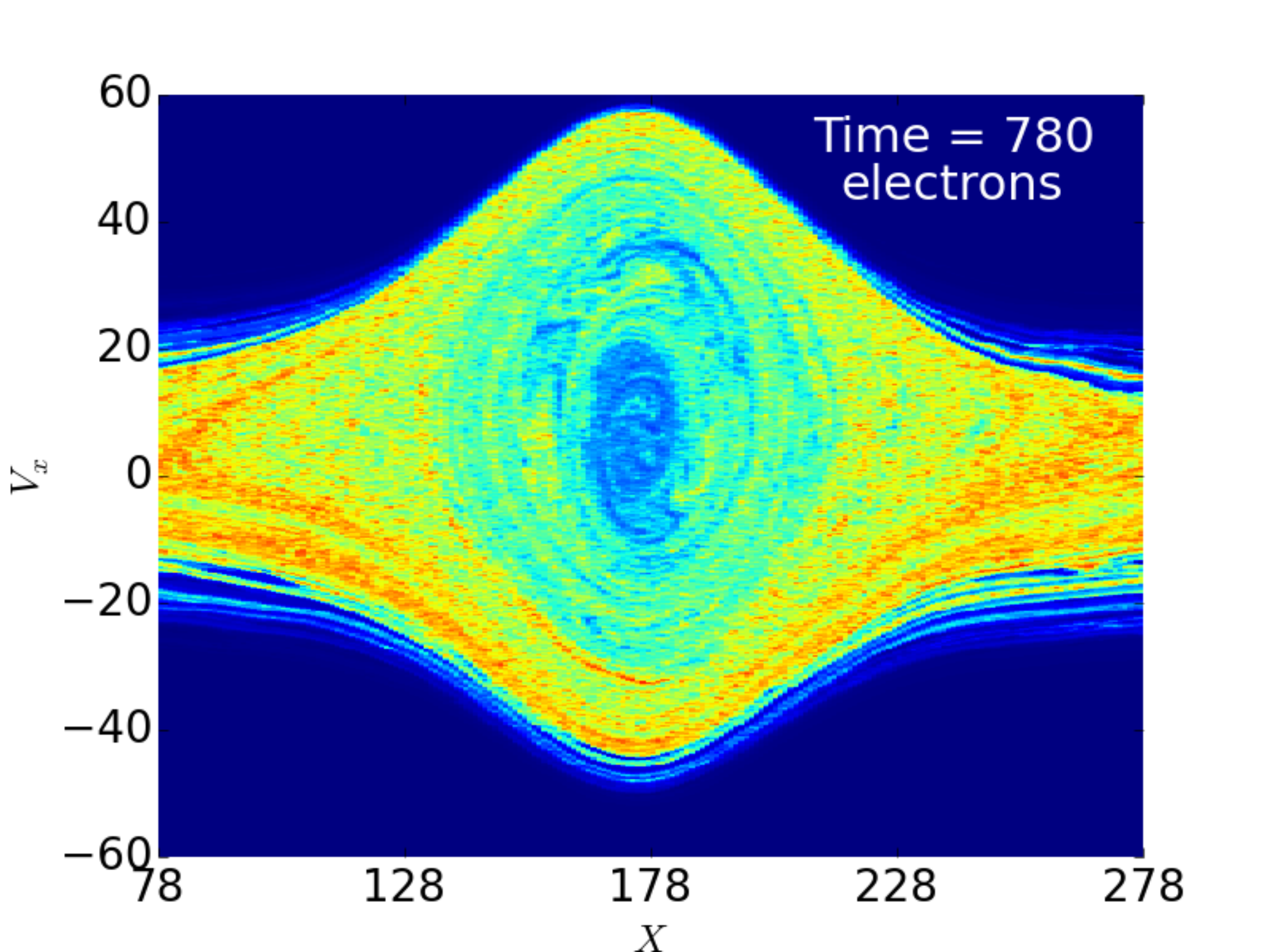} }
 \subfloat{\includegraphics[width=0.24\textwidth]{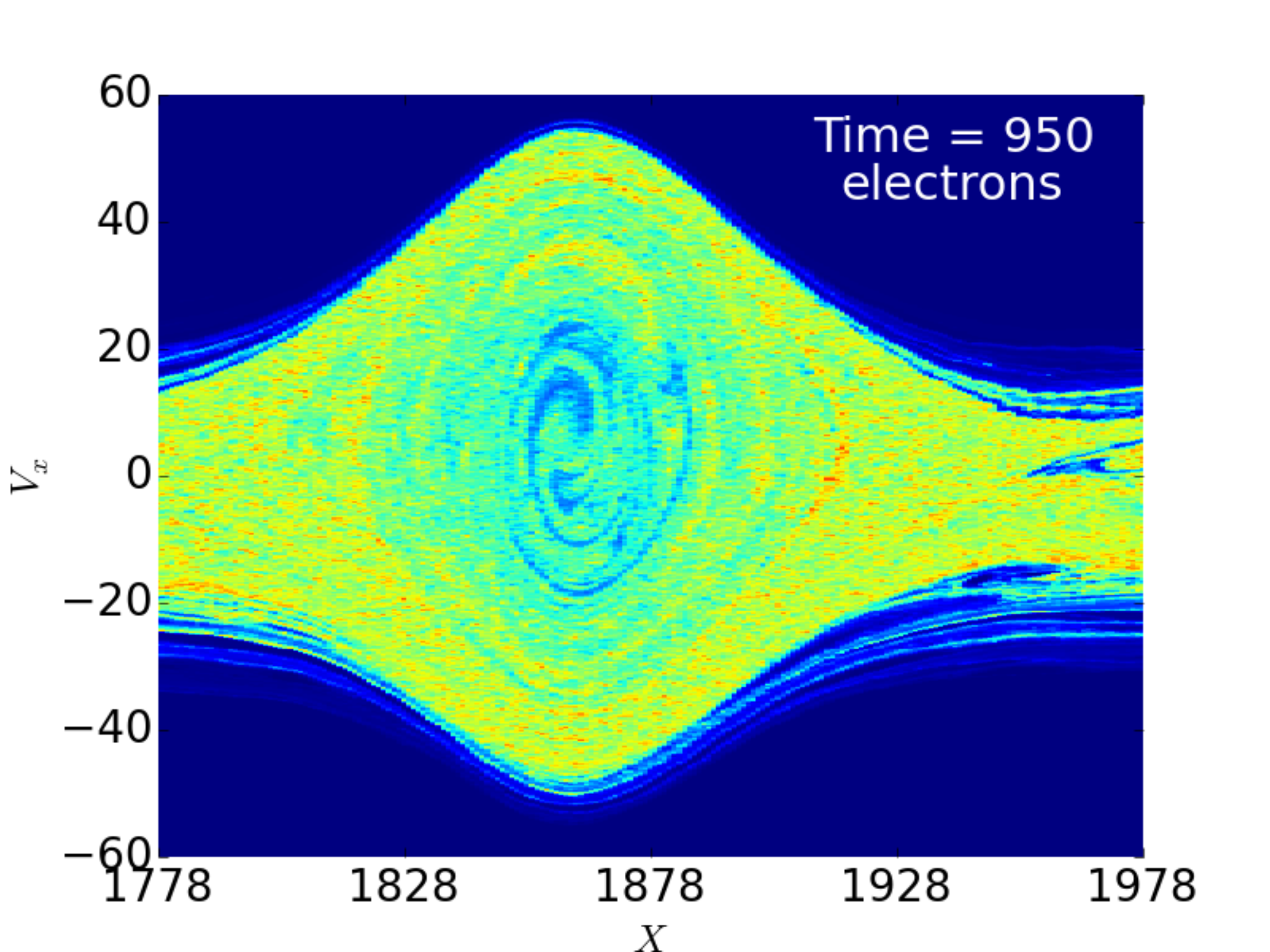} }\\
\end{tabular}
 \caption{For the case of large IDP ($\psi=0.2$ and $\Delta=500$) and $\beta = -0.1$, the hollow in the electrons' distribution function accompanying 
 the right-propagating first/dominant IA soliton are shown in the phase space. 
 The phase space structure of the trapped population 
 is presented before and after 
 first ($\tau = 180, 250)$,
 second ($\tau=380, 450$),
 third ($\tau=580, 650$) 
 and fourth($\tau=780, 900$) triple collisions respectively, 
 starting from the top left corner.
 The size and shape of hollows stay the same for all the figures, confirming the stability.
 However, the symmetry of the hollow (in the phase space) are distorted increasingly as the number of collisions increases.
 \textbf{See Supplemental Material at [\textit{URL will be inserted by publisher for ``1st\_IA\_Soliton\_formation\_beta\_negative.avi''}] 
 for the early stage development of of the hollow, i.e. $\tau<180$.}
 }
 \label{Fig_Internal_BN01}
\end{figure}

For the kinetic level study, the temporal evolution of the distribution functions of plasma species
are focused upon. 
Fig. \ref{Fig_Internal_BN01} displays the phase space structure of the electron distribution function
at the same times as Fig. \ref{Fig_amplitude}.
The size and shape of the distribution function hollow accompanying the IA soliton remain intact,
confirming the stability of velocity-direction features of IA solitons against mutual collisions. 
However, the symmetry of the distribution function inside the hollow is changing 
into a more and more chaotic form, as the IA soliton passes through more and more collisions.

\begin{figure}
 \begin{tabular}{c}
  \subfloat{\includegraphics[width=0.5\textwidth]{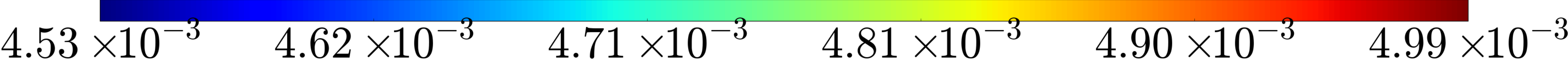} } \\
  \subfloat{\includegraphics[width=0.24\textwidth]{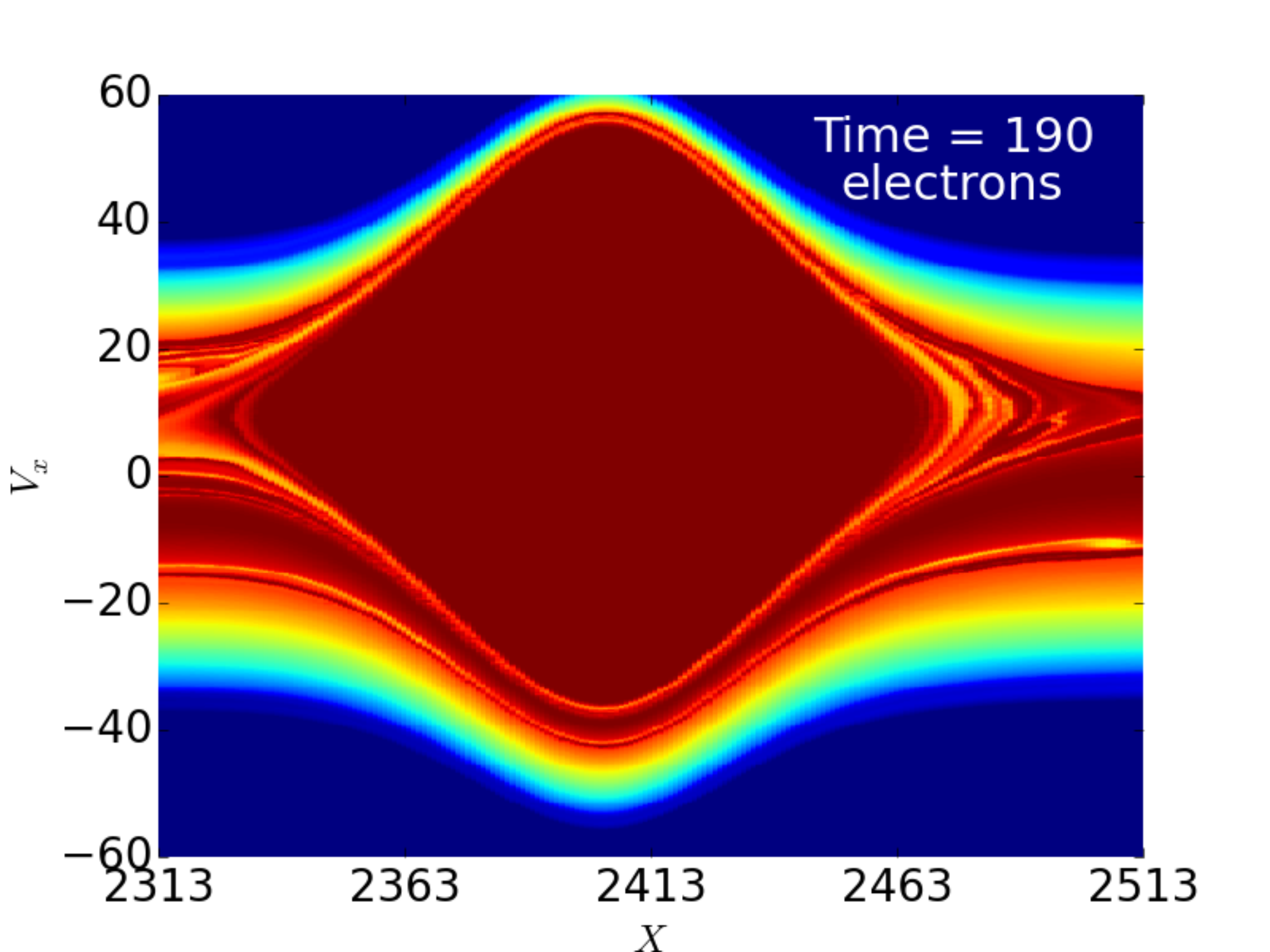} }
  \subfloat{\includegraphics[width=0.24\textwidth]{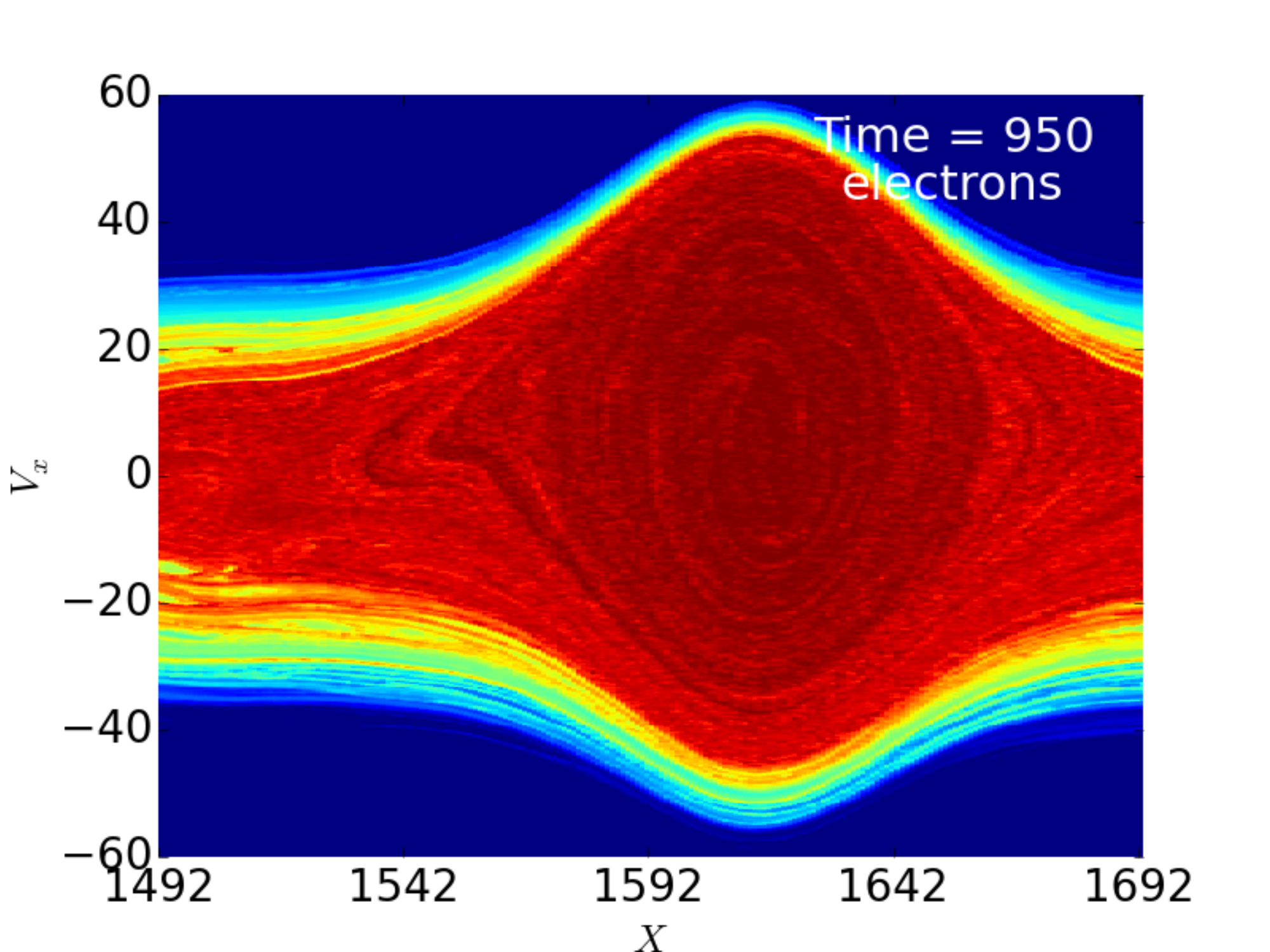} }\\
  \\
  \subfloat{\includegraphics[width=0.5\textwidth]{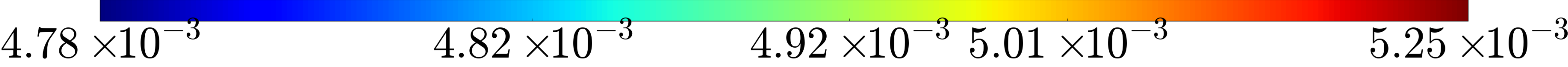} } \\
  \subfloat{\includegraphics[width=0.24\textwidth]{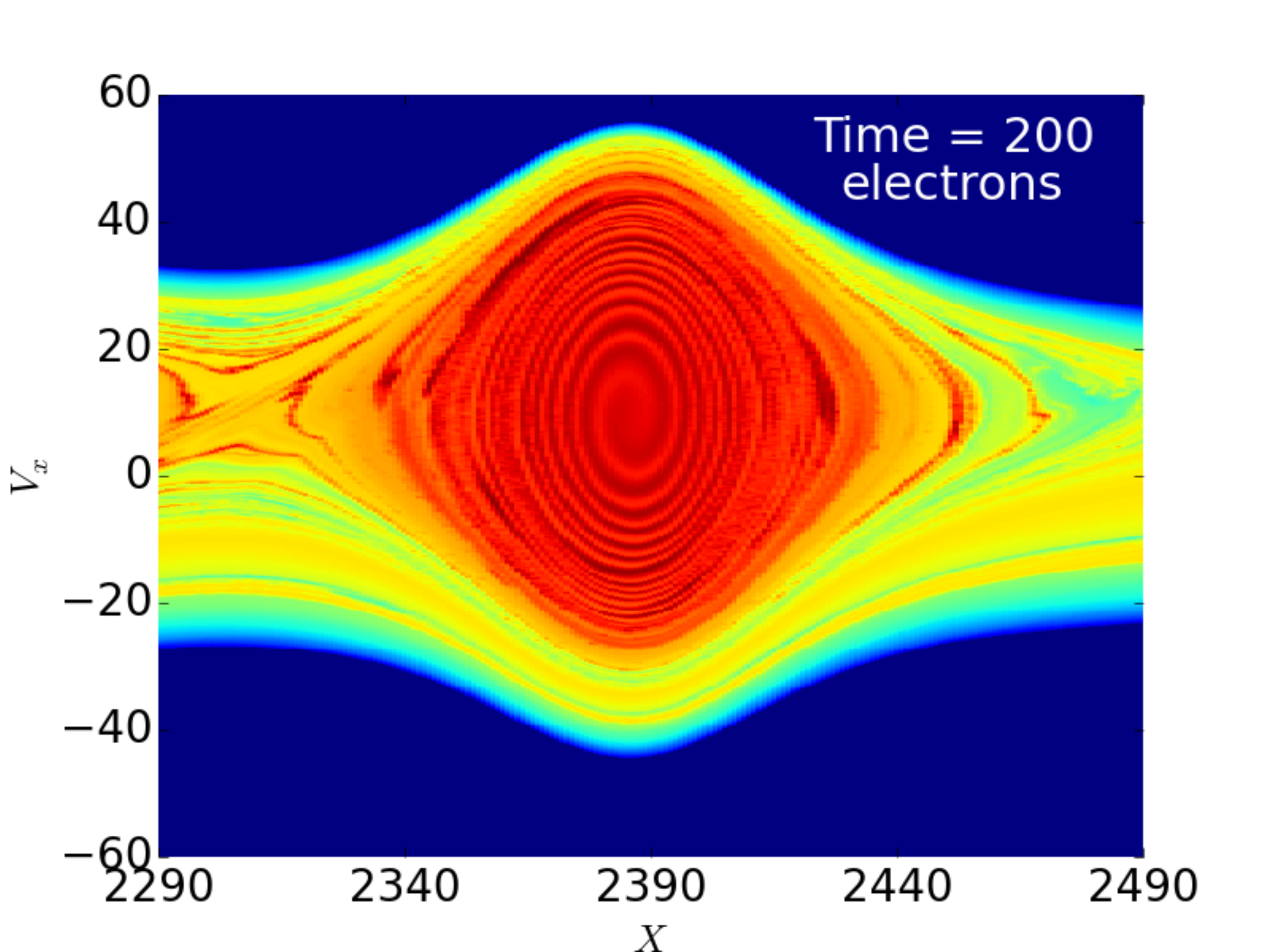} }
  \subfloat{\includegraphics[width=0.24\textwidth]{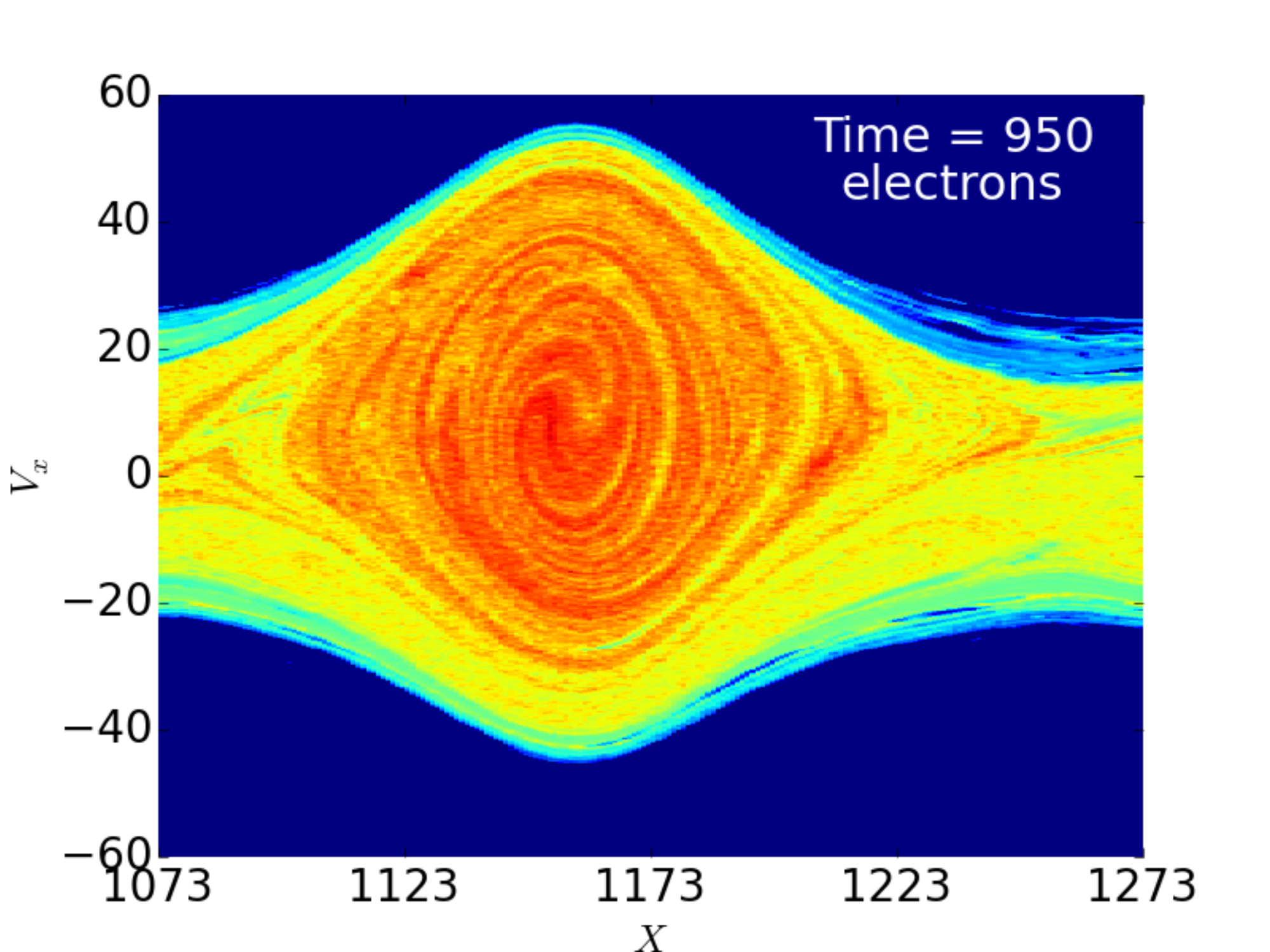} }\\
 \end{tabular}
 \caption{For the case of large IDP, i.e. $\psi=0.2$ and $\Delta=500$, with $\beta = 0.0$ ($\beta = 0.2$),
 the electrons' distribution function are shown in the phase space in the two top (bottom) figures.
 These figures represent the trapping area associated with 
 the right-propagating first/dominant IA soliton. 
 The phase space structure of the plateau ($\beta = 0.0$) and the hump ($\beta = 0.2$)
 is presented before the first collision ($\tau = 190, 200$ respectively) and after 
 the final collision ($\tau = 950)$.
 Stability of IA solitons can be confirmed by comparing the size and shape of nonlinear structures.
 However, their symmetry of distorted due to number of mutual collisions.
 Furthermore, plateau structure prove to be more robust than the other two shapes. 
 }
 \label{Fig_Internal_B0_B02}
\end{figure}

Fig. \ref{Fig_Internal_B0_B02} provides the same results as of Fig. \ref{Fig_Internal_BN01} 
for two other cases, i.e. $\beta = 0.0$ and $\beta = 0.2$.
The same tendency can be witnessed, trapped population becomes more chaotic at the end of simulation
compared to the initial step before the first collision. 
However, the internal structure of plateau structure ($\beta =0$) displays less interruption compared to the two other forms, 
i.e. hollows and humps. 
Hence, we conclude that the plateau structure shows more resilience on the kinetic level during mutual collisions.
\textbf`{This is due to the constant value of distribution function inside the plateau which can hide the spiral movement inside it.}
Moreover, the final snapshot of the hollow and humps at the end of simulations (after a few collisions)
resemble the plateau distribution ($\beta = 0$) due to the increasing distortion in the trapped populations.

The same comparisons have been carried out for $-1<\beta<1$,
specifically $\beta =$ -1.0, -0.5, 0.5 and 1.0.
This range of $\beta$ covers all the three regimes  proposed by Schamel \cite{schamel_3}
as well as all the three possible shapes of the trapped electrons distribution function.
Hence, the stability of IA solitons,
in presence of trapped electrons,
against successive mutual collisions is confirmed which consequently proves Schamel's theory. 

  \subsection{Collision process on kinetic level}\label{section_process}
\begin{figure*}
 \begin{tabular}{c||c}
   
    \subfloat{\includegraphics[width=0.22\textwidth]{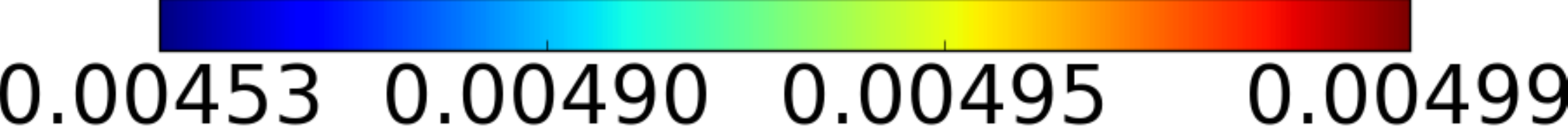} } \ \ \
    \subfloat{\includegraphics[width=0.2\textwidth]{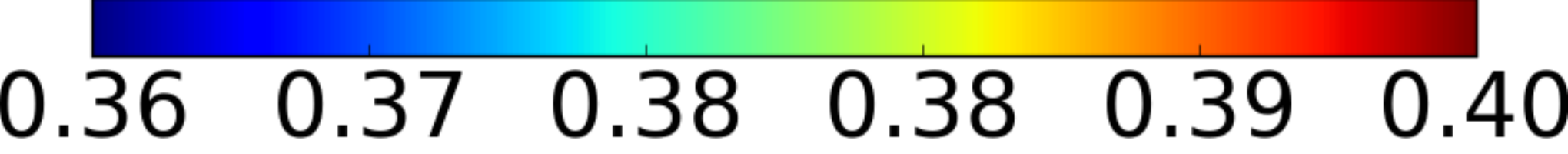} }& 
    \subfloat{\includegraphics[width=0.24\textwidth]{9_13_colorbar_electrons.pdf} } \ \ \ 
    \subfloat{\includegraphics[width=0.2\textwidth]{9_13_colorbar_ions.pdf} }\\
    
    \subfloat{\includegraphics[width=0.24\textwidth]{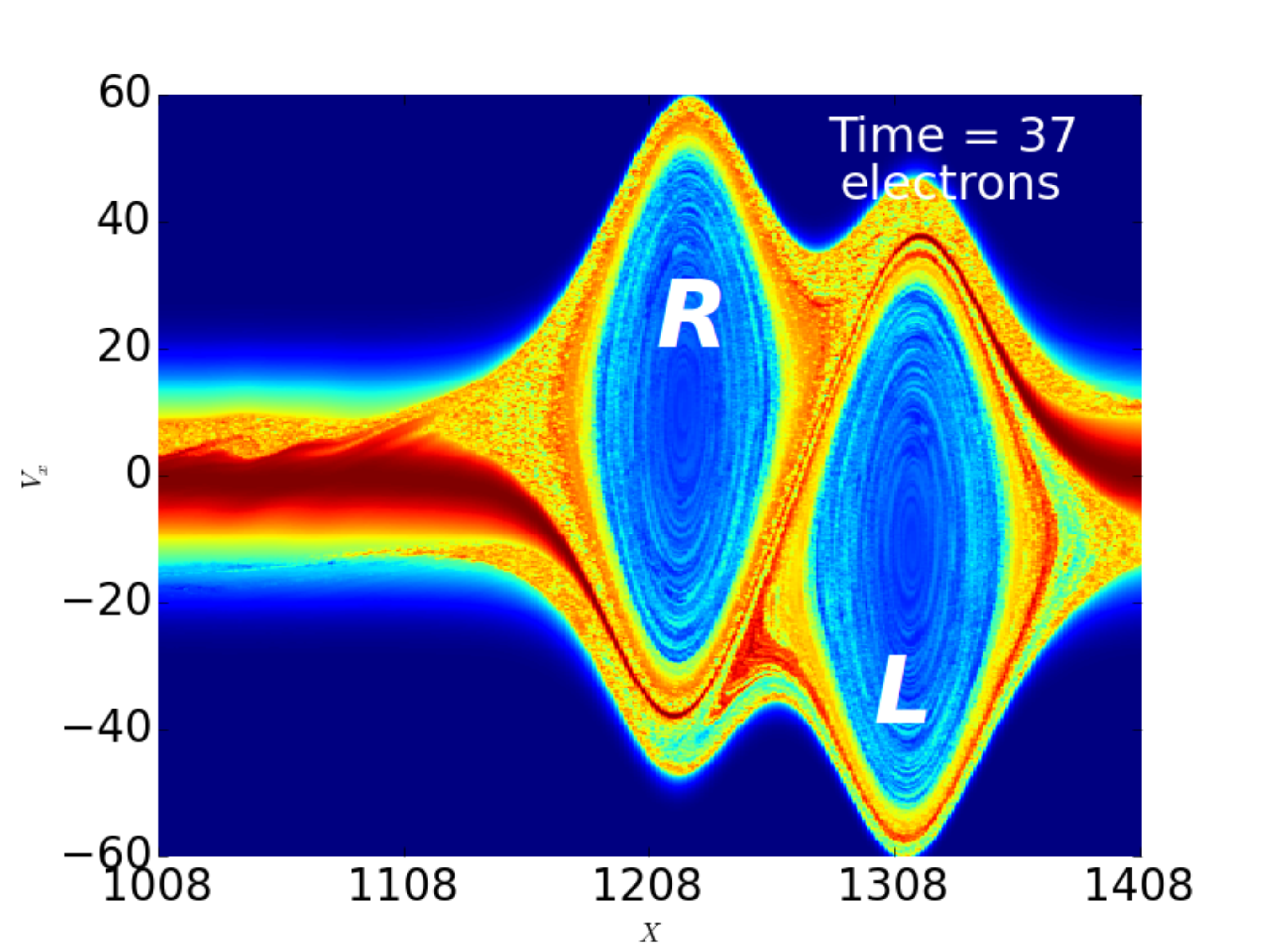} }
    \subfloat{\includegraphics[width=0.24\textwidth]{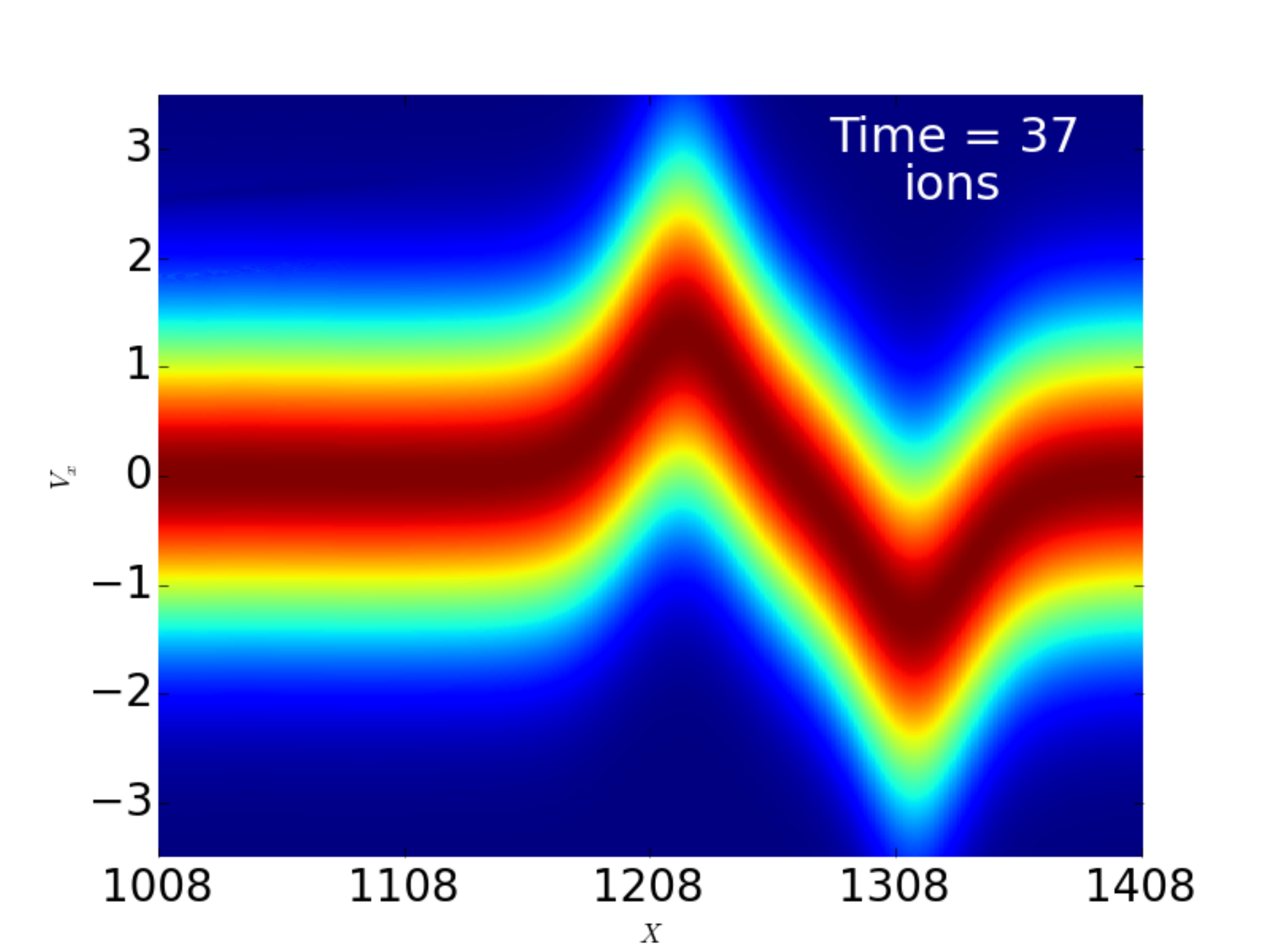} }& 
    \subfloat{\includegraphics[width=0.24\textwidth]{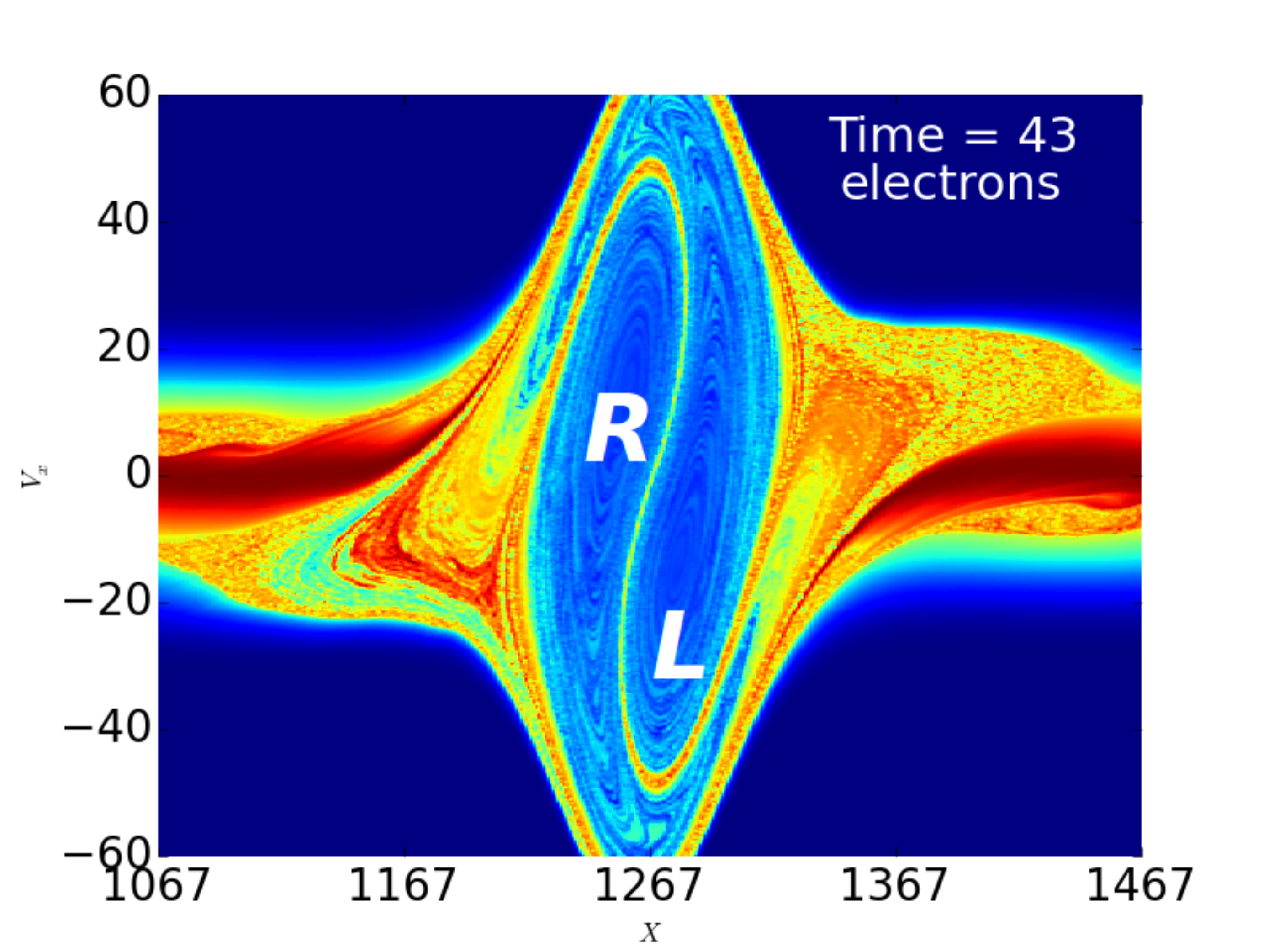} }
    \subfloat{\includegraphics[width=0.24\textwidth]{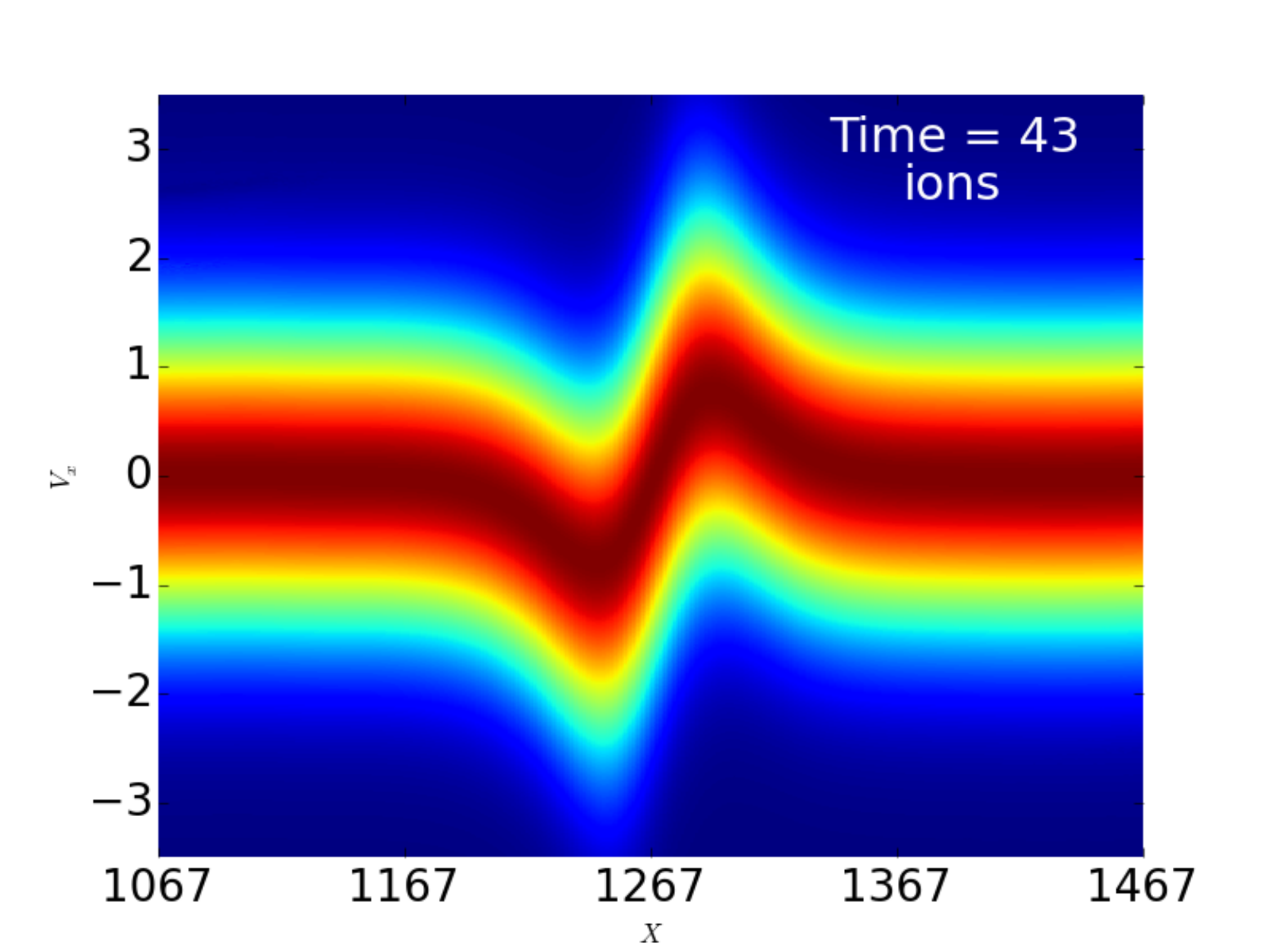} }\\
 
 \subfloat{\includegraphics[width=0.24\textwidth]{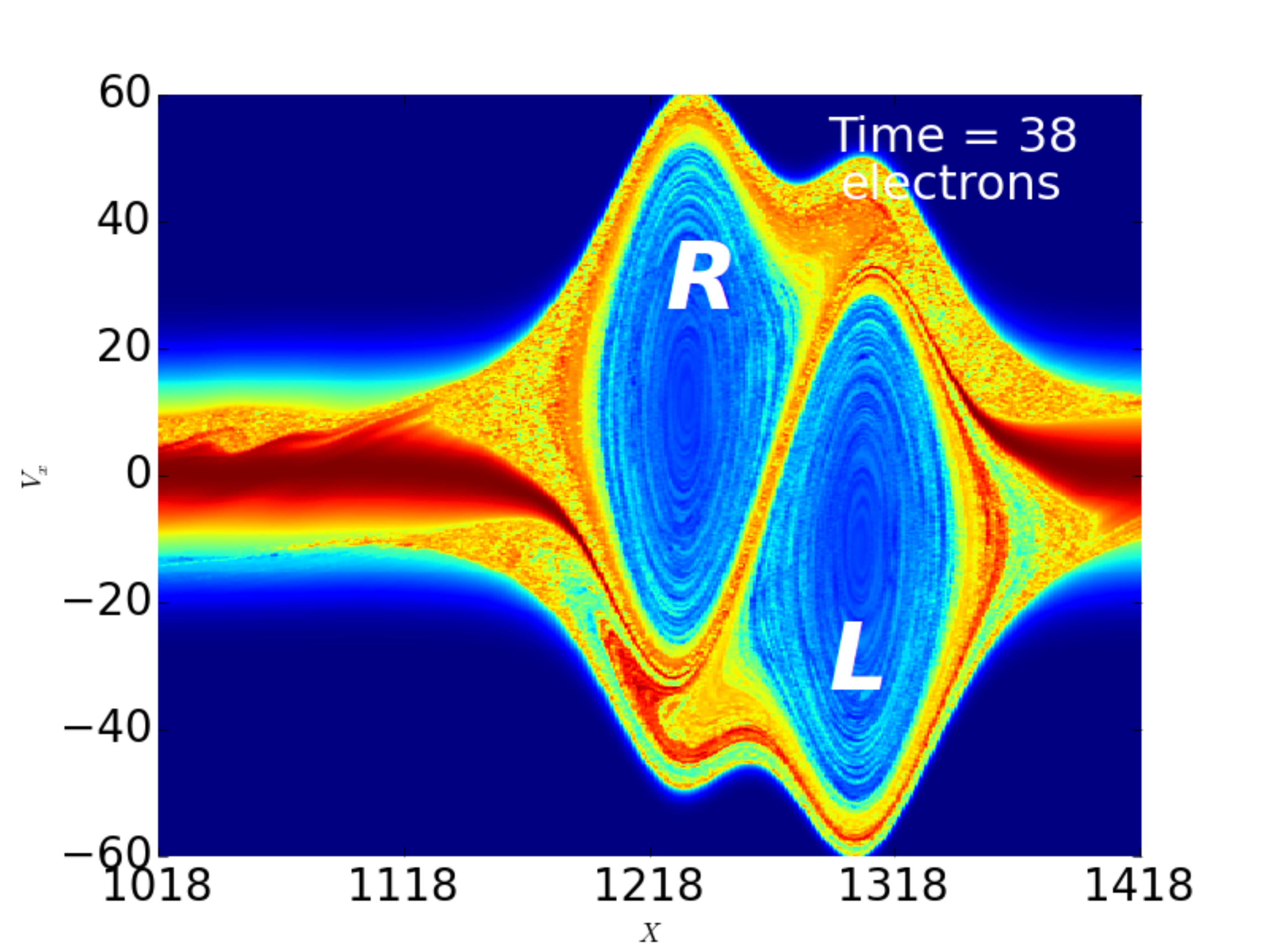} }
 \subfloat{\includegraphics[width=0.24\textwidth]{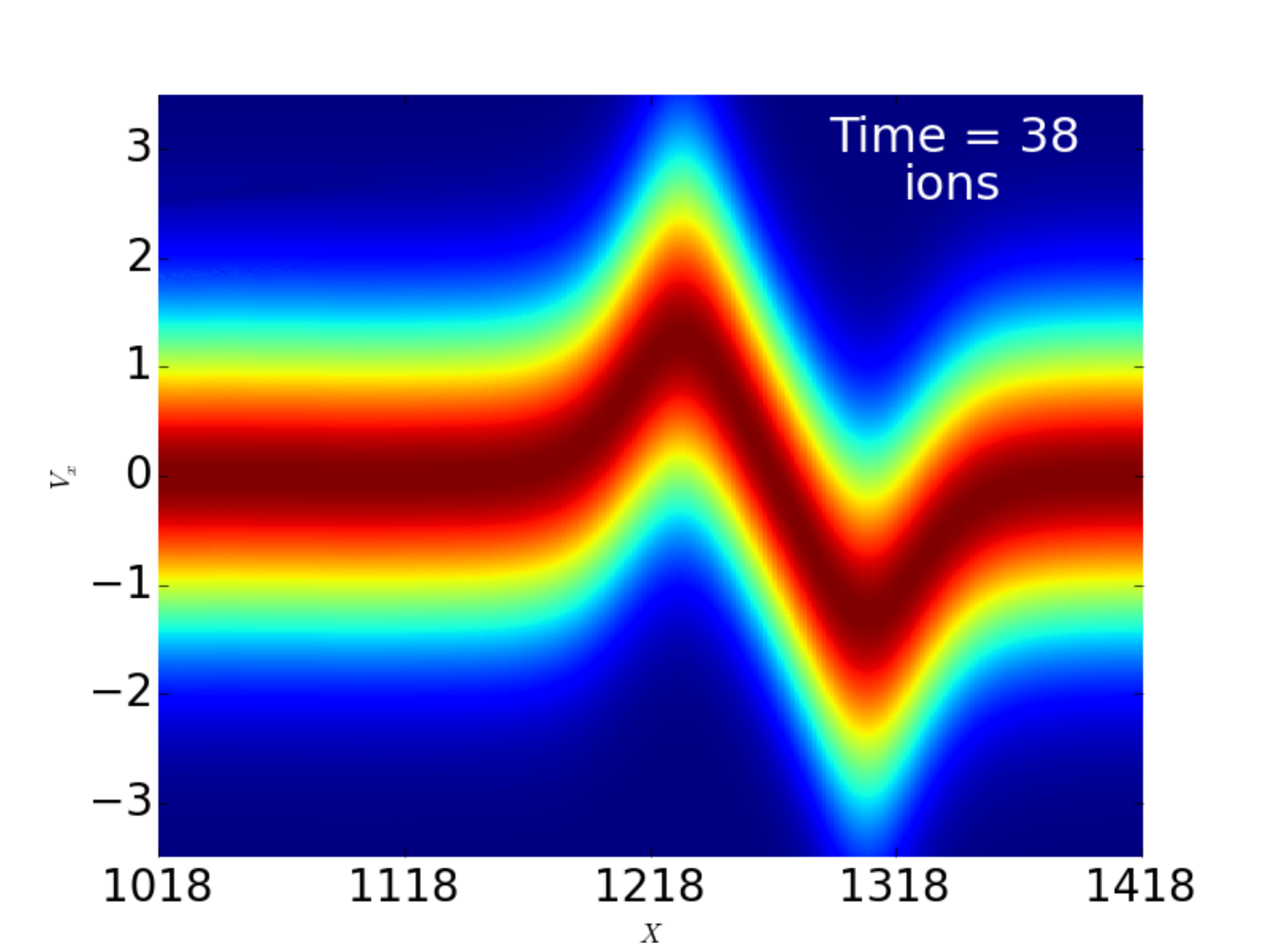} }& 
  \subfloat{\includegraphics[width=0.24\textwidth]{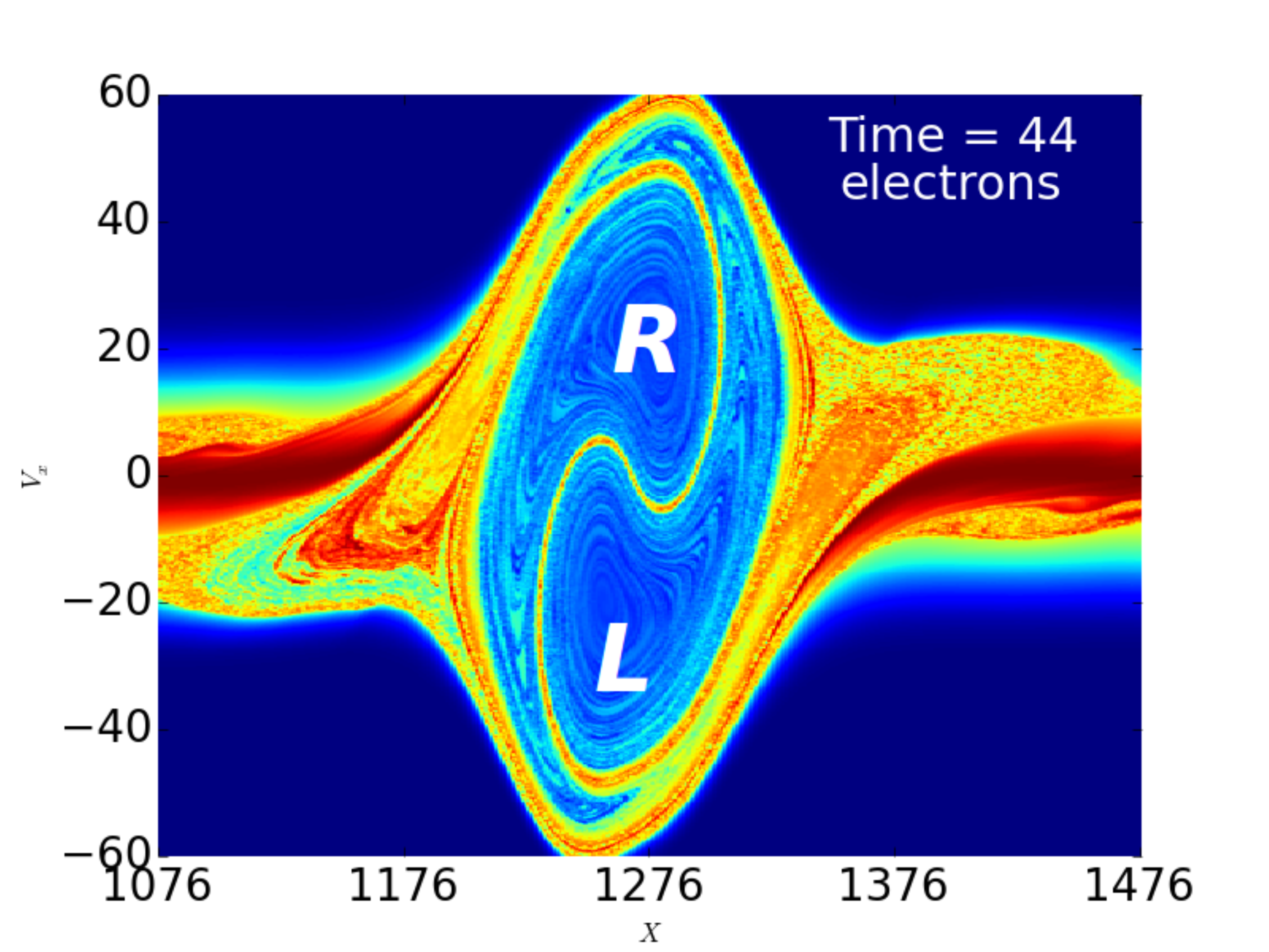} }
 \subfloat{\includegraphics[width=0.24\textwidth]{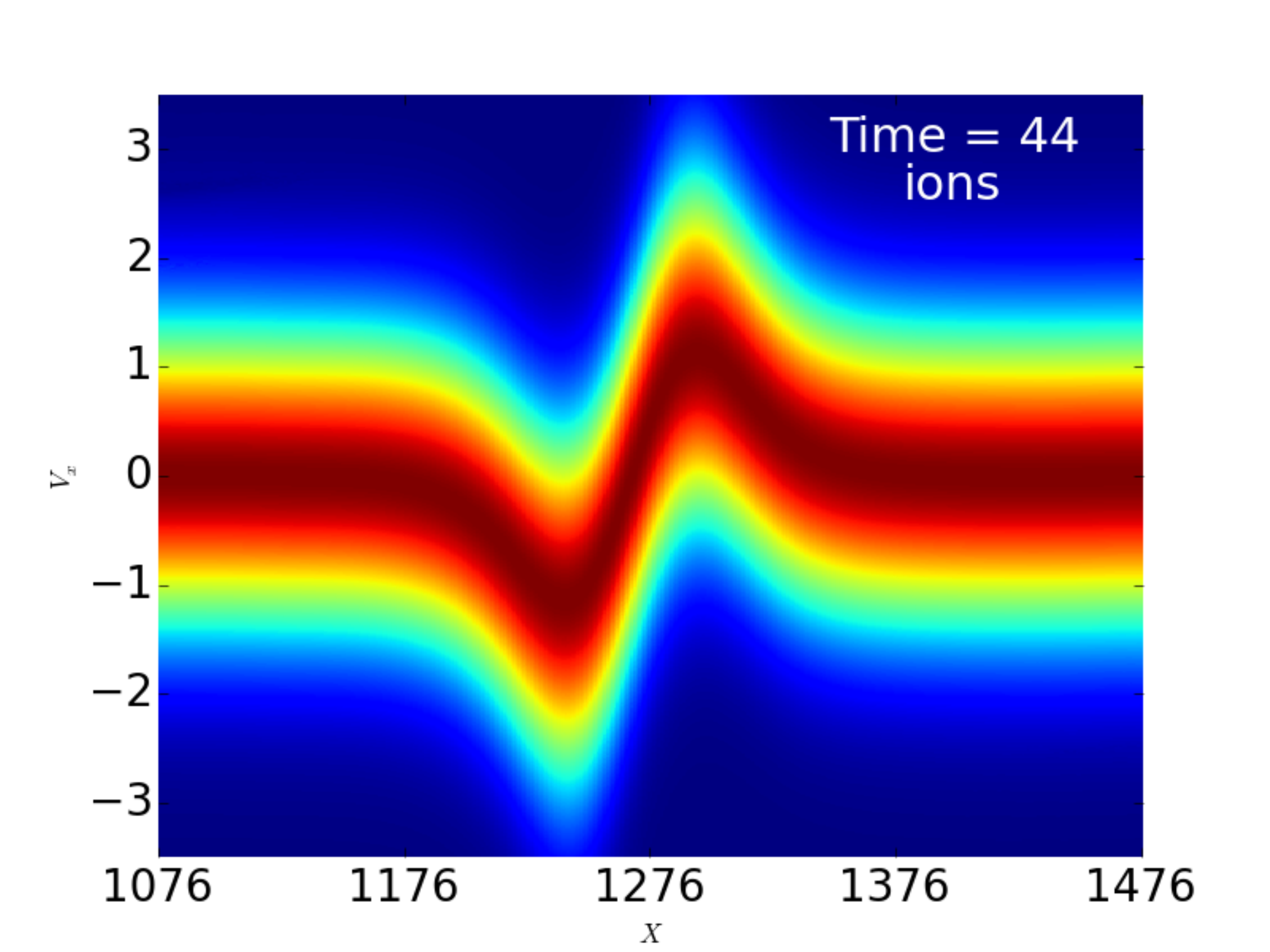} }\\
 
 \subfloat{\includegraphics[width=0.24\textwidth]{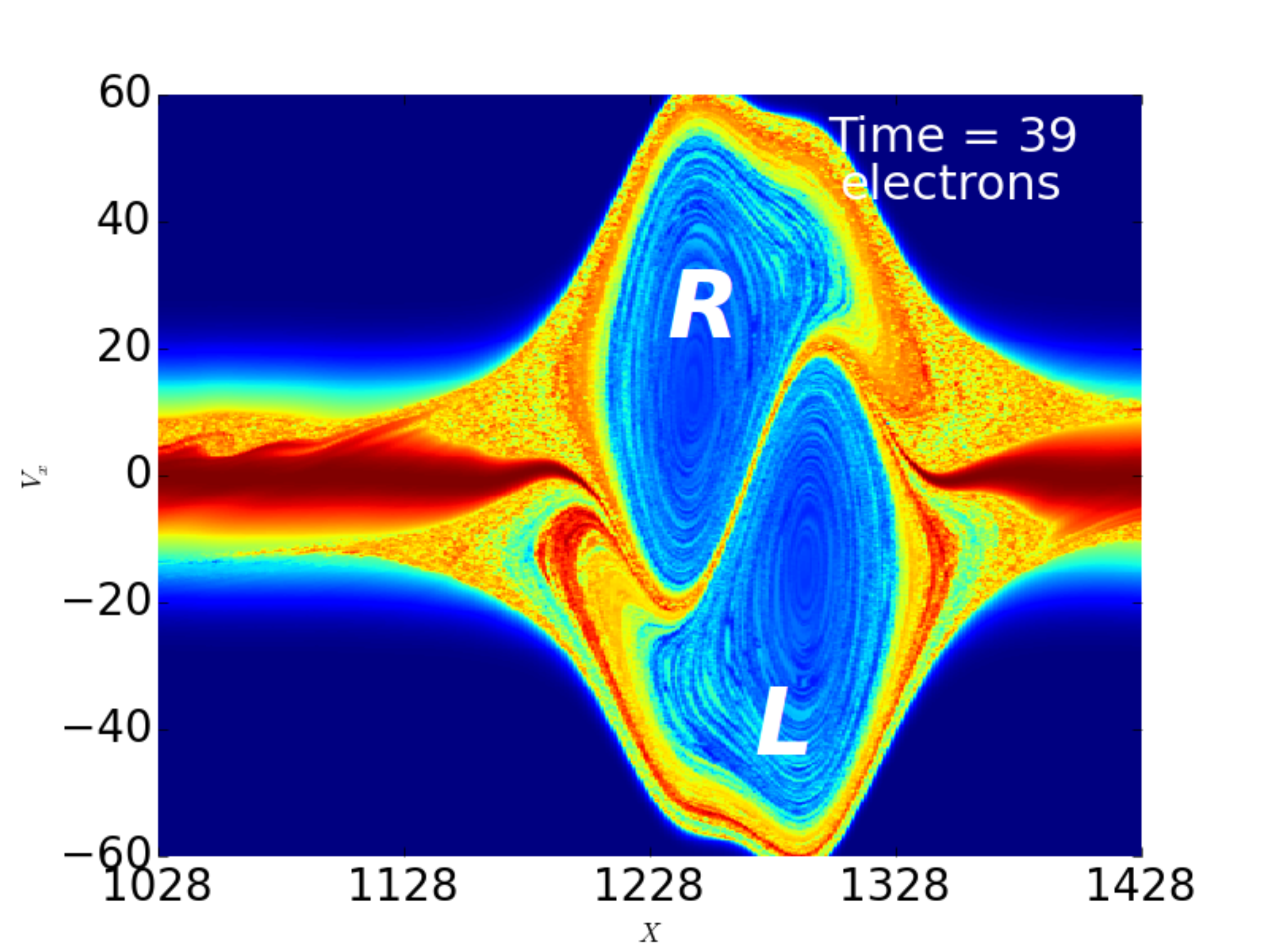} }
 \subfloat{\includegraphics[width=0.24\textwidth]{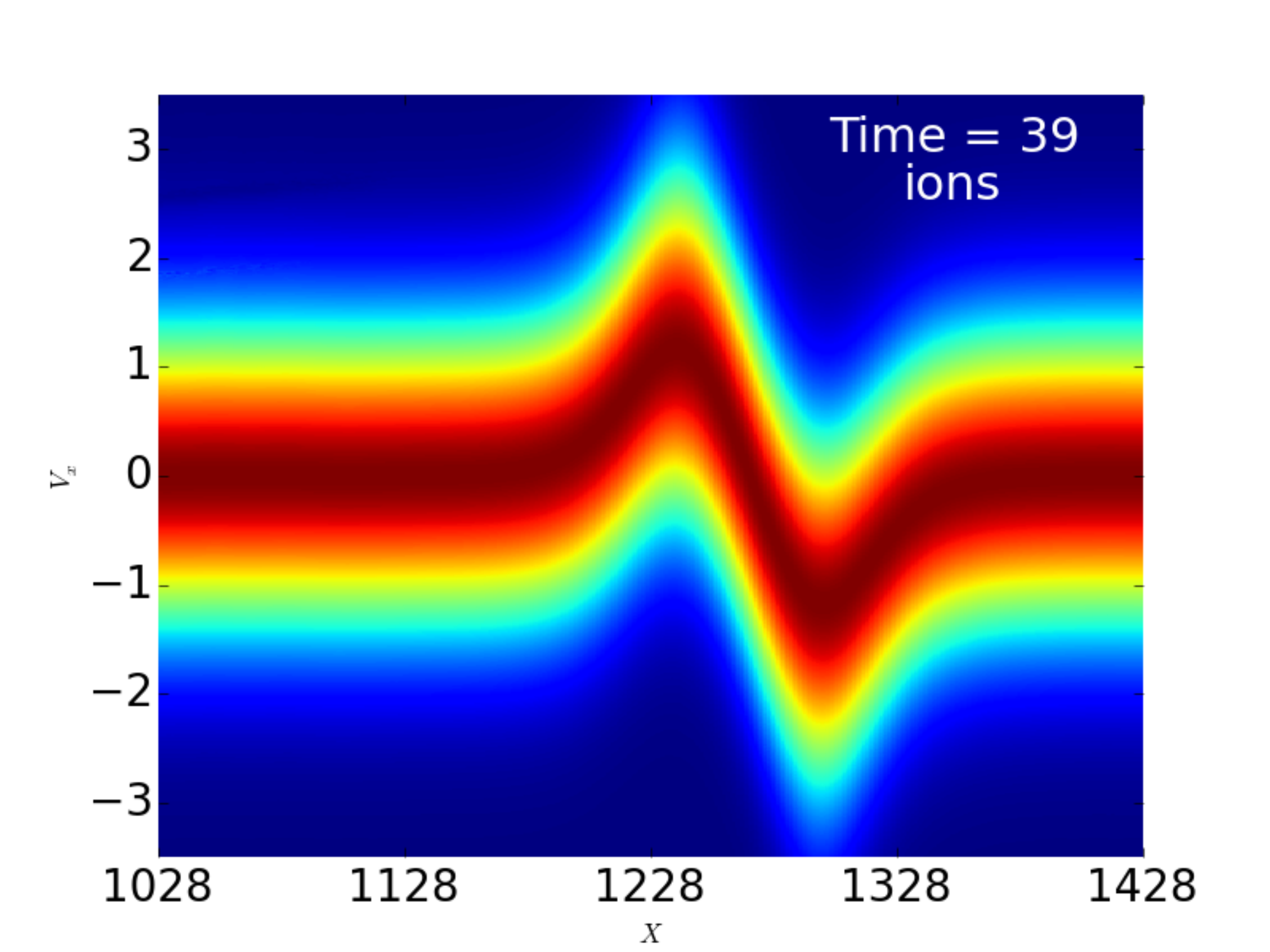} }& 
  \subfloat{\includegraphics[width=0.24\textwidth]{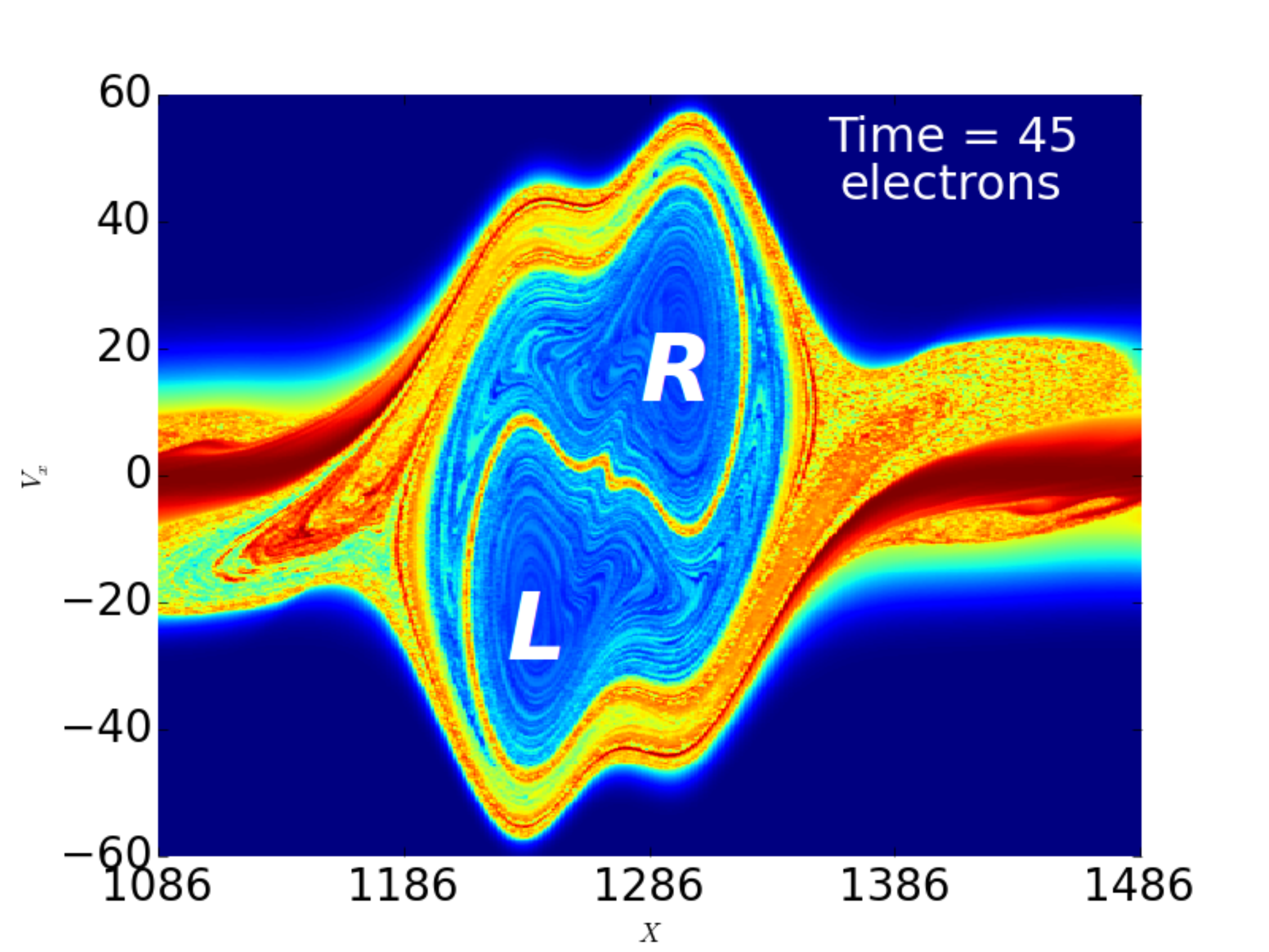} }
 \subfloat{\includegraphics[width=0.24\textwidth]{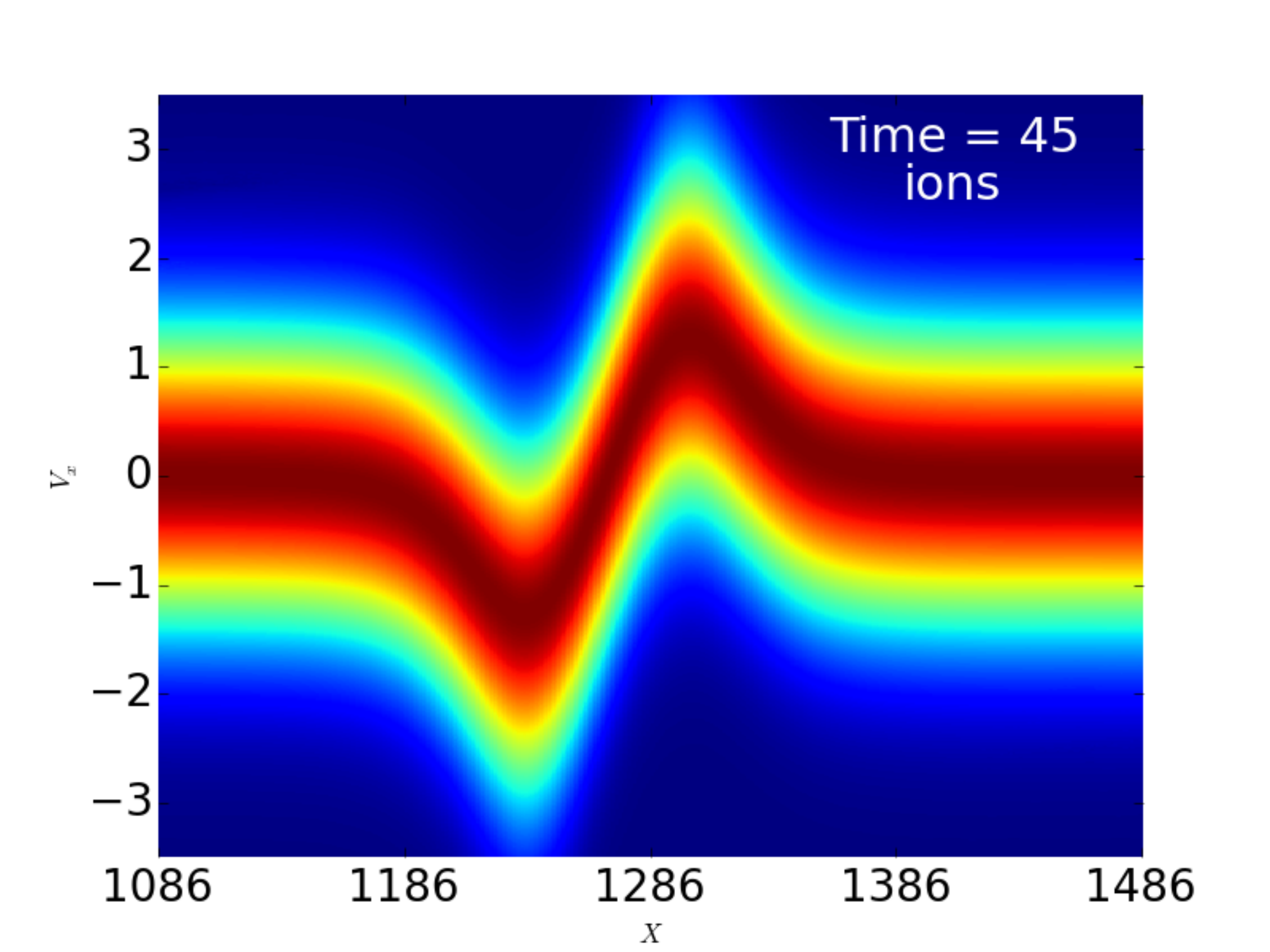} }\\

 \subfloat{\includegraphics[width=0.24\textwidth]{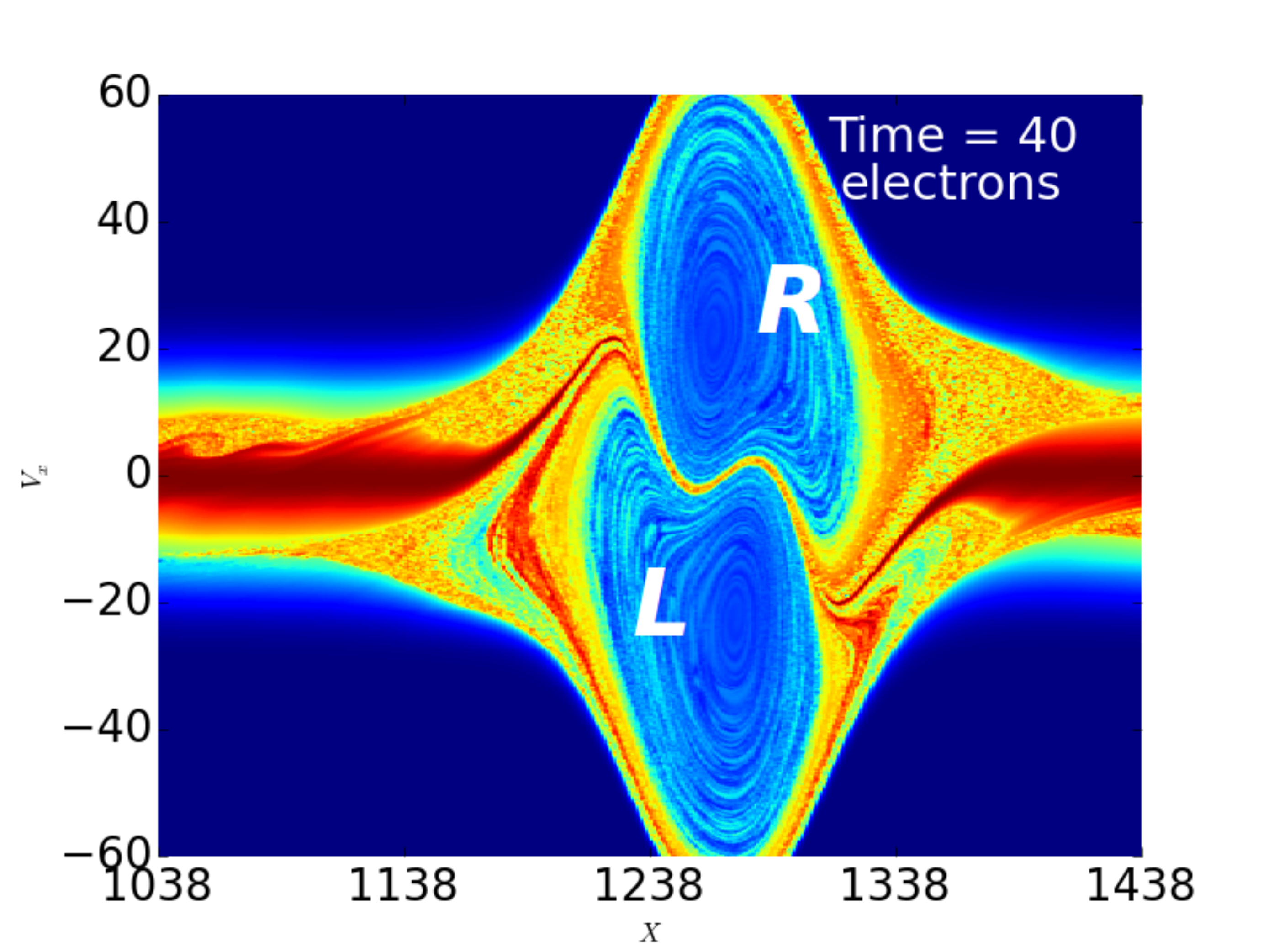} }
 \subfloat{\includegraphics[width=0.24\textwidth]{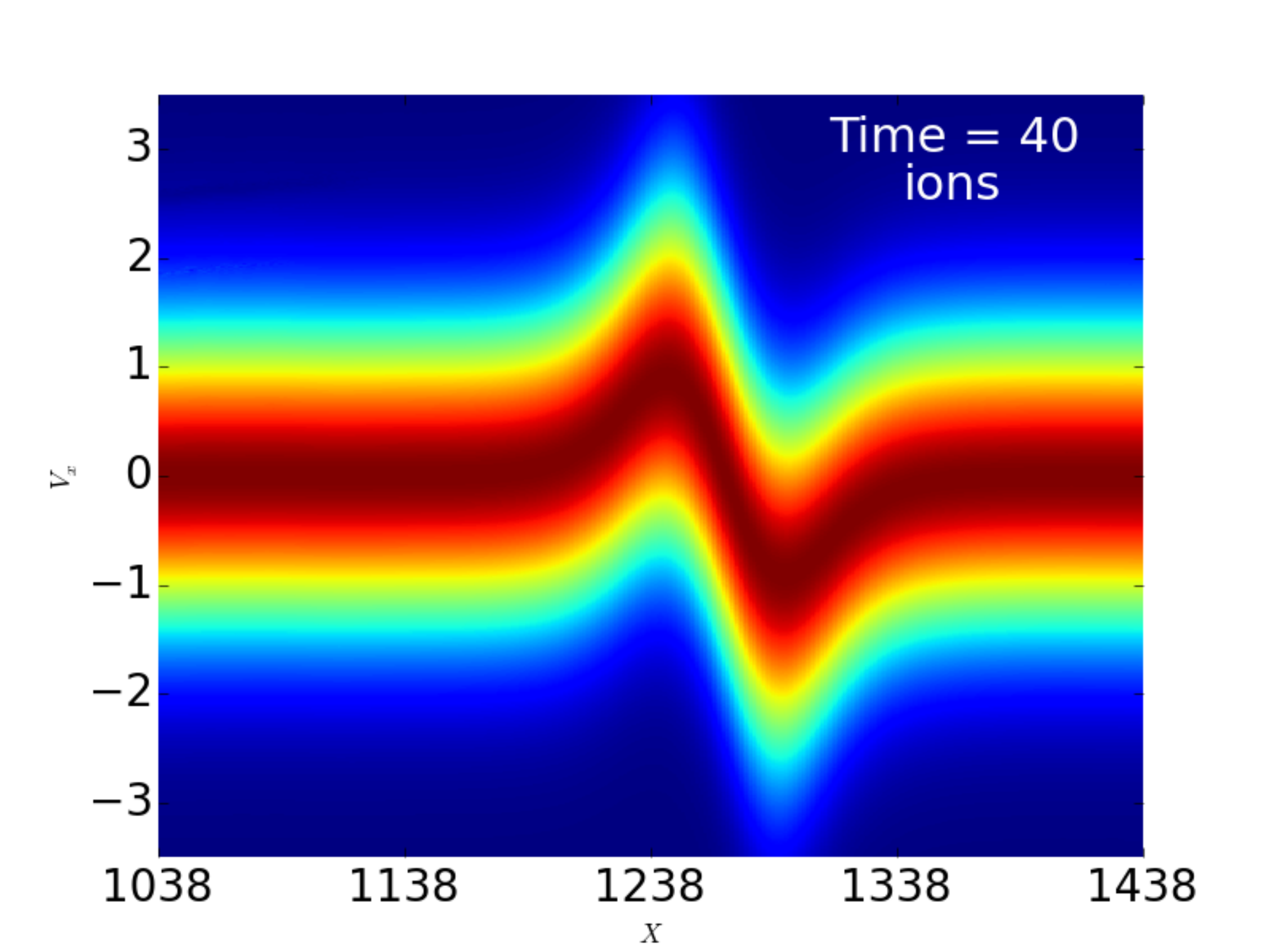} }&
  \subfloat{\includegraphics[width=0.24\textwidth]{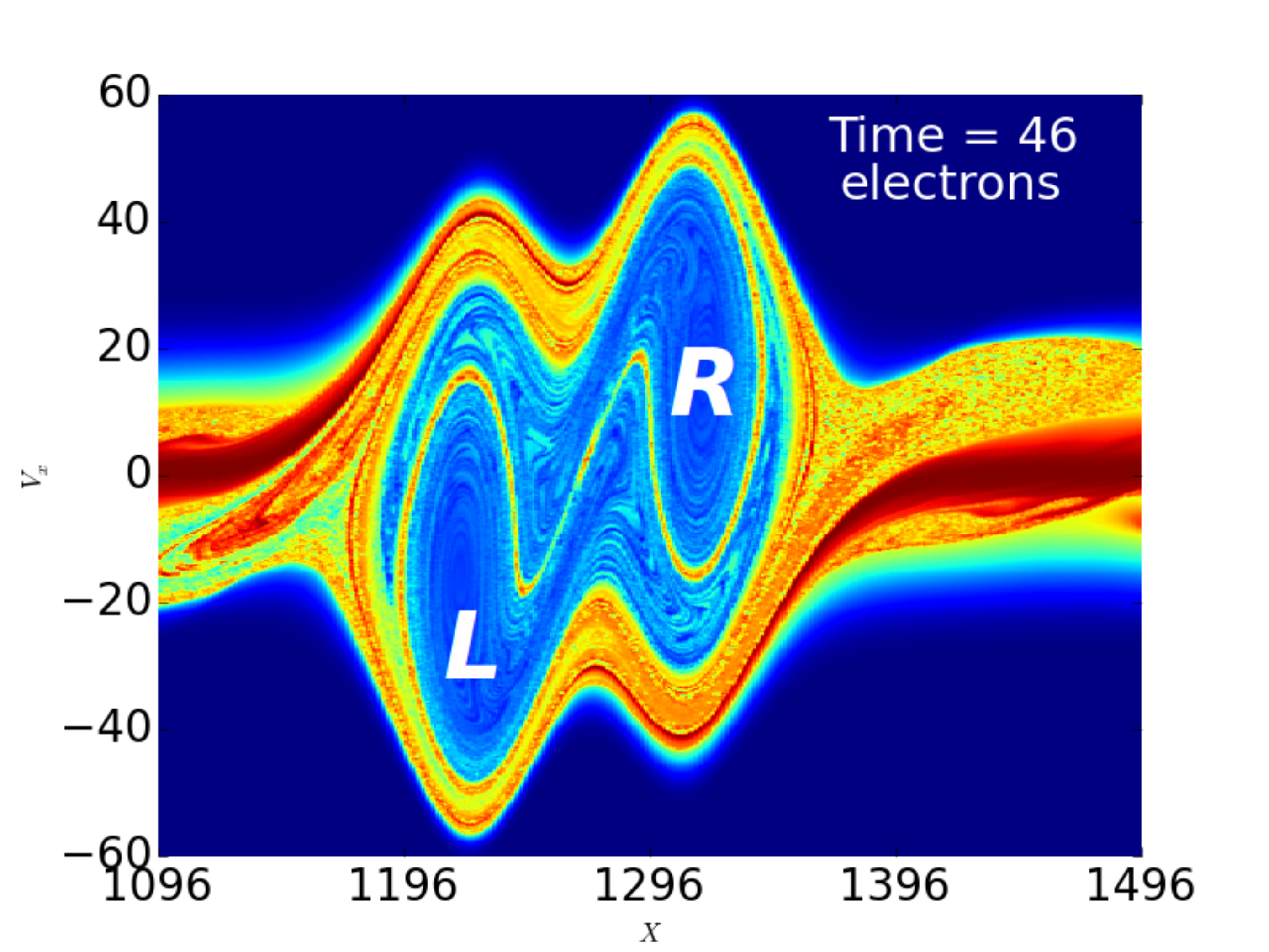} }
 \subfloat{\includegraphics[width=0.24\textwidth]{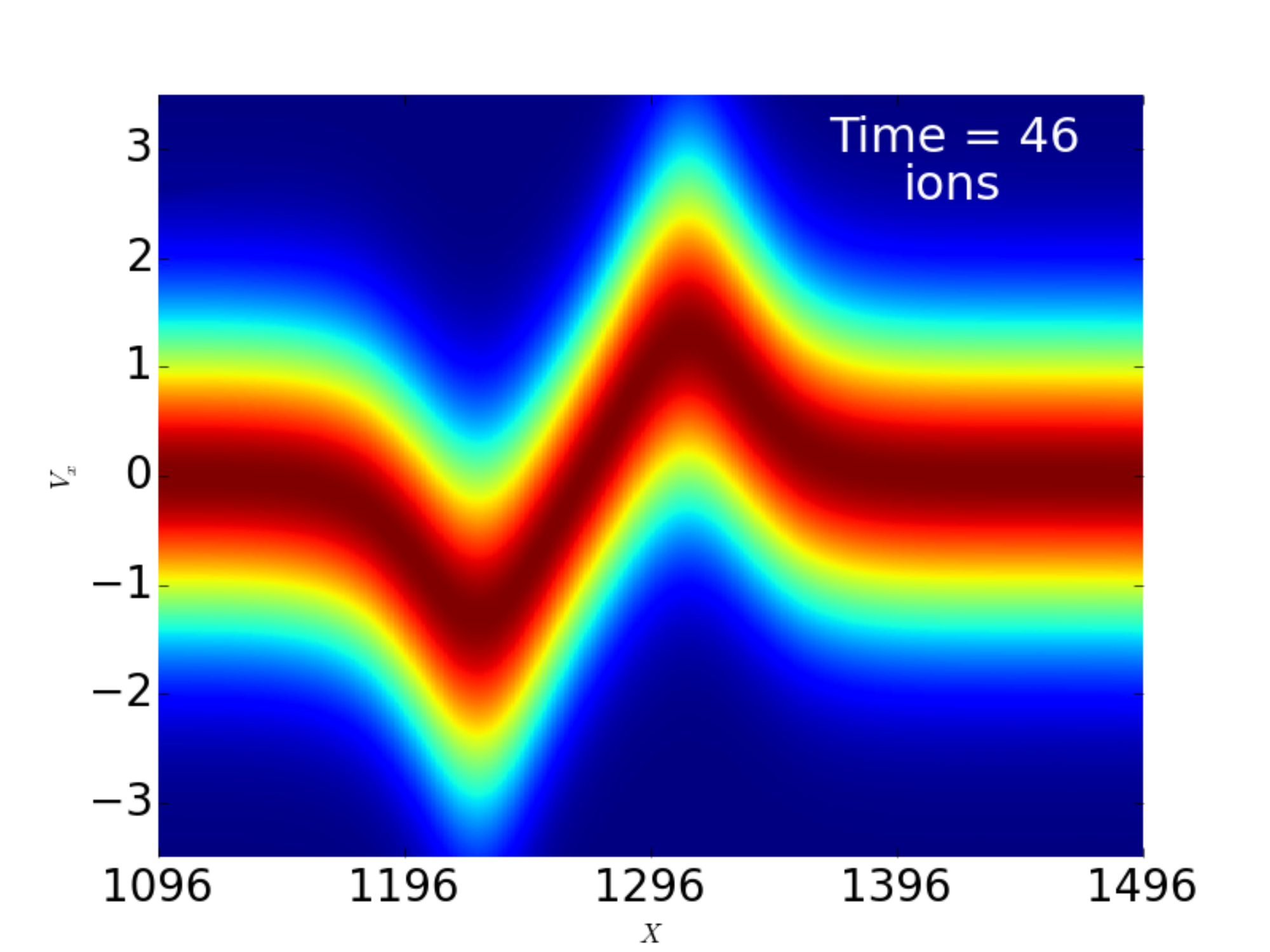} }\\
 
 \subfloat{\includegraphics[width=0.24\textwidth]{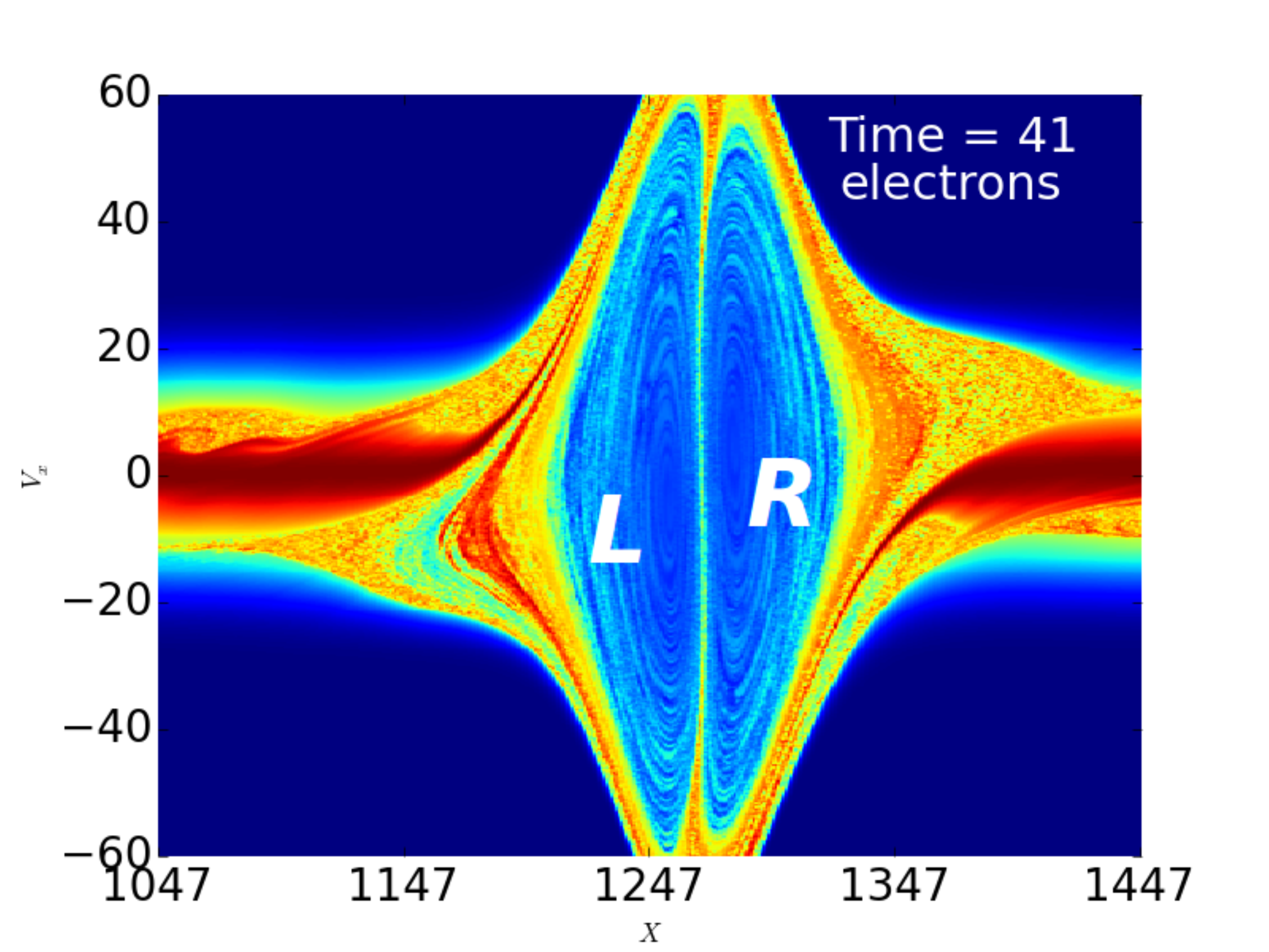} }
 \subfloat{\includegraphics[width=0.24\textwidth]{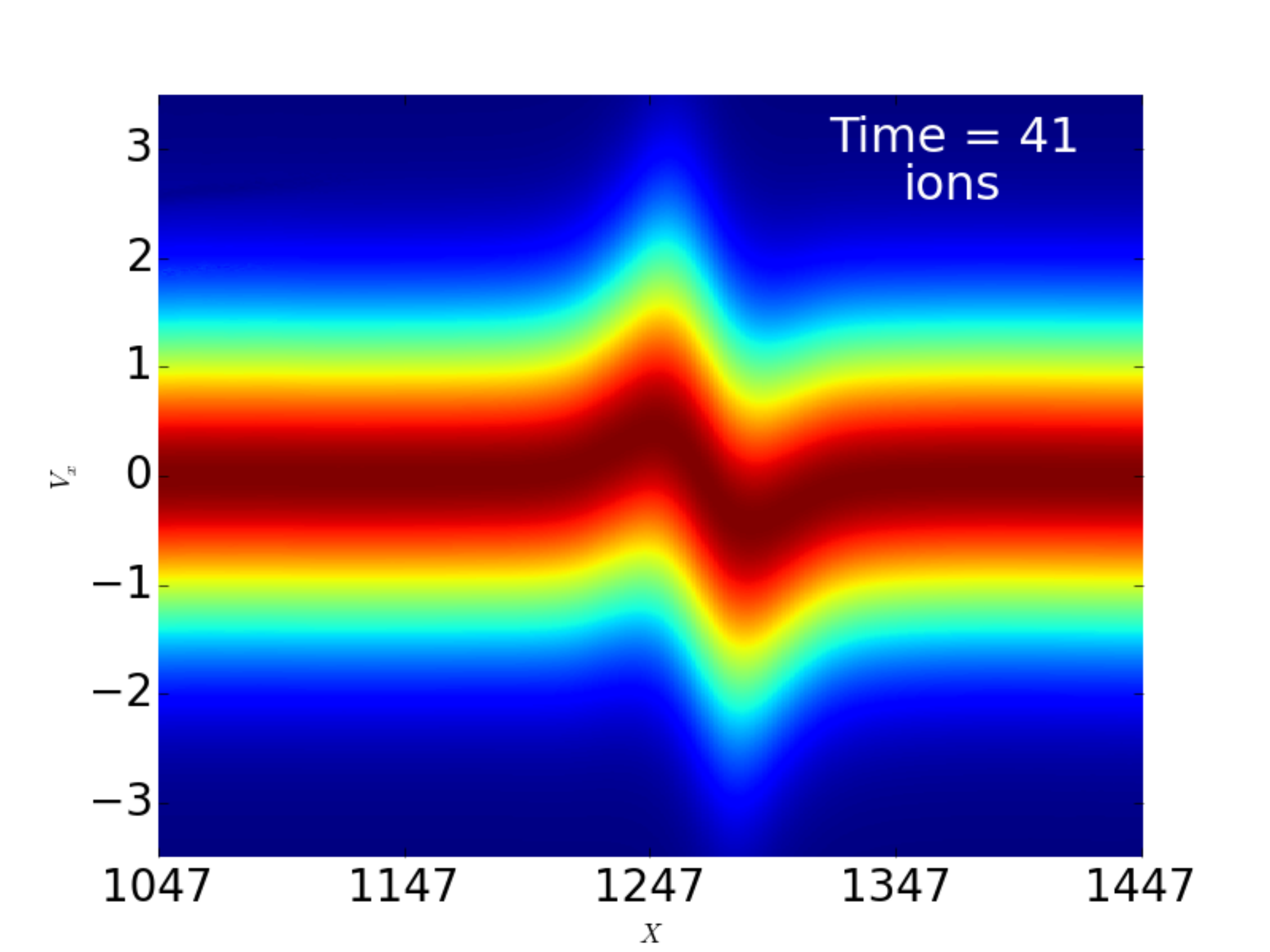} }&
  \subfloat{\includegraphics[width=0.24\textwidth]{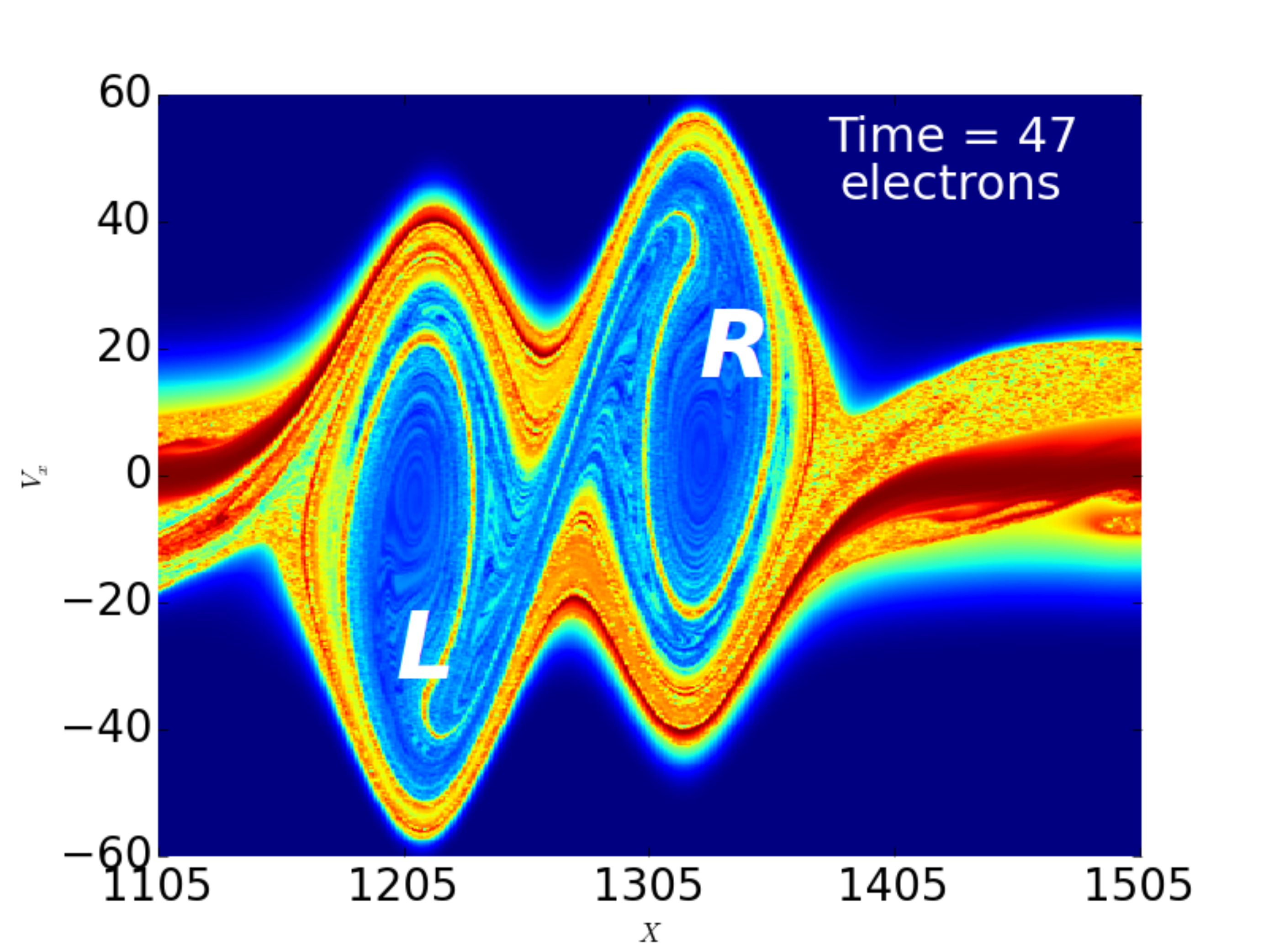} }
 \subfloat{\includegraphics[width=0.24\textwidth]{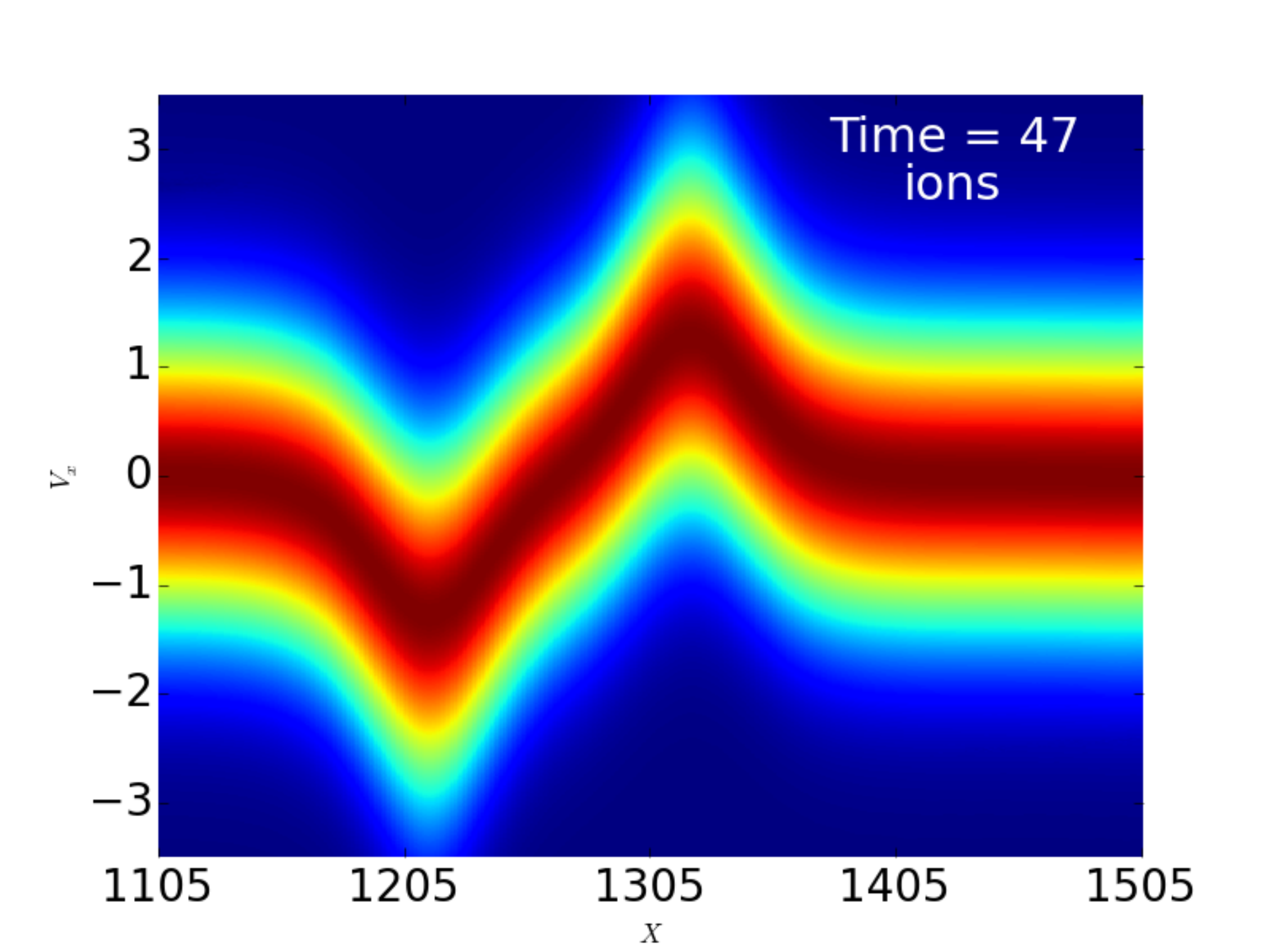} }\\
 
 \subfloat{\includegraphics[width=0.24\textwidth]{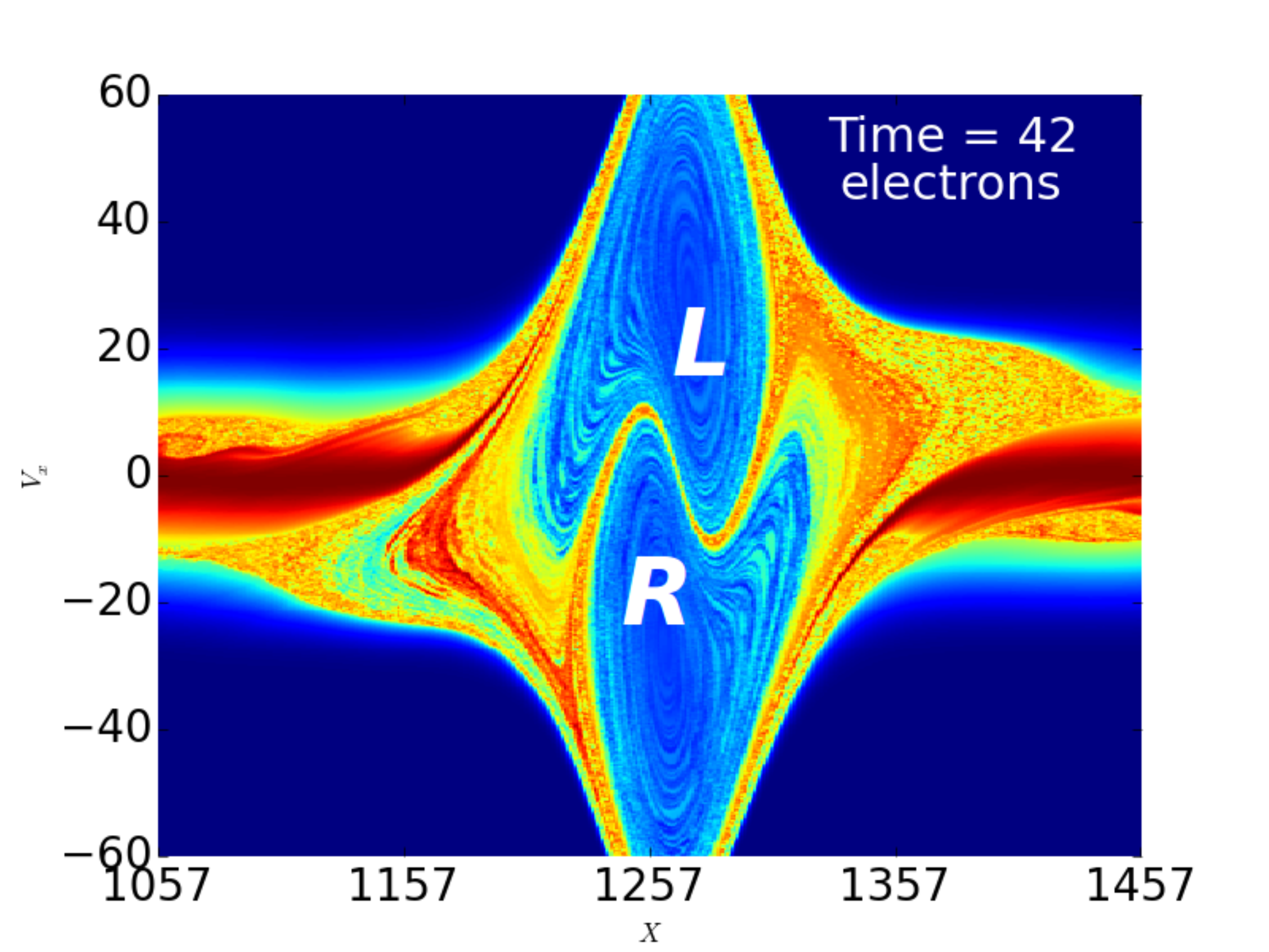} }
 \subfloat{\includegraphics[width=0.24\textwidth]{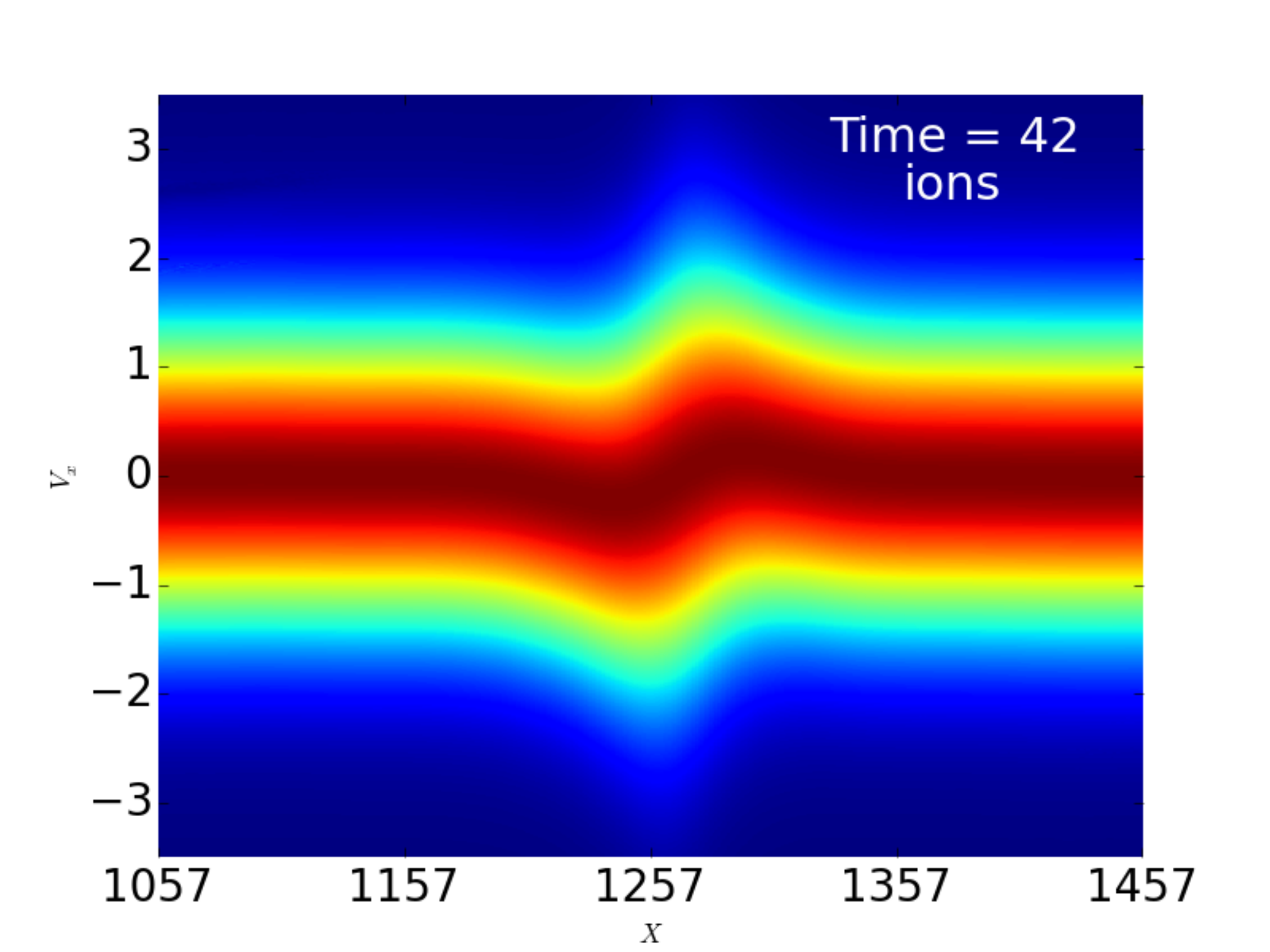} }&
  \subfloat{\includegraphics[width=0.24\textwidth]{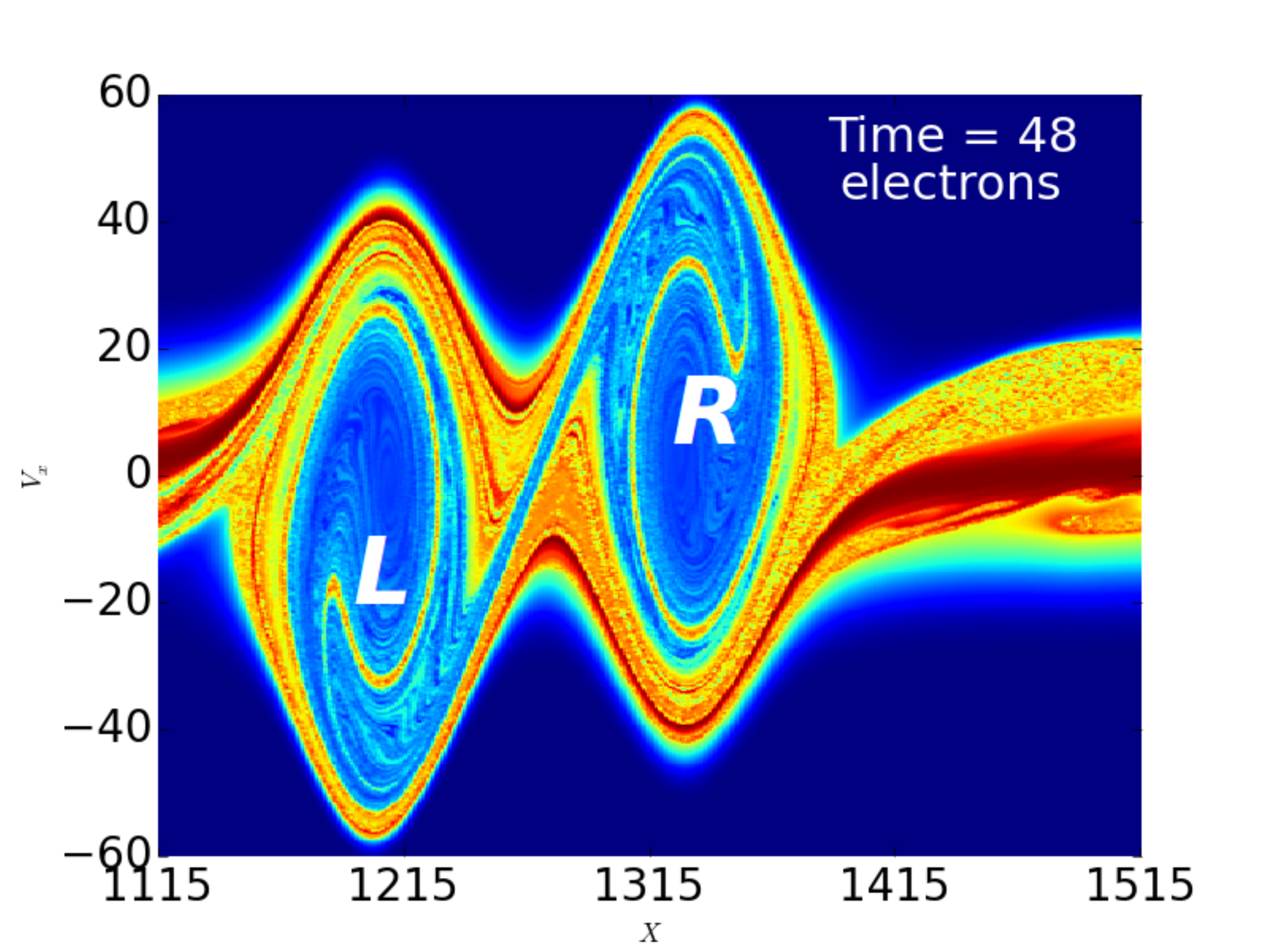} }
 \subfloat{\includegraphics[width=0.24\textwidth]{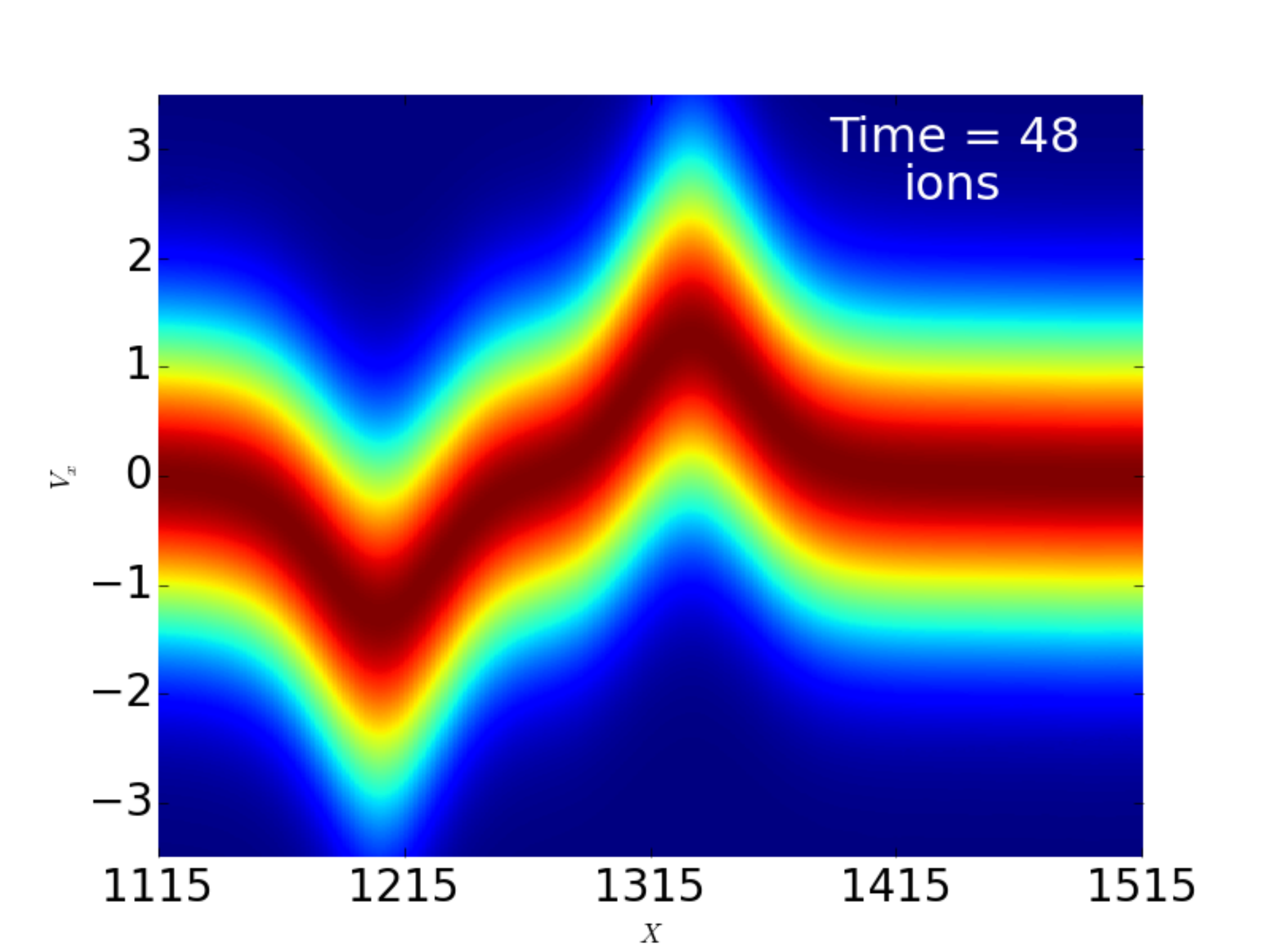} }\\
 \end{tabular}
 
 \caption{\textbf{collision of hollows}; for the case of $\beta = -0.1$, 
 the distribution functions of electron and ions are presented during a collision between 
 right (\textbf{R}) and left (\textbf{L}) propagating IA solitons. 
 The frame is moving along side the right-propagating IA soliton
 and shows the times from $\tau=37$ upto $\tau=48$ (arranged from top left corner).
 The collision for the electrons appears as a rotation of both phase-space hollows around their collective center of mass. 
 Furthermore two solitons exchange a portion of their trapped population during collision. 
 However for the ions, the collision simply consists of two displacements moving through each other. 
 \textbf{See Supplemental Material at [\textit{URL will be inserted by publisher for ``Collision\_two\_negative\_beta.avi''}] 
 for complete successive steps of this collision, i.e. $0<\tau<130$.}
 }
 \label{collision_hollow}
\end{figure*}
In order to show the kinetic details of a collision between two solitons, 
we have carried out another set of simulations, in which the two oppositely propagating solitons are 
isolated from the chain formation simulation and are introduced into a new simulation box. 
Hence, the collision happens purely in pairs.
This removes the effect of secondary phenomena coming from the chain formation process
(such as wavepackets, dribs and other solitons)
from the collisions. 
\textbf{The time of extraction of the first soliton from the chain formation simulation is chosen $\tau <200$,
just before the first set of collisions.}

Fig.\ref{collision_hollow} presents the simulation results for 
the first collision between the two first/dominant IA solitons (marked as \textbf{R} and \textbf{L}) propagating oppositely for the case of $\beta = -0.1$, $\psi=0.2$ and $\Delta=500$. 
Due to the negative value of trapping parameter, IA solitons are accompanied by a \textit{hollow} in electron distribution function.
The right and left propagating IA solitons collide at time $37<\tau<42$.
During this time they undergo one rotation around their collective center of mass. 
During collision, this rotation
has been observed for all the other collisions of the first set of simulations (see. Fig. \ref{Fig_amplitude}) as well,
e.g. $180<\tau<250$, $390<\tau<400$, $590<\tau<600$ and $790<\tau<800$.
The rotation of the hollows in phase space around each other has been witnessed before in context of beam instability. 
Especially in the case of the two-beam instability,
hollows in the phase space would attract each other, rotate and merge together two by two until they form one hollow\cite{krasovsky1999interaction}. 
Fig.\ref{collision_hollow} also reports the phase space of ions 
and their behavior during collision, which is rather simple compared to electrons. 

\begin{figure}
 \subfloat{\includegraphics[width=0.5\textwidth]{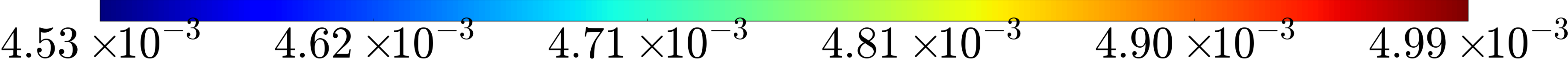} } \\
 \subfloat{\includegraphics[width=0.24\textwidth]{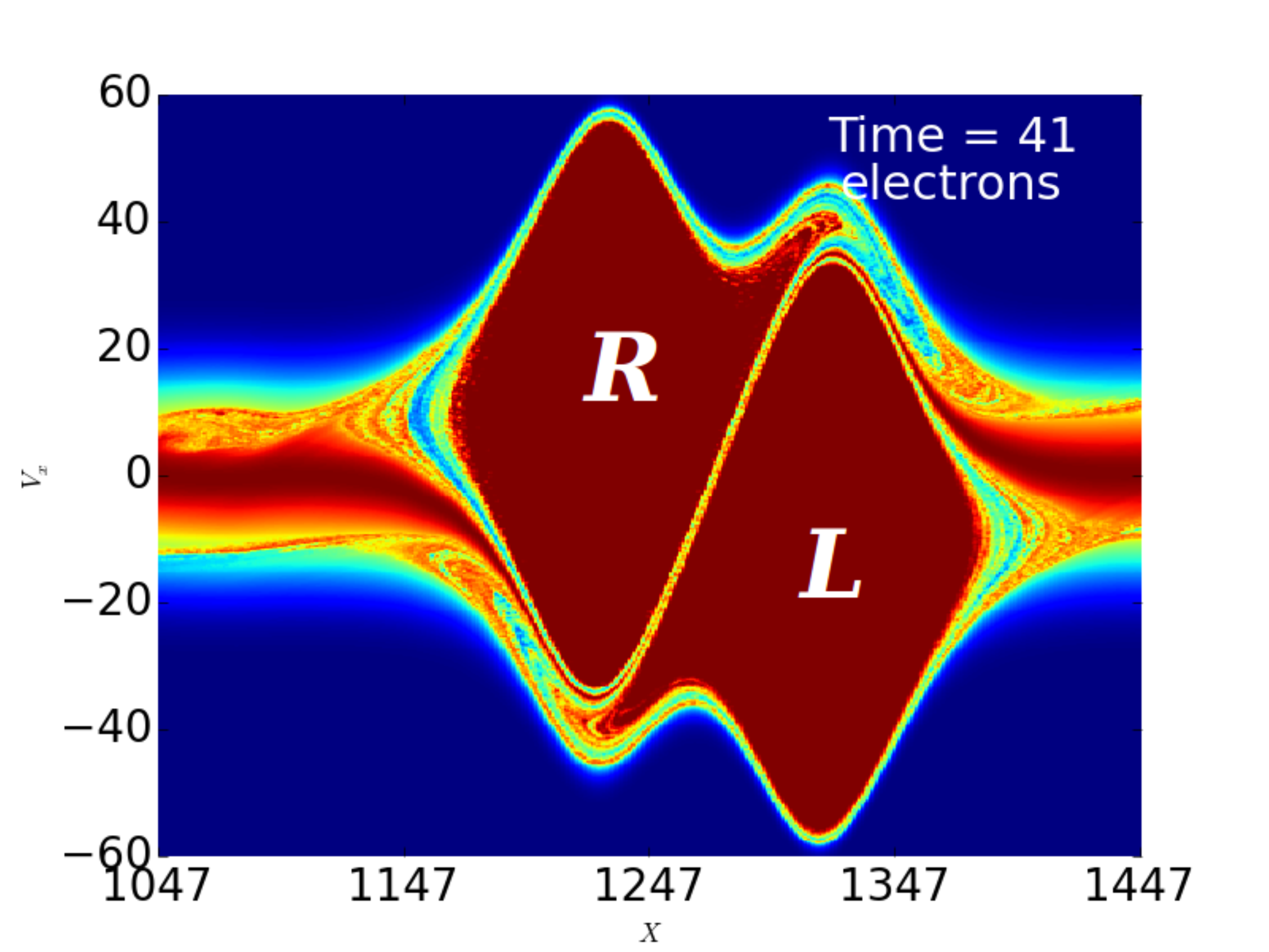} }
 \subfloat{\includegraphics[width=0.24\textwidth]{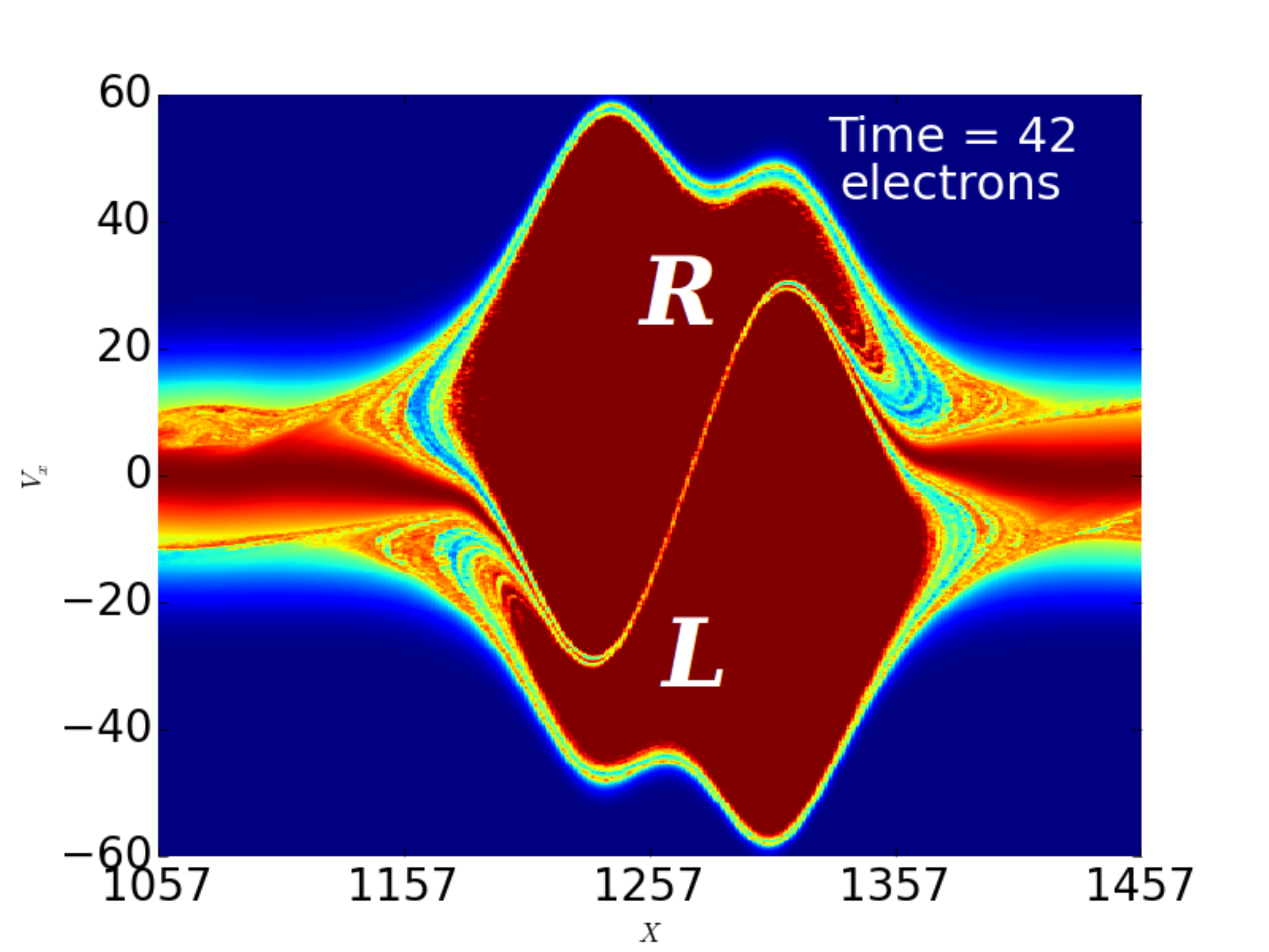} }\\  
 \subfloat{\includegraphics[width=0.24\textwidth]{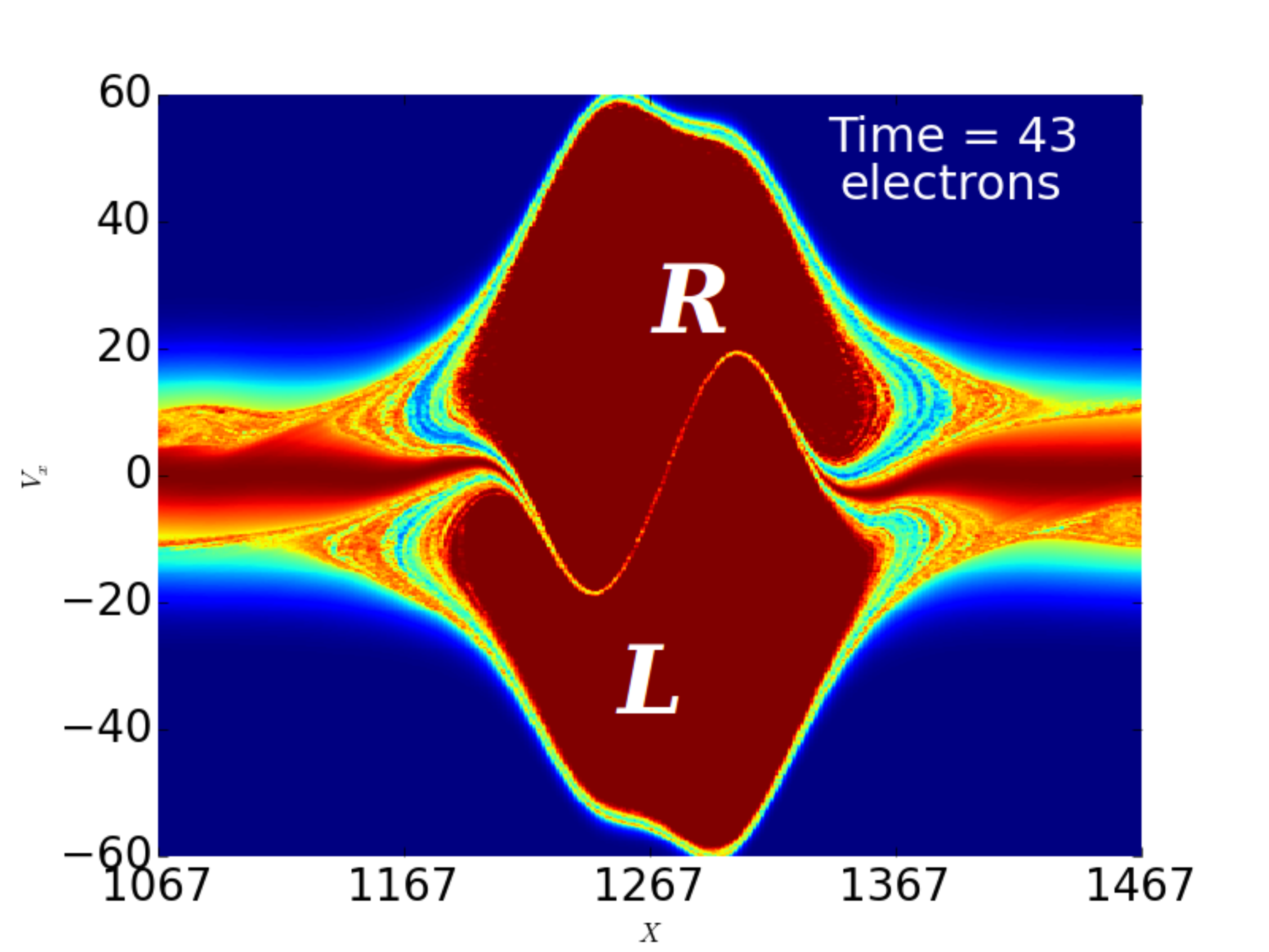} }
 \subfloat{\includegraphics[width=0.24\textwidth]{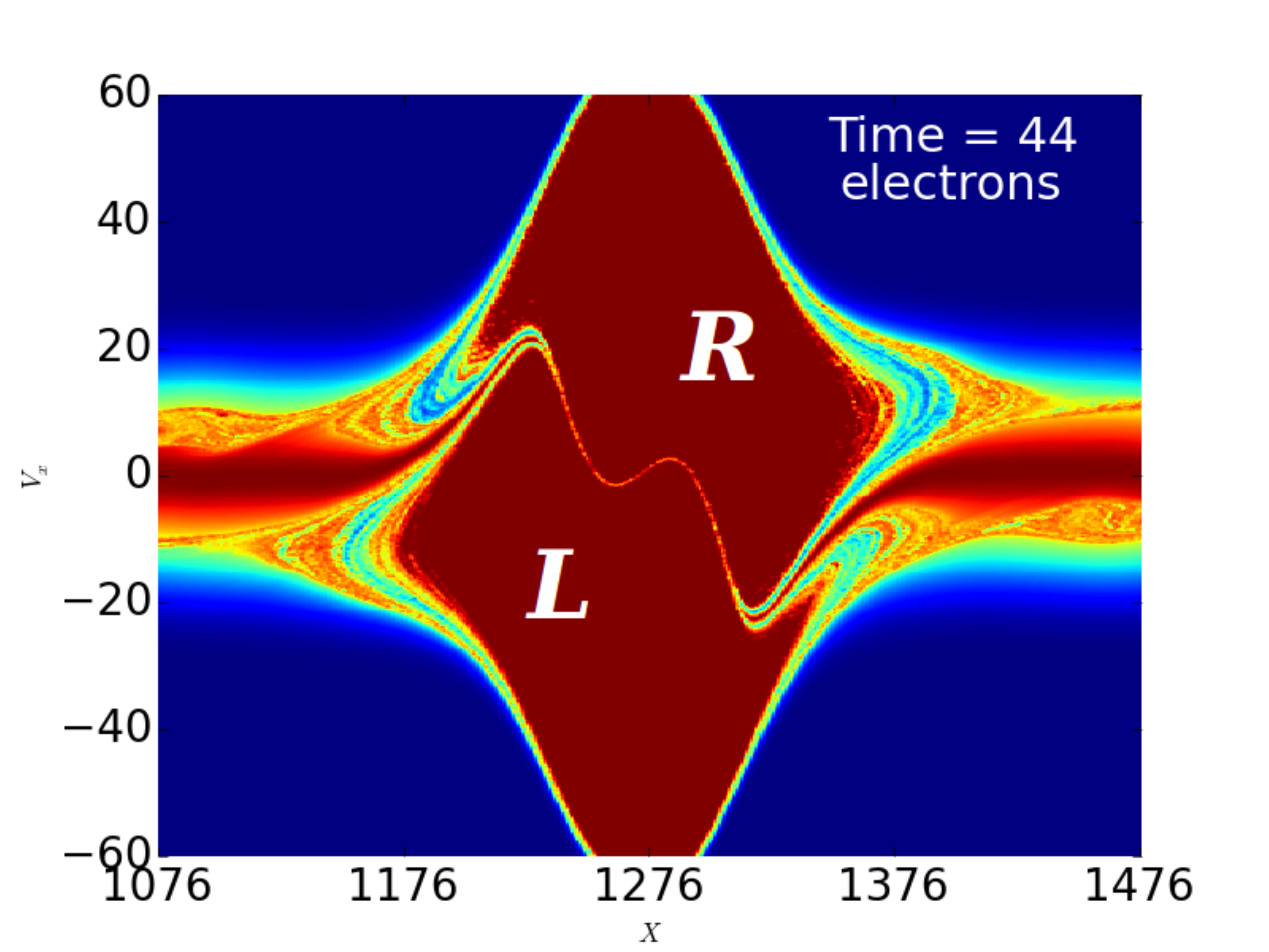} }\\
 \subfloat{\includegraphics[width=0.24\textwidth]{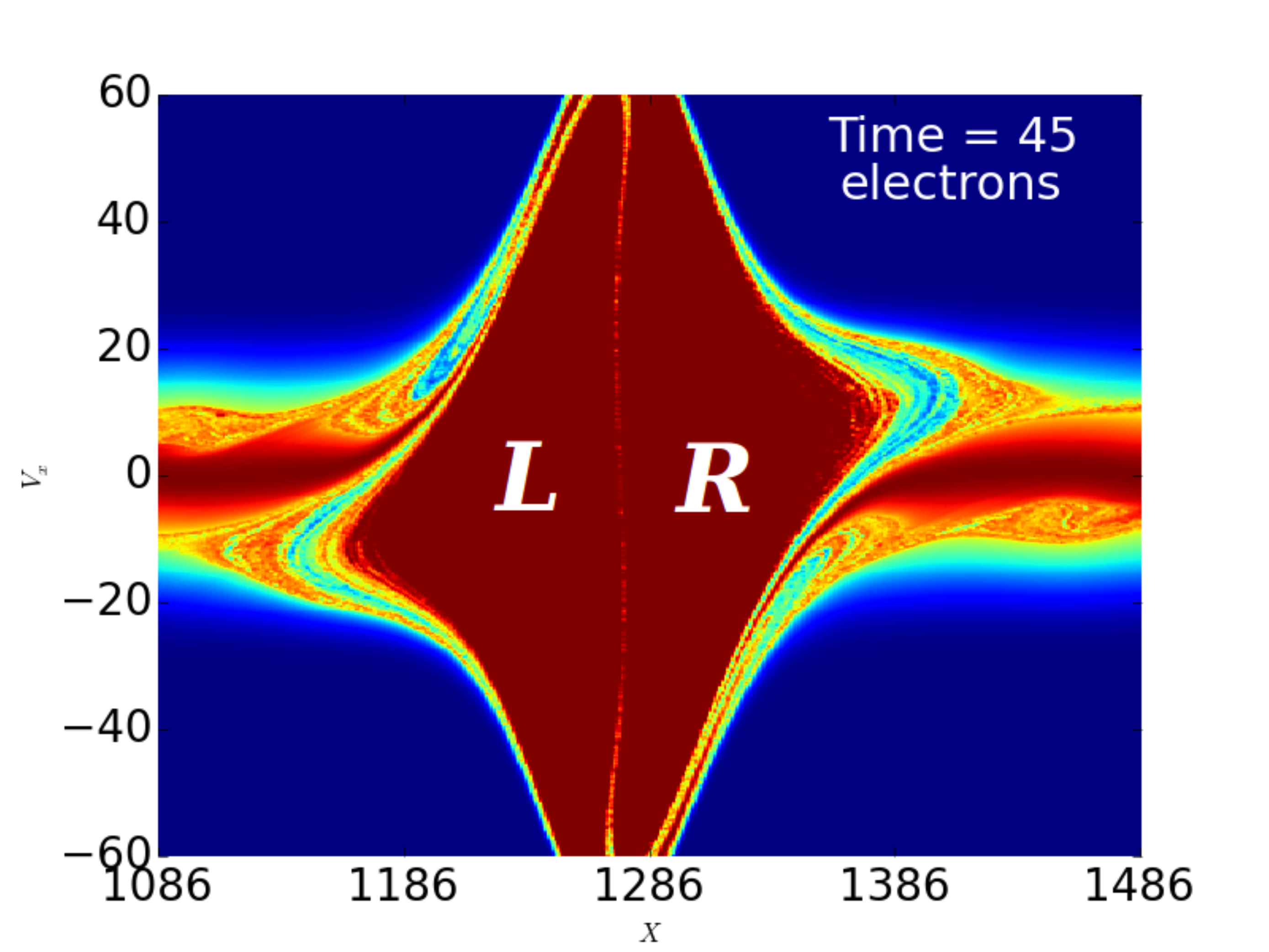} }
 \subfloat{\includegraphics[width=0.24\textwidth]{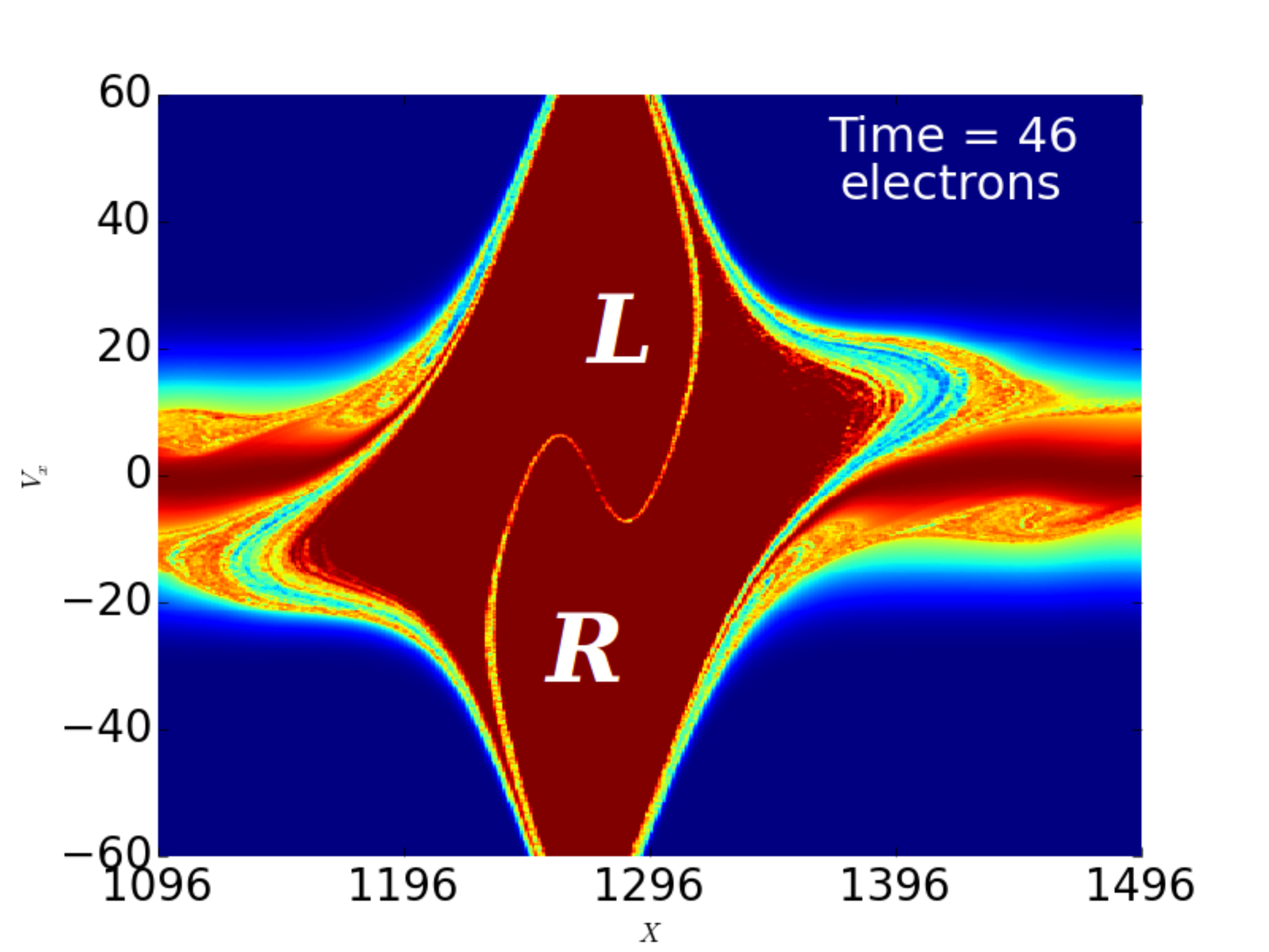} }\\
 \subfloat{\includegraphics[width=0.24\textwidth]{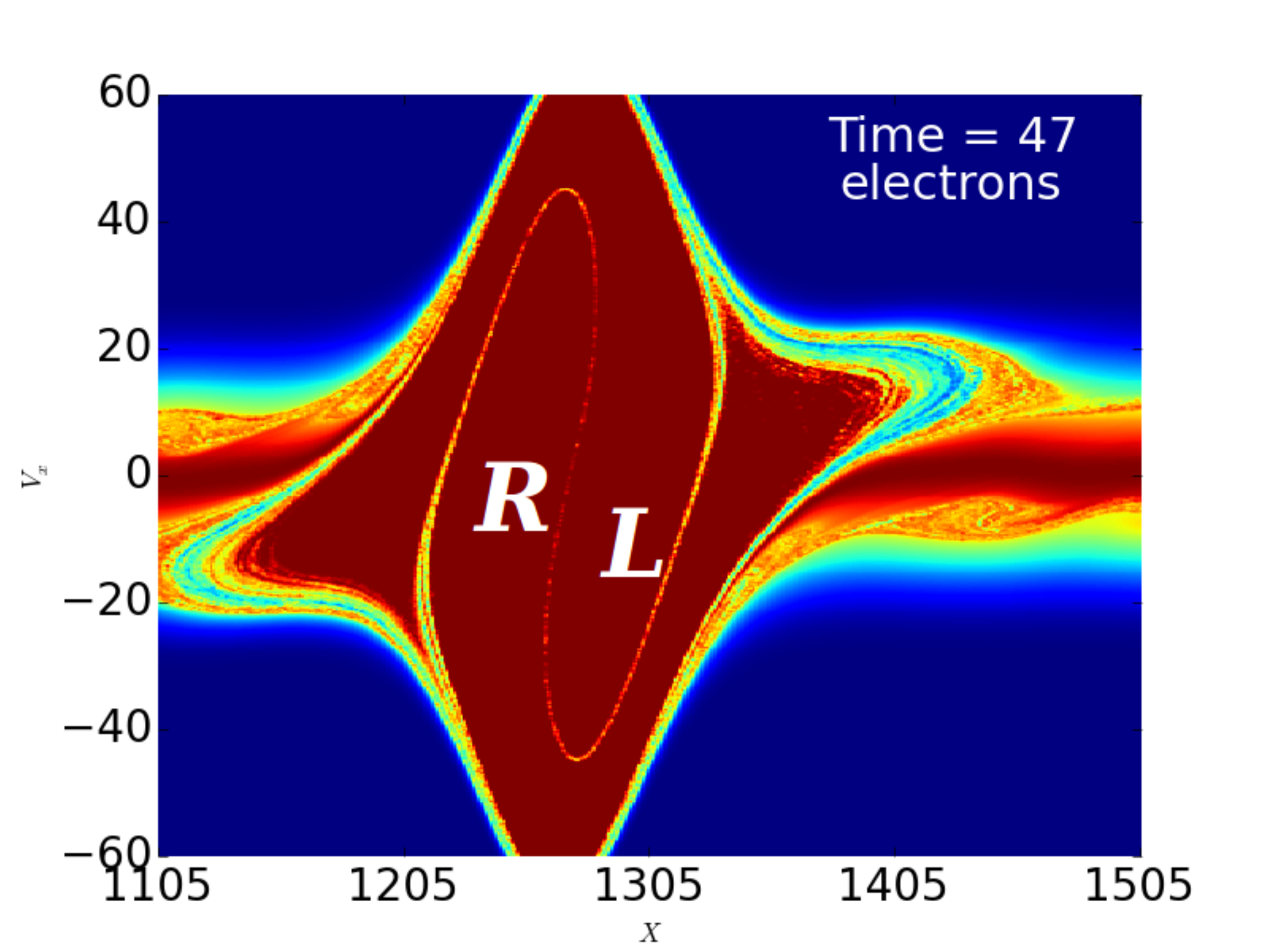} }
 \subfloat{\includegraphics[width=0.24\textwidth]{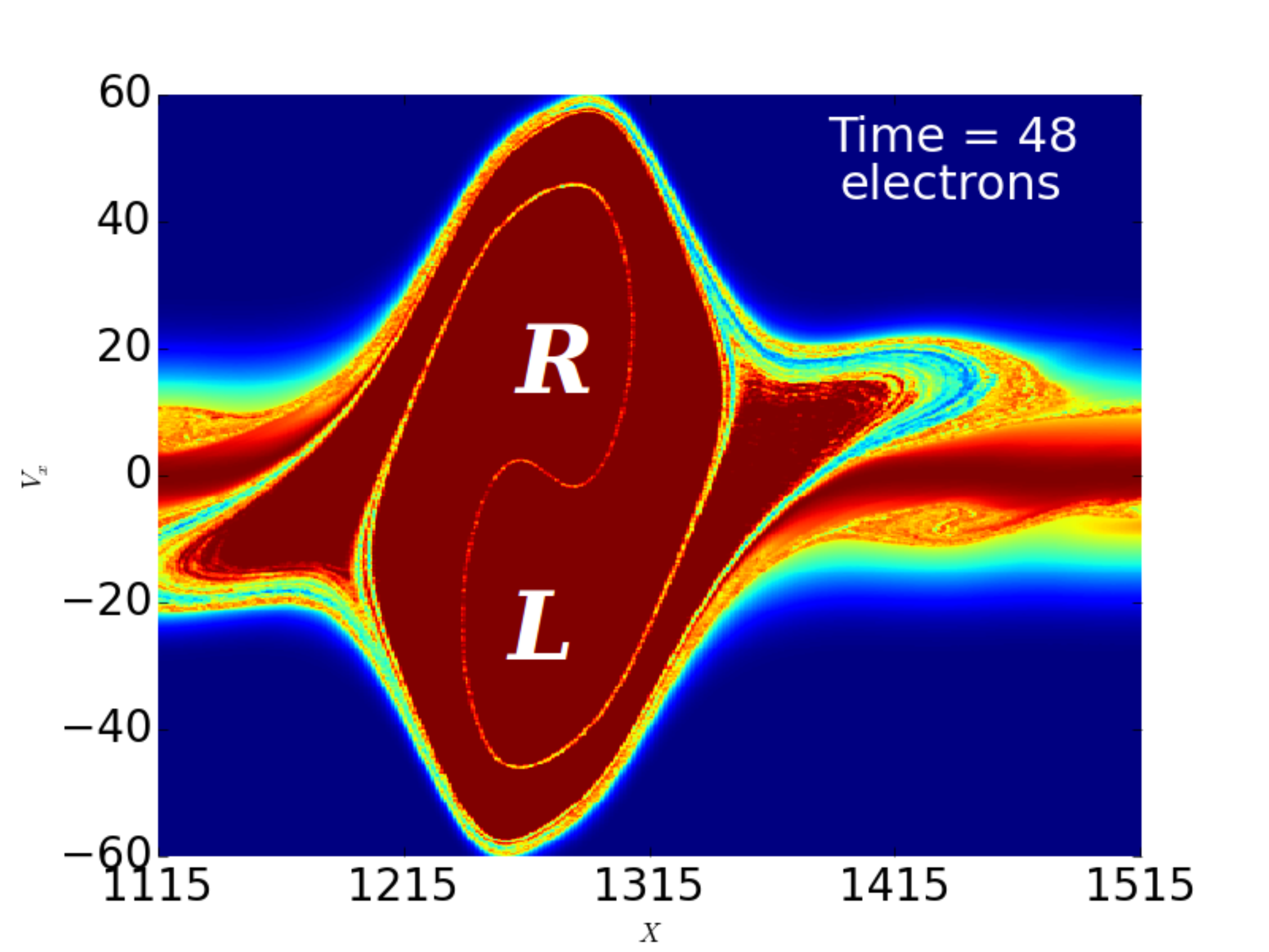} }\\
 \subfloat{\includegraphics[width=0.24\textwidth]{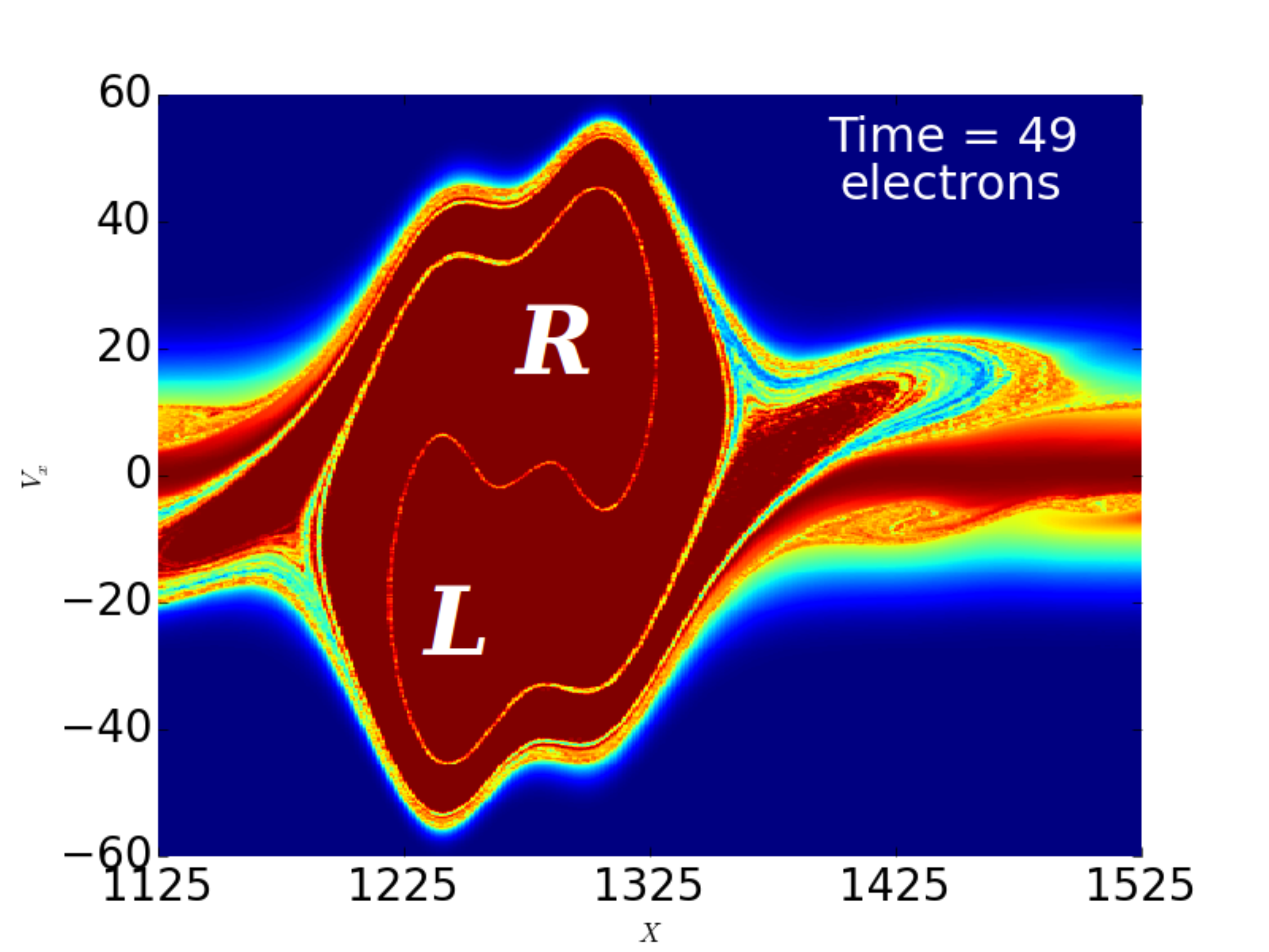} }
 \subfloat{\includegraphics[width=0.24\textwidth]{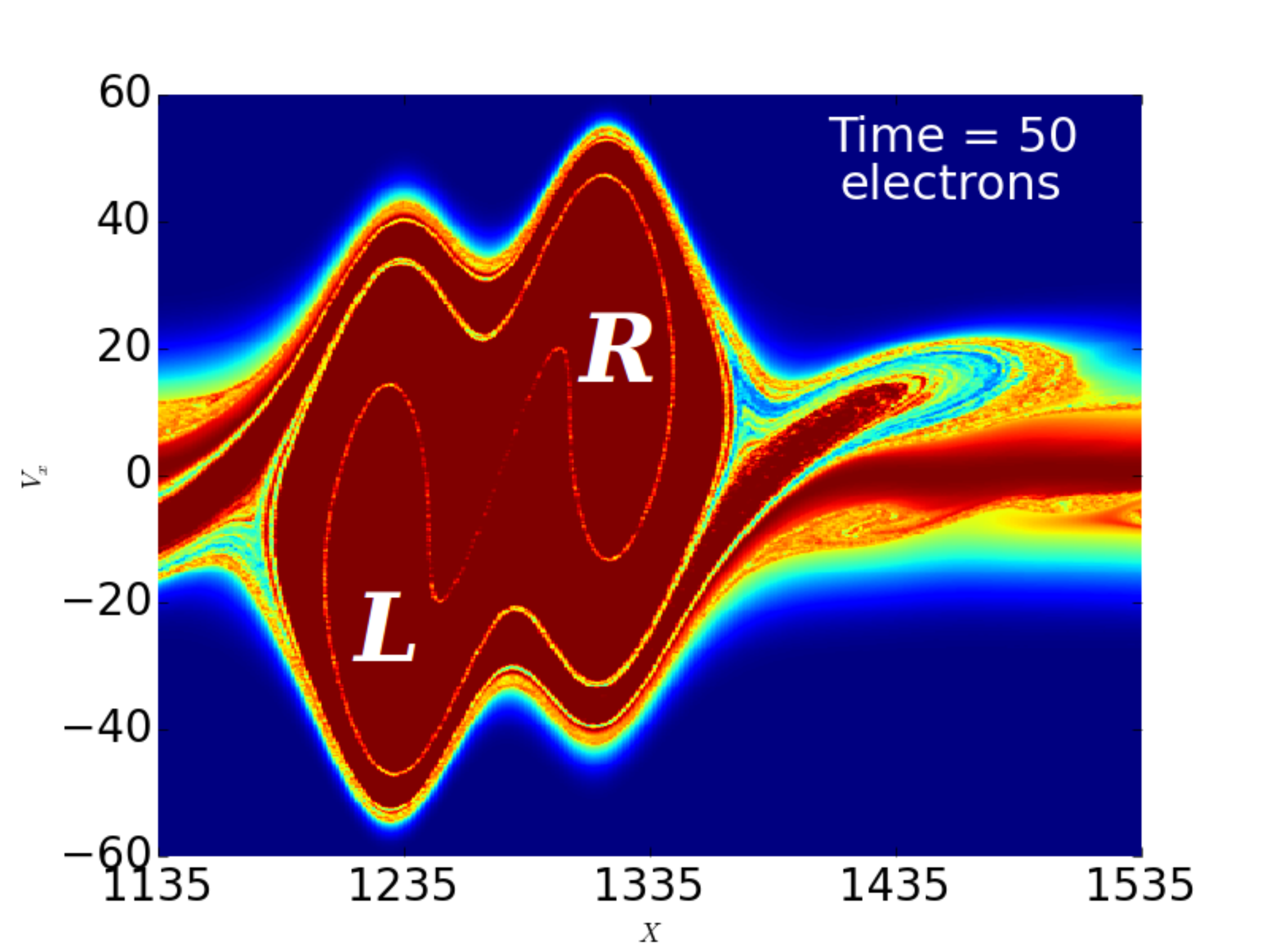} }\\
 \subfloat{\includegraphics[width=0.24\textwidth]{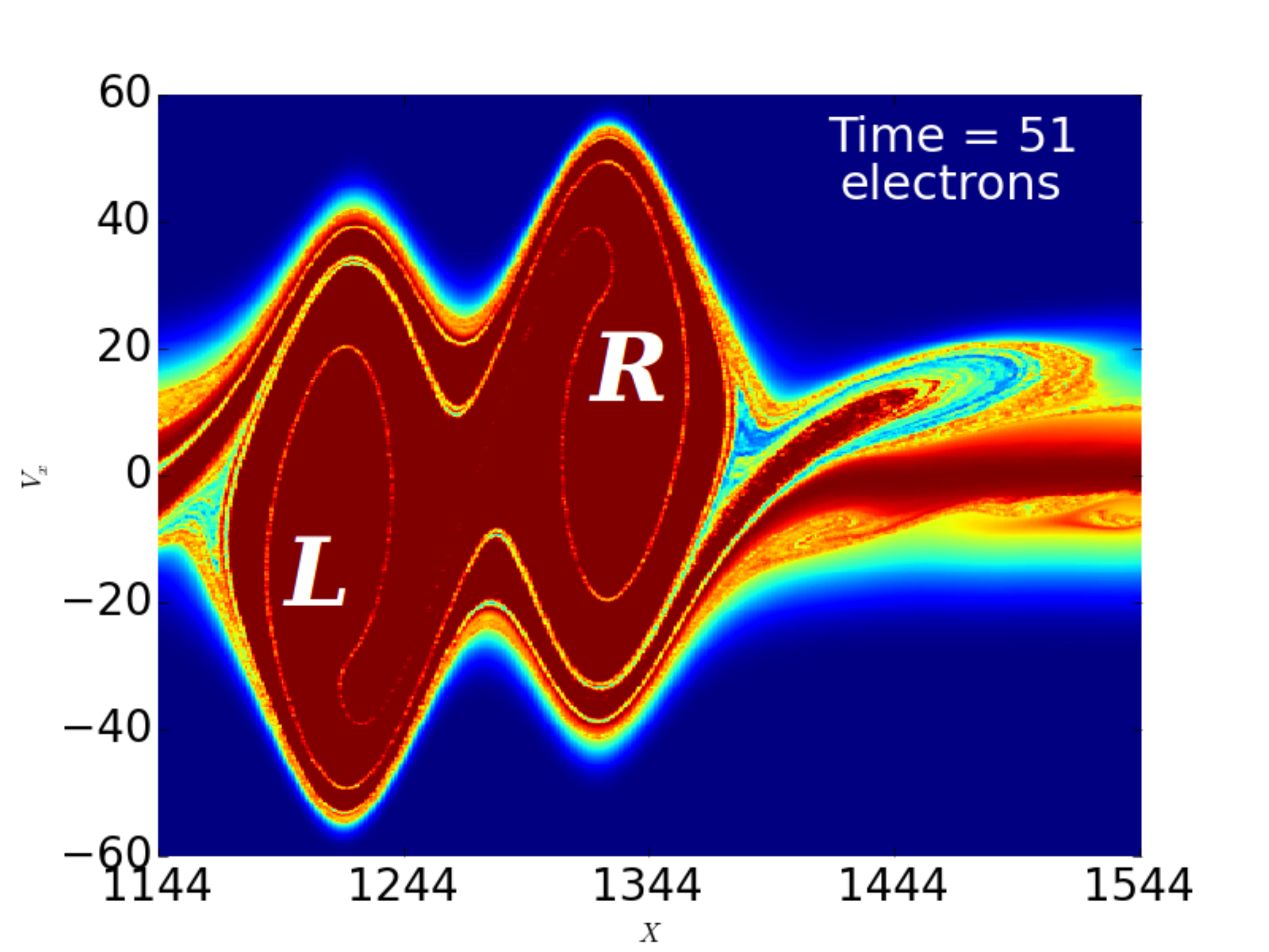} }
 \subfloat{\includegraphics[width=0.24\textwidth]{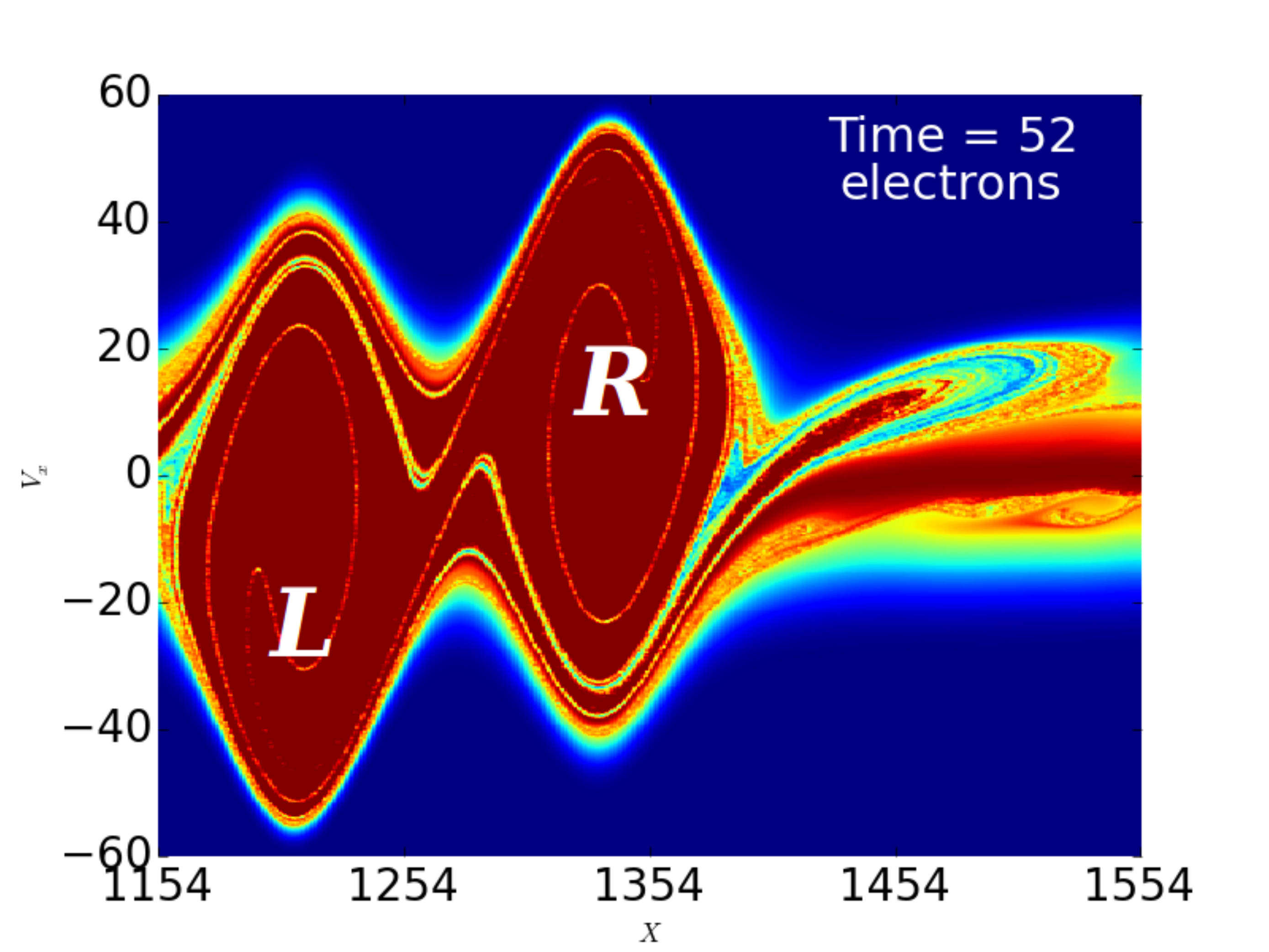} }
 \caption{\textbf{collision of plateaus}; A collision between two IA solitons for the case of $\beta =0.0$ 
 is presented in the phase space of electrons from $\tau = 41$ up to $\tau=52$ (starting from the top left corner).
 The collision takes place between right  (\textbf{R}) and left (\textbf{L}) propagating solitons.
 The frame follows the right-propagating (\textbf{R}) soliton. 
 The collision causes the two plateaus in the phase space to rotate once around their collective center and exchange some 
 parts of their populations.}
 \label{collision_plateau}
\end{figure}

For a \textit{plateau} accompanying the IA solitons,
i.e. $\beta = 0$,
the same pattern has been witnessed during their collisions as well (see Fig.\ref{collision_plateau}).

\begin{figure}
 \subfloat{\includegraphics[width=0.5\textwidth]{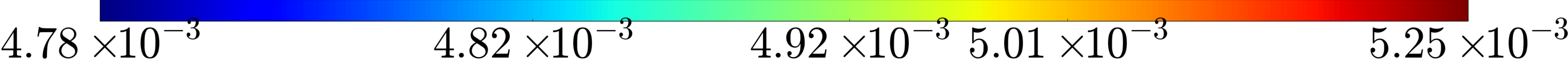} }\\
 \subfloat{\includegraphics[width=0.24\textwidth]{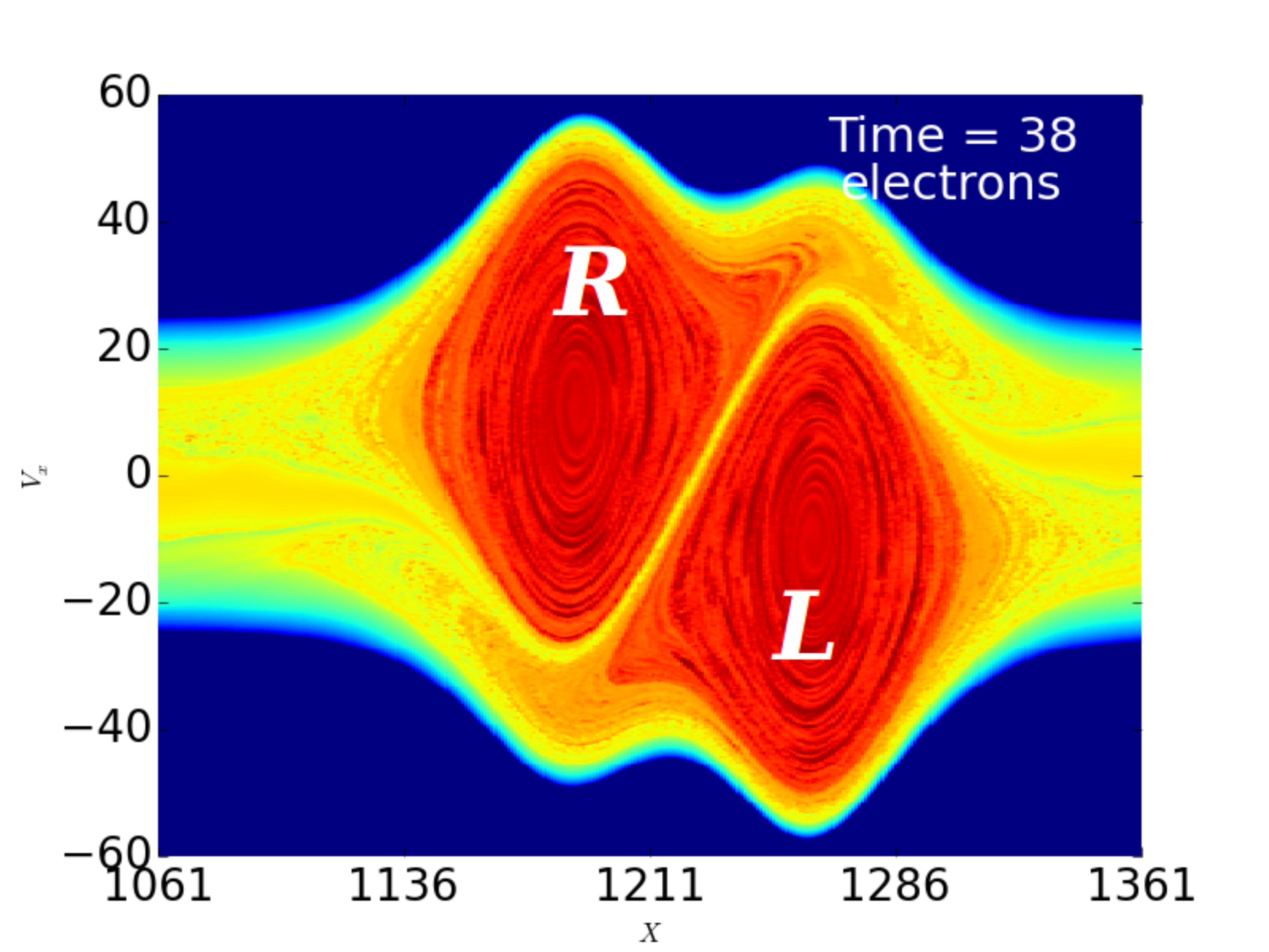} }
 \subfloat{\includegraphics[width=0.24\textwidth]{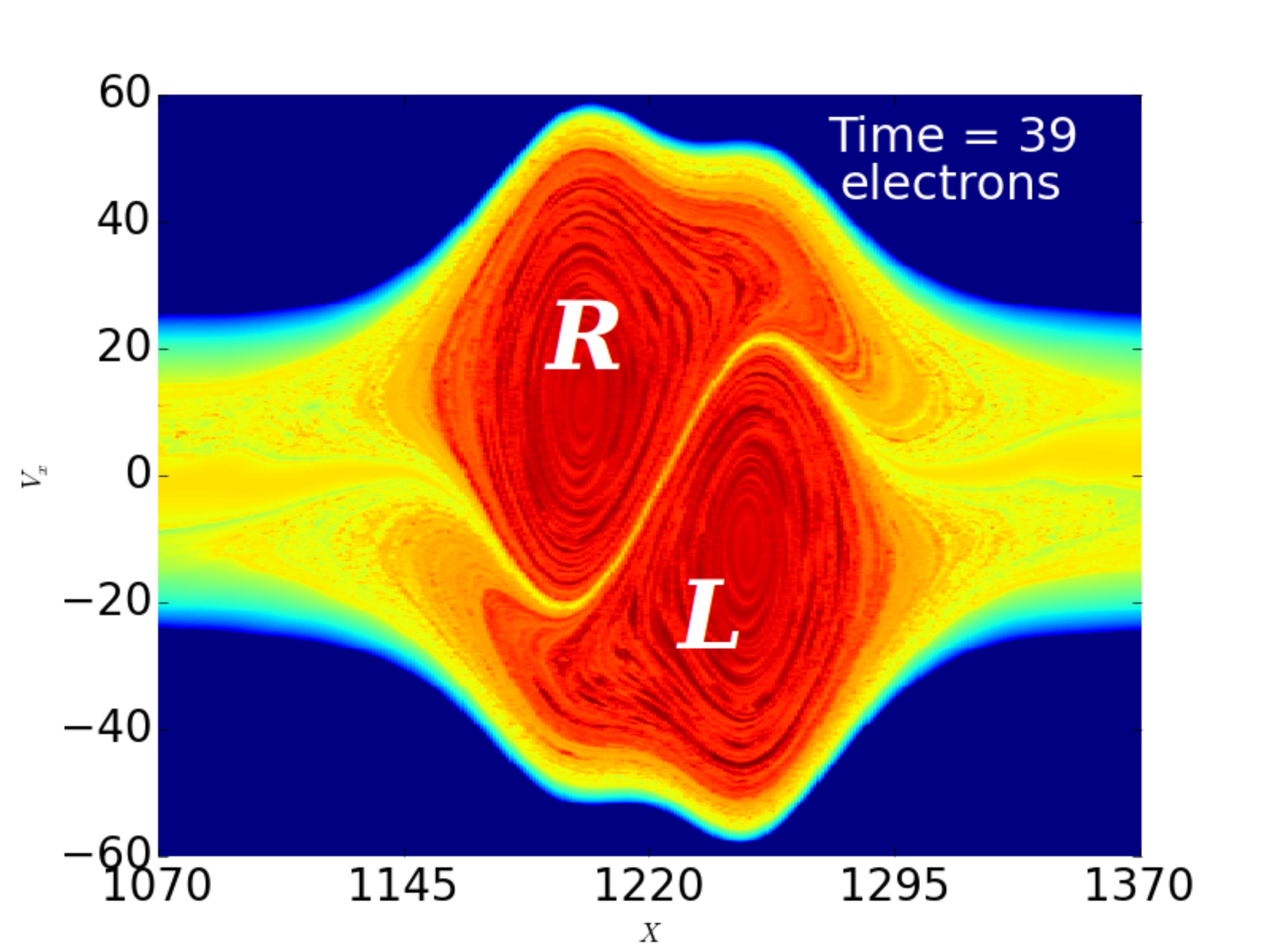} }\\  
 \subfloat{\includegraphics[width=0.24\textwidth]{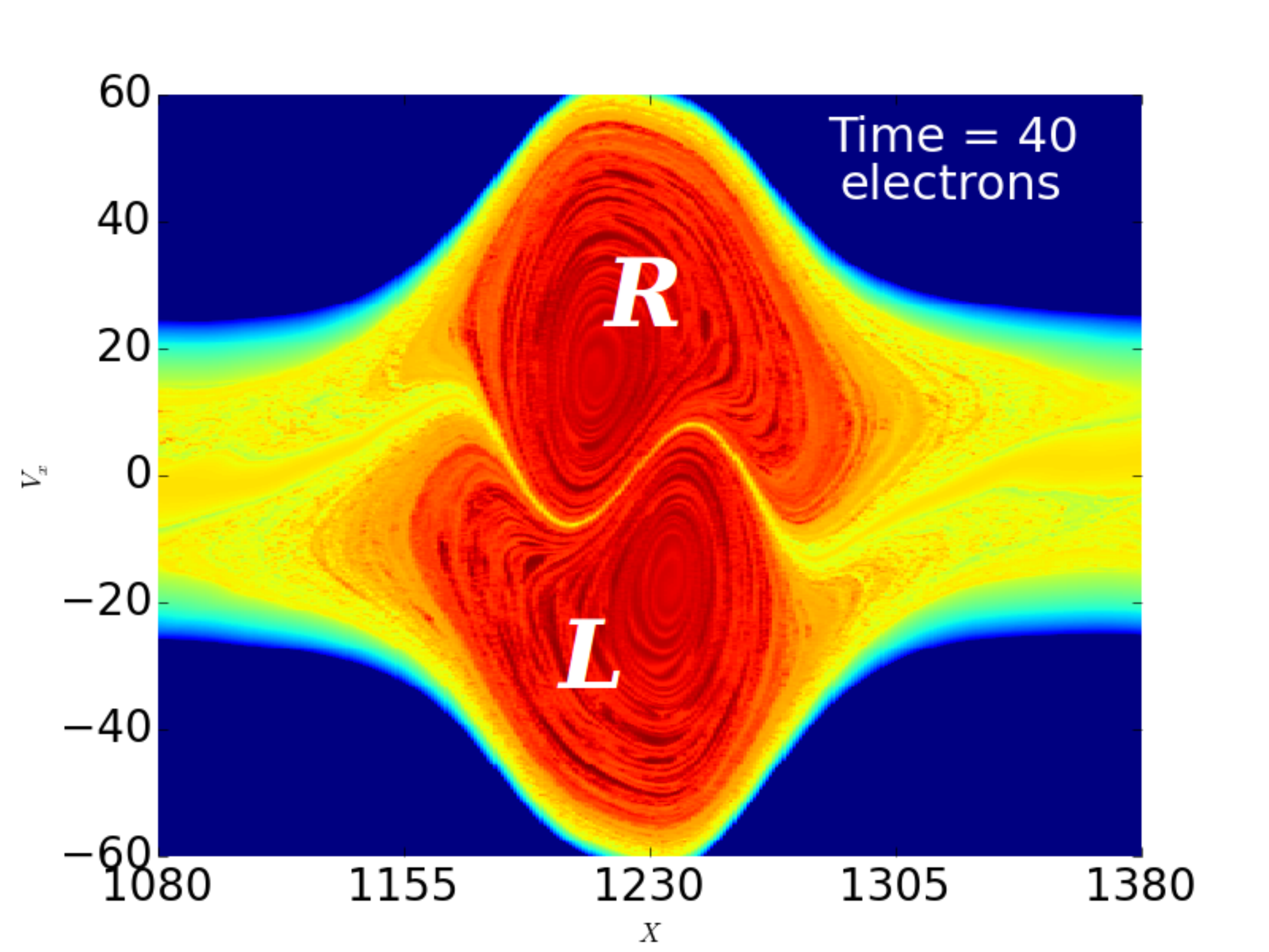} }
 \subfloat{\includegraphics[width=0.24\textwidth]{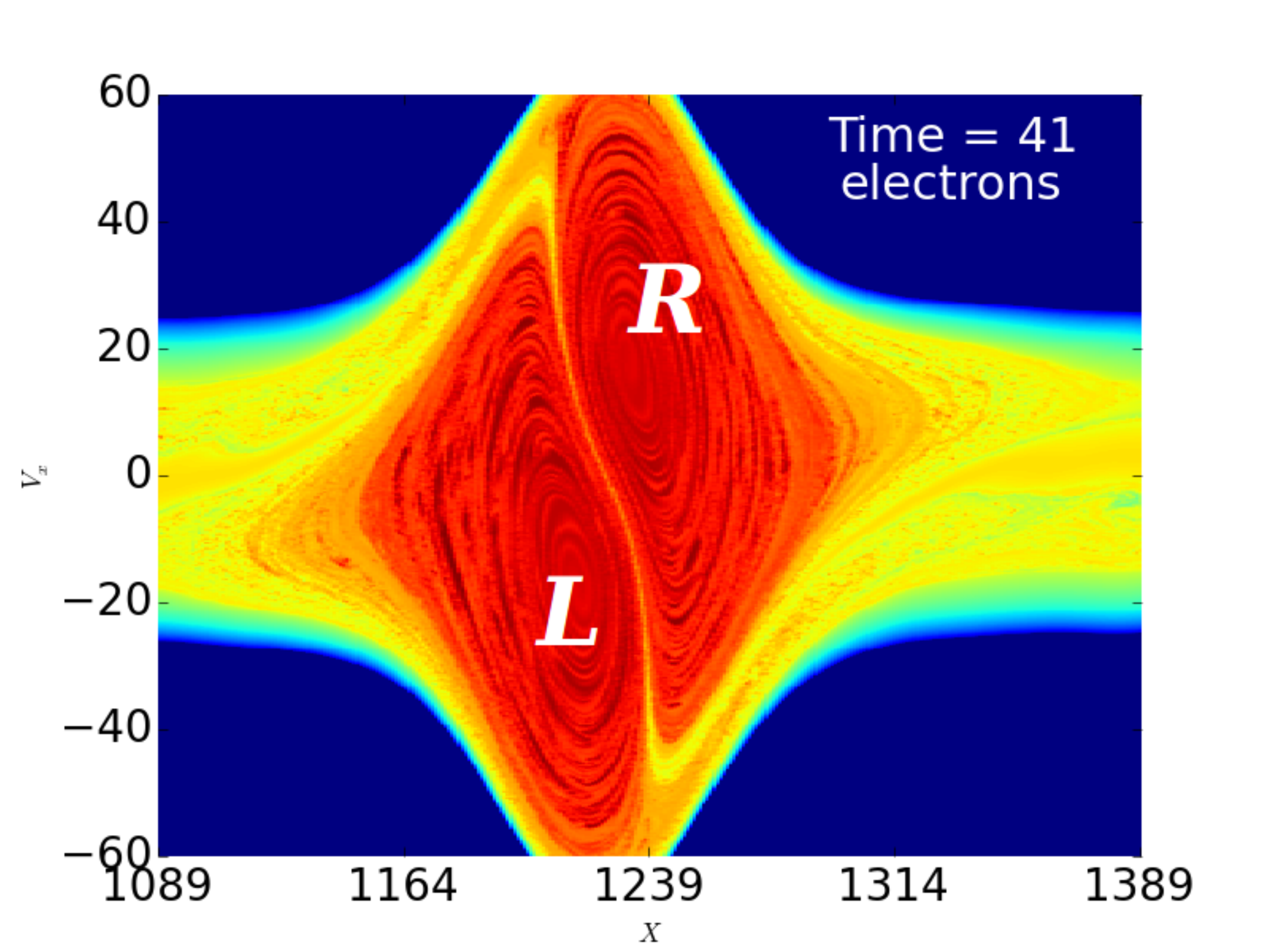} }\\
 \subfloat{\includegraphics[width=0.24\textwidth]{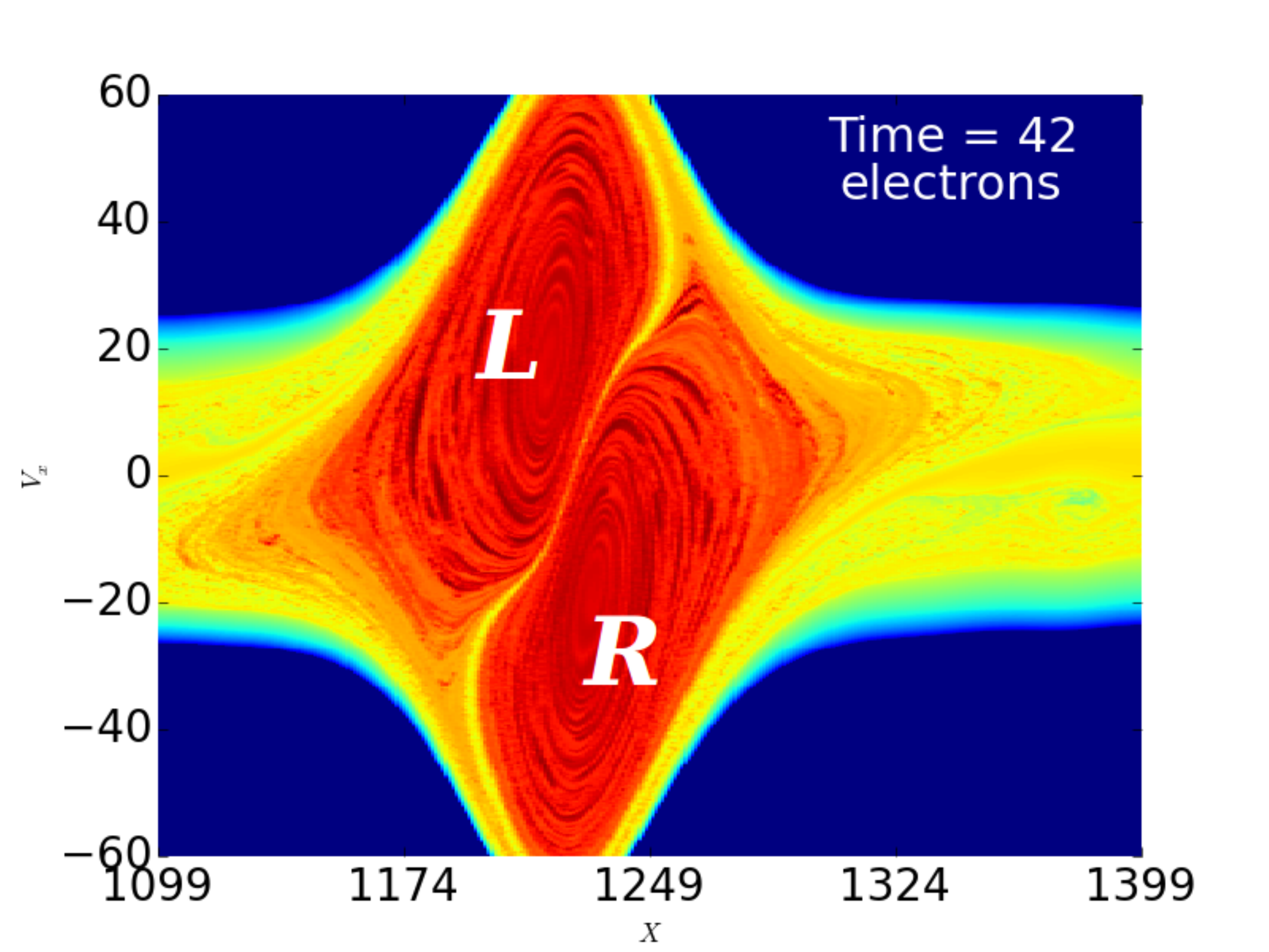} }
 \subfloat{\includegraphics[width=0.24\textwidth]{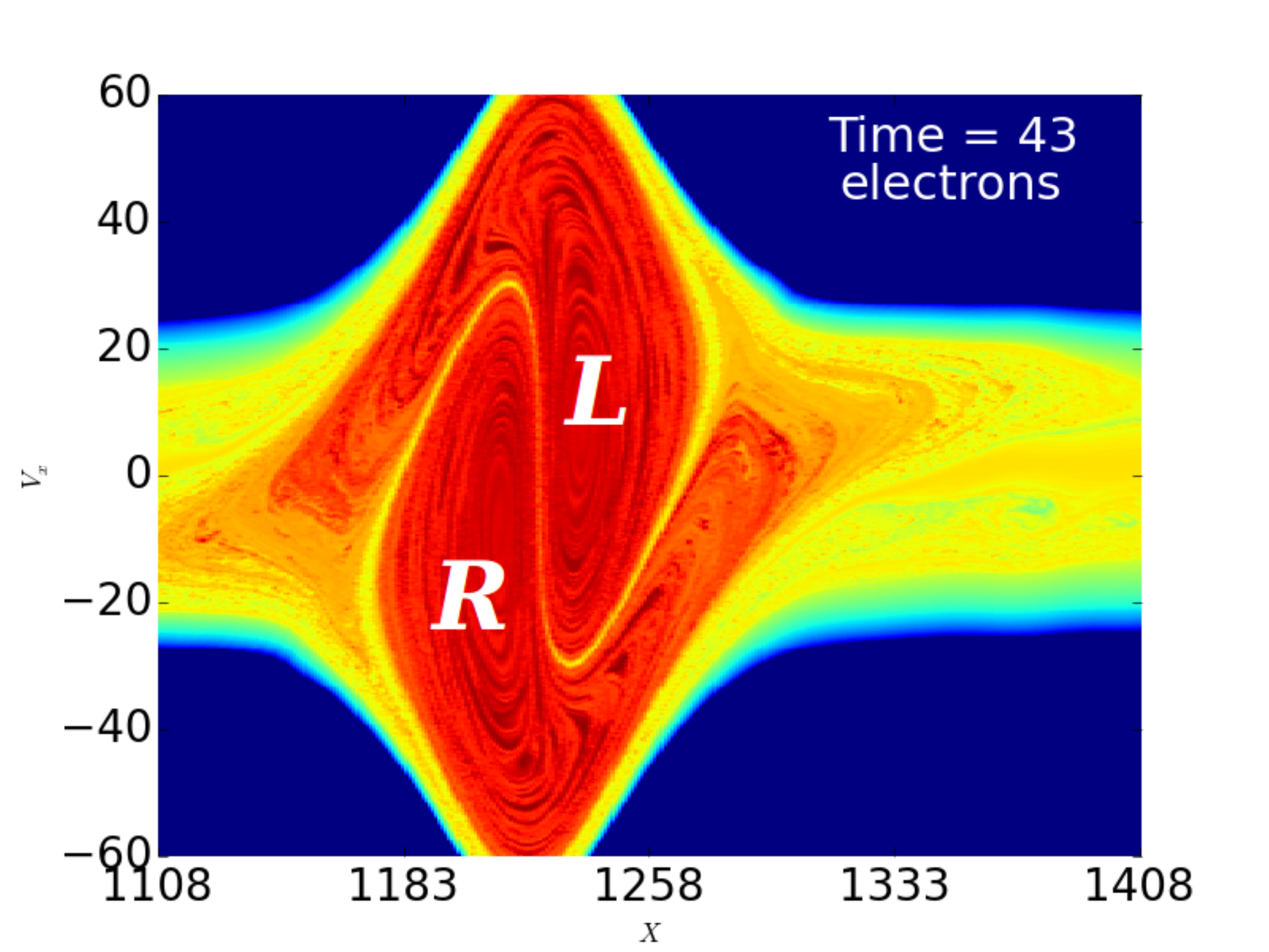} }\\
 \subfloat{\includegraphics[width=0.24\textwidth]{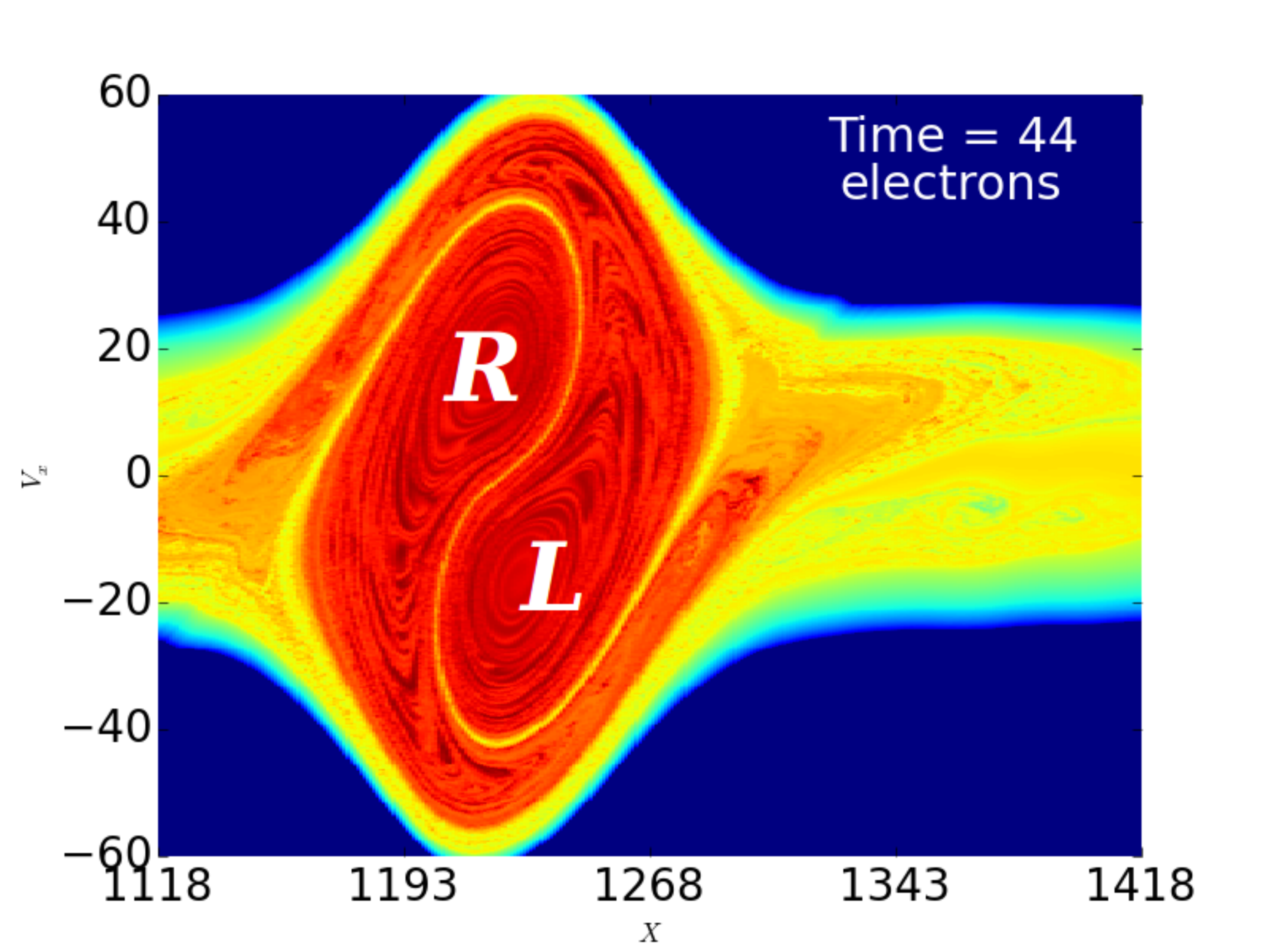} }
 \subfloat{\includegraphics[width=0.24\textwidth]{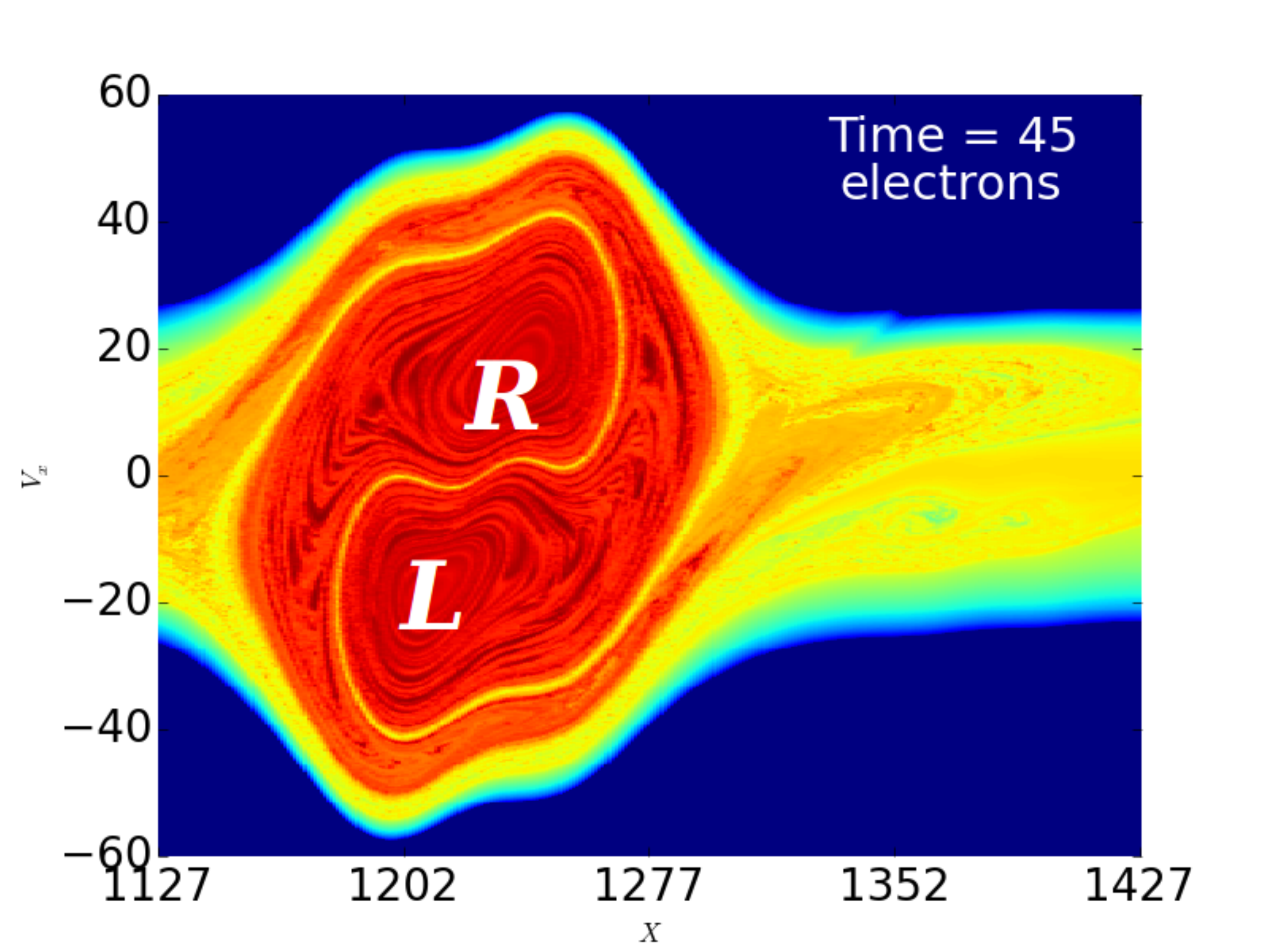} }\\
 \subfloat{\includegraphics[width=0.24\textwidth]{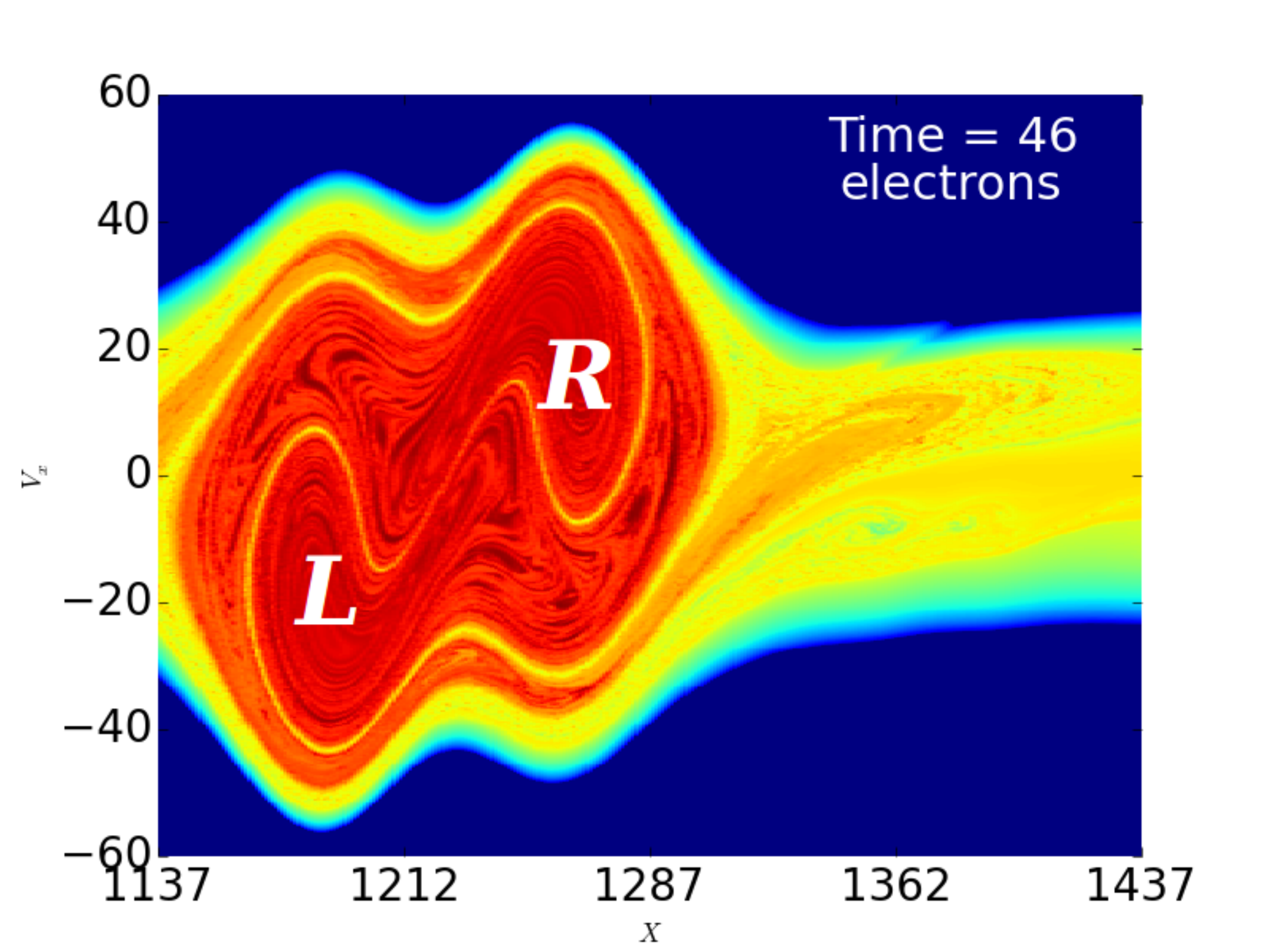} }
 \subfloat{\includegraphics[width=0.24\textwidth]{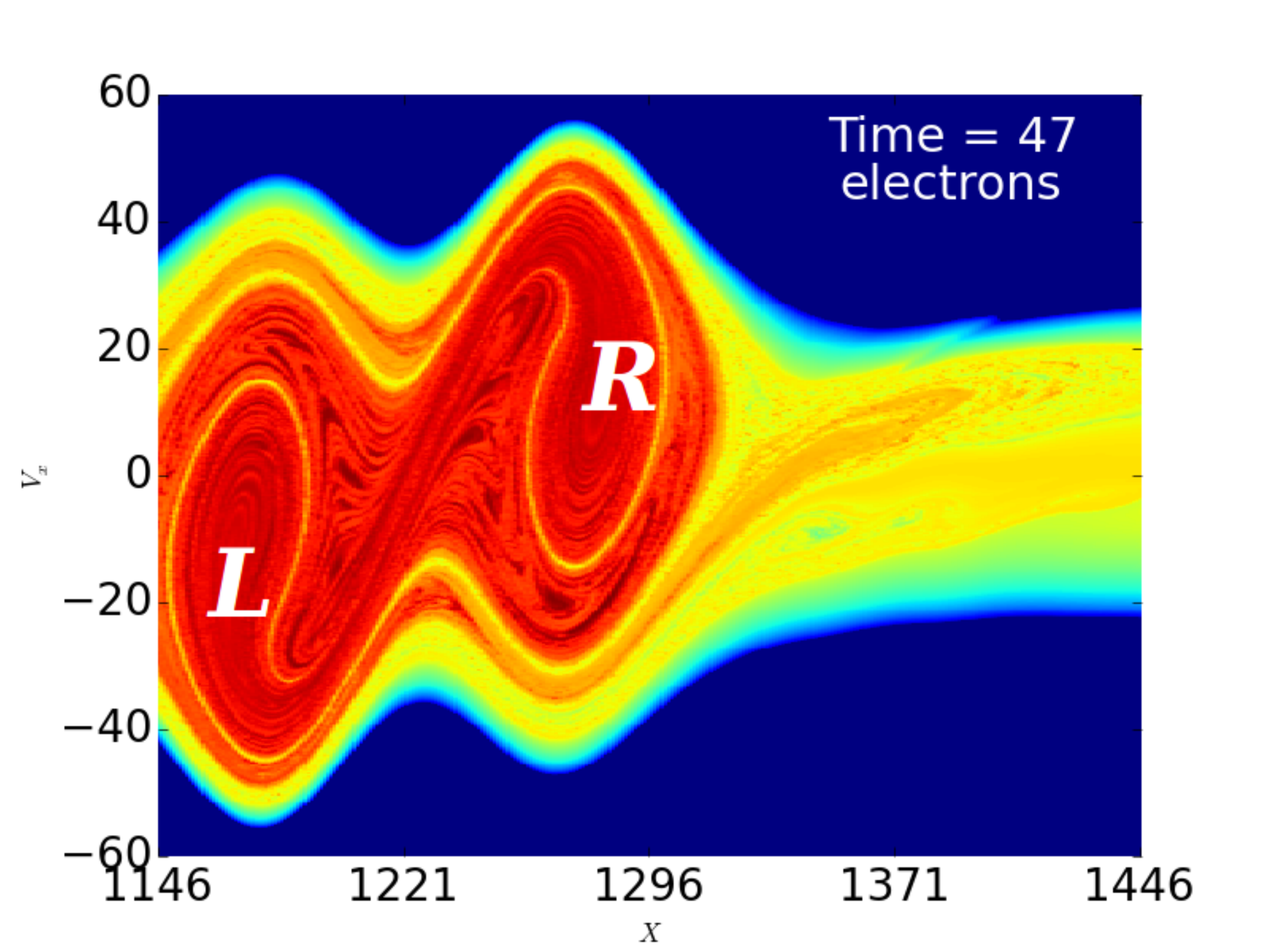} }\\
 \subfloat{\includegraphics[width=0.24\textwidth]{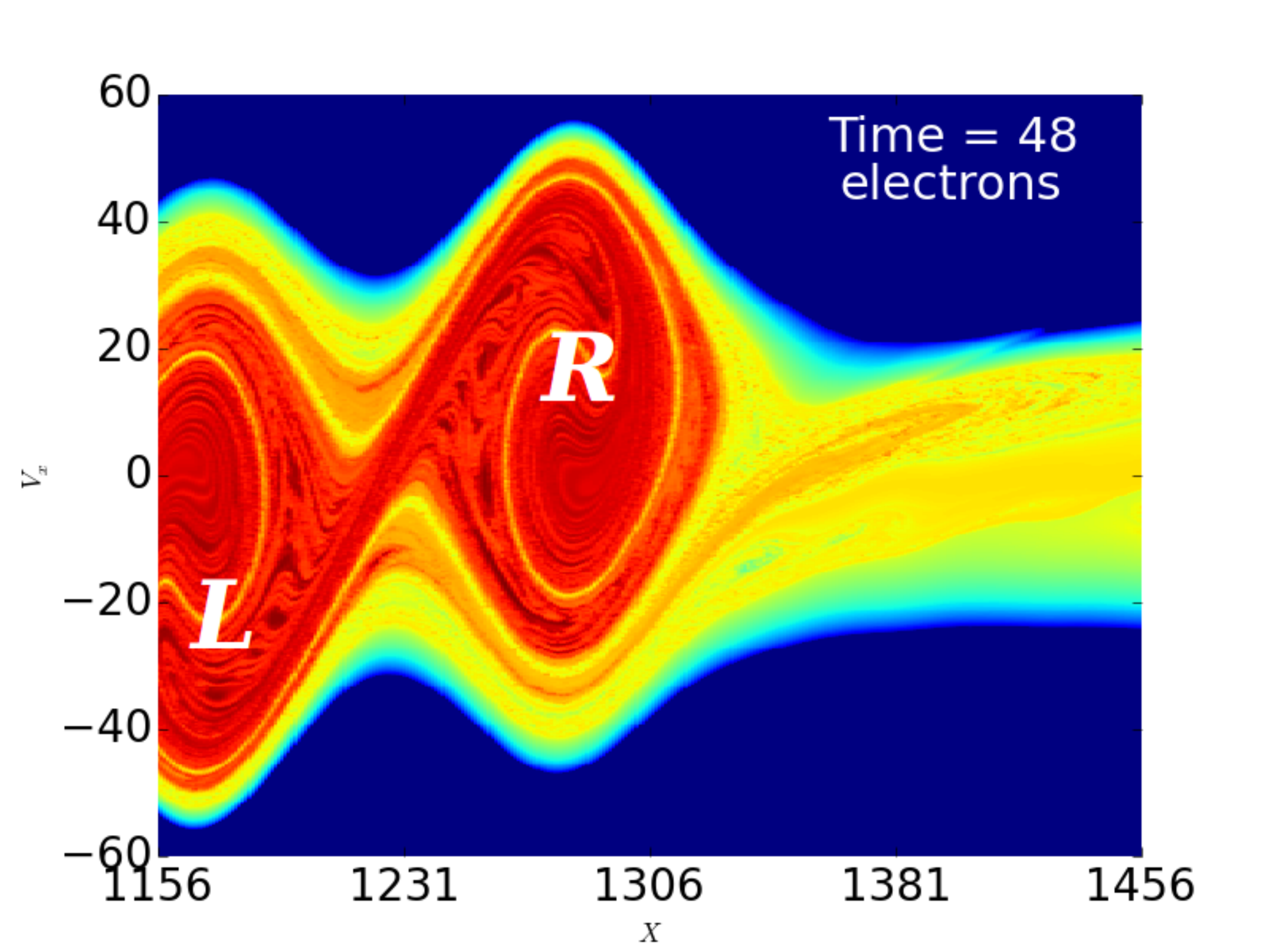} }
 \subfloat{\includegraphics[width=0.24\textwidth]{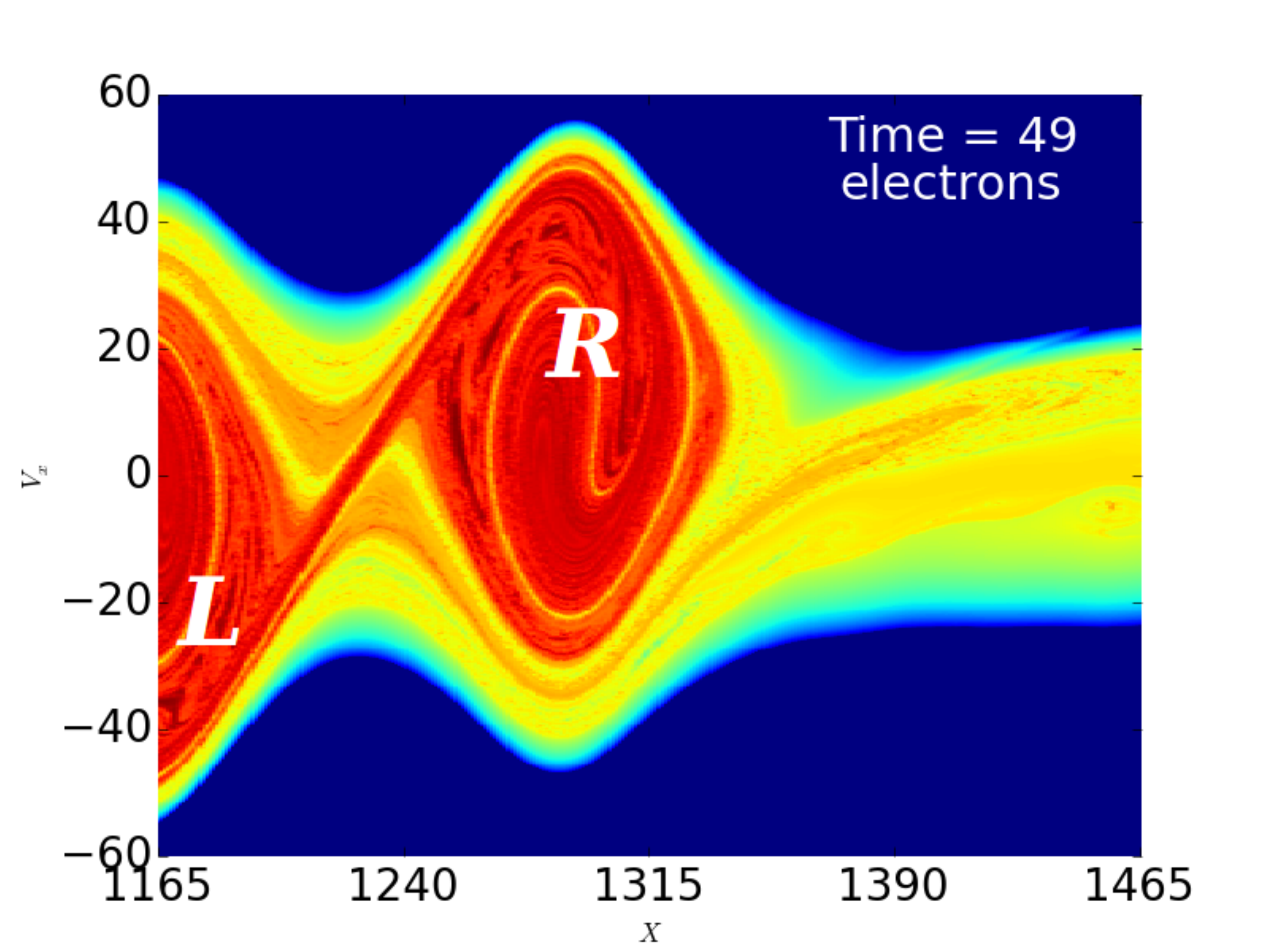} }
  \caption{\textbf{collision of humps}; for the case of $\beta = 0.2$, 
 a collision of two oppositely propagating IA solitons is shown in the phase space of electrons. 
 The collision happens from $\tau = 40$ up to $\tau = 49$ (starting from the top left corner).
 The phase space structure accompanying the IA solitons are humps here, hence the red color.
 One rotation around the collective center and the exchange of trapped population takes place during collision
 between the two humps.}
 \label{collision_hump}
\end{figure}
Fig. \ref{collision_hump} presents the results for the case of positive value of $\beta = 0.2$,
when there is a hump in the electrons' distribution function following IA solitons. 
In cases of the other value of $-1<\beta<1$ (as far as considered here)
the same pattern has been observed during the mutual collisions.
We conclude that this rotational behavior of trapped populations 
stays independent from the value of trapping parameter ($\beta$).

Furthermore, in all the cases shown ($\beta = -0.1, 0, 0.2$) and studied ($-1.0<\beta<1.0$), 
during collision, the trapped populations of the two solitons are exchanged and shared during each collisions. 
This causes the internal structure of the accompanying nonlinear structures 
to change and reemerges more chaotic after each mutual collision (see Figs. \ref{Fig_Internal_BN01} and \ref{Fig_Internal_B0_B02}).
However, the smoothing (which is due to the discretization of the phase space 
on the both spatial and velocity direction)
contributes to this phenomenon as well.

\section{Conclusions} \label{Conclusions}

  A fully kinetic simulation approach is utilized to verify the Schamel's theoretical predictions
concerning ion-acoustic (IA) solitons in the presence of trapped electrons. 
This study confirms the stability of different features of these solitons against mutual collisions.
Hence, this study conclude that kinetic effects such as electron trapping
don't destroy IA solitons at least in the range of $\beta$ (trapping parameter) considered here. 
To the best of our knowledge, this study stands as the first-ever attempt to address this issue
purely based on kinetic theory.

The collisions 
(limited to the encounter of trapped electrons with the same trapping parameter)
have been studied here on two levels, i.e. fluid and kinetic.
On the fluid level, we have established that the IA solitons reemerge from the successive mutual collision intact. 
Four features of them including hight, width, velocity and shape have been focused upon.
The results of the analysis for $\beta = -0.1$ are reported here in which each soliton has experienced 12 mutual collisions. 
The constancy of these characteristic, under $10\%$ fluctuation around the average values, 
implies the stability of IA solitons against mutual collisions.

On the kinetic level, it is presented that the overall shape and 
width of the trapped population accompanying the IA solitons does not change
after a few mutual collisions. 
However, the internal structure of the trapped population changes after each collision
without any traceable impact on the fluid-level characteristics. 
These changes push the phase space structures of the trapped populations to become more chaotic. 
We have noted that for $\beta \neq 0$, more prominent alteration has been witnessed.

Furthermore, the collision process itself, on the kinetic level,
displays a more complicated behavior than what has been observed on the fluid level
i.e., 
two solitons simply passing through each other.
Two main phenomena have been witnessed, 
i.e. rotation of the trapped populations around their collective center, 
and the partial exchange of their trapped populations.
These two procedures are shown to be independent from the value of trapping parameter ($\beta$).
We have carried out simulations in which solitons are isolated from the chain formation process
and have been introduced into a new simulation box in order to remove all secondary effects from collision process.
The results confirm the same pattern of behavior as the first set of simulations.
The exchange of populations affects the internal arrangement of the trapped population and causes the 
alteration which have been reported here. 
Nonetheless, the effect of smoothing contributes to the alteration.

Comparison of these results with the theoretical predictions for the collision of phase-space electron hollows
should reveal the dynamical process behind the rotation during collisions.  
However, such a comparison and study stay beyond the scope of this paper.
It is under consideration and will be communicated elsewhere.
But this much can be mentioned here that the results presented here are limited to moderate-size IA solitons and $-1.0<\beta<1.0$. 
In case of high-amplitude IA solitons, the collision of IA soliton might be affected by the kinetic effects
more strongly and 
the number of rotations (here equals one) might change.

\acknowledgments
 We are grateful to the anonymous referees for their helpful comments and constructive hints
  which improved the paper extensively.
  This work is based upon research supported by the National Research Foundation 
and Department of Science and Technology.
Any opinion, findings and conclusions or recommendations expressed in this 
material are those of the authors and therefore the NRF and DST do not accept 
any liability in regard thereto.


%

\end{document}